%% file: LineBootstrap.tex
\RequirePackage{pdf14}

\documentclass[letterpaper,10pt]{article}
\pdfoutput=1
\usepackage[utf8]{inputenc}
\usepackage[usenames,dvipsnames,table]{xcolor}
\usepackage{graphicx}
\usepackage{caption}
\usepackage{subcaption}
\usepackage{array,setspace,mathrsfs,amsthm,amsfonts,amsmath,colonequals,mathtools,bbm}
\usepackage{./auxi/jheppub}
\usepackage{afterpage}
\usepackage{xspace}
\usepackage[normalem]{ulem}
\usepackage{cancel}
\usepackage{cleveref}
\usepackage{hyperref}

\input{./auxi/topmatter}

\title{Bootstrapping the half-BPS line defect}

\author[1]{Pedro Liendo,}
\author[2]{Carlo Meneghelli,}
\author[3]{Vladimir Mitev}

\affiliation[1]{DESY Hamburg, Theory Group, Notkestrasse 85, D–22607 Hamburg, Germany}
\affiliation[2]{Mathematical Institute, University of Oxford, Woodstock Road, Oxford, OX2 6GG, UK}
\affiliation[3]{PRISMA Cluster of Excellence, Institut f\"ur Physik,
JGU Mainz,
Staudingerweg 7, 55128 Mainz, Germany}

\emailAdd{pedro.liendo@desy.de}
\emailAdd{carlo.meneghelli@maths.ox.ac.uk}
\emailAdd{vmitev@uni-mainz.de}

\preprint{DESY 18-087, MITP/18-043}

\bigskip
\abstract{
We use modern bootstrap techniques to study half-BPS line defects in  $4d$ $\mathcal{N}=4$ superconformal theories. 
Specifically, we consider the $1d$ CFT with $\text{OSP}(4^*|4)$  superconformal symmetry living on such a defect. 
Our analysis is general and based only on symmetries, it includes however important examples like Wilson and 't Hooft lines in $\mathcal{N}=4$ super Yang-Mills.
We present several numerical bounds on OPE coefficients and conformal dimensions. Of particular interest is a numerical island obtained from a mixed correlator bootstrap that seems to imply a unique solution to crossing. The island is obtained if some assumptions about the spectrum are made, 
and is consistent with Wilson lines in planar $\mathcal{N}=4$ super Yang-Mills at strong coupling. 
We further analyze the vicinity of the strong-coupling point by calculating perturbative corrections using analytic methods.
This perturbative solution 
has the sparsest spectrum and is expected to saturate the numerical bounds, explaining some of the features of our numerical results.
}


\begin{document}
\setcounter{tocdepth}{2}
\maketitle
\setcounter{page}{1}


\input{sections/intro}
\input{sections/preliminaries}
\input{sections/gauge_theory}

\input{sections/crossing}
\input{sections/numerical_results}
\input{sections/analytical_results}

\input{sections/conclusions}

\input{sections/acknowledgements}


\appendix

\input{sections/appendix}


\bibliography{./auxi/biblio}
\bibliographystyle{./auxi/JHEP}

\end{document}

%% file: auxi/topmatter.tex
\newcommand{\beqa}{\begin{eqnarray}}
\newcommand{\eeqa}{\end{eqnarray}}
\newcommand{\beq}{\begin{equation}}
\newcommand{\eeq}{\end{equation}}

\newcommand{\vac}[1]{\ensuremath{\left< \, #1\, \right>}}

\newcommand{\CB}{g^{1d}}
\newcommand{\block}{\mathcal{G}}

\newcommand{\calW}{\mathcal{W}}
\newcommand{\calA}{\mathcal{A}}
\newcommand{\calN}{\mathcal{N}}
\newcommand{\calD}{\mathcal{D}}
\newcommand{\calO}{\mathcal{O}}
\newcommand{\rep}{\mathcal{R}}
\newcommand{\RR}{\text{R}}
\newcommand{\BS}{\mathbb{B}}
\newcommand{\SU}{\text{SU}}

\newcommand{\sX}{\mathfrak{X}}
\newcommand{\sXt}{\tilde{\mathfrak{X}}}

\renewcommand{\b}{\beta}

\makeatletter
\def\maketag@@@#1{\hbox{\m@th\normalfont\normalsize#1}}
\makeatother

\def\Am{{\mathcal{A}}}

\def\Nm{{\mathcal{N}}}

\def\Dm{{\mathcal{D}}}

\def\b{{\beta}}

\def\nn{\nonumber}

%% file: sections/intro.tex

\section{Introduction}

Since the revival of the bootstrap program \cite{Rattazzi:2008pe}, our understanding of conformal field theory (CFT) dynamics has improved significantly.  
There has been a huge amount of progress, numerical as well as analytical, that has severely constrained the landscape of CFTs, and has also brought us closer to solving individual models (see \cite{Poland:2018epd} for a recent comprehensive review). However, most of the works during this first decade of modern bootstrap research have focused on correlation functions of local operators, in particular by using the crossing symmetry of the four-point functions.

Extended objects, or \textit{defects}, are an important class of observables in CFT that can also be studied using the bootstrap approach. In the presence of defects, the information associated to a system is enlarged: it includes the standard data associated to bulk quantities, but also data associated to the defect itself and to the interaction between the defect and the bulk. This new data is otherwise inaccessible if one is restricted to bulk correlation functions. Hence, the addition of defects and the study of their interplay with the bulk is necessary if we want a complete understanding of the dynamics of a theory.

The majority of the defect bootstrap studies done so far usually consider local bulk operators in the presence of a defect. 
The conformal blocks for correlation functions in the presence of a boundary, i.e.~a defect of codimension one, were obtained in \cite{McAvity:1995zd}, and in \cite{Billo:2016cpy,Lauria:2017wav} (see also \cite{Guha:2018snh}) for defects of higher codimension. 
Here we should point out that in this setup the crossing equations in general lack a certain positivity property which is necessary for the numerical bootstrap of \cite{Rattazzi:2008pe}. There is an alternative (underexplored) approach by Gliozzi that does not require positivity \cite{Gliozzi:2013ysa}, and is therefore better suited for the defect bootstrap. Both approaches have been used in the context of boundary CFTs and have given reasonable results \cite{Liendo:2012hy,Gliozzi:2015qsa,Gliozzi:2016cmg}.

In the case of defects of codimension higher than one, it is possible to extract analytic information from the crossing equations. The analysis mimics what is called the ``analytic'' or ``lightcone'' bootstrap 
\cite{Komargodski:2012ek,Fitzpatrick:2012yx,Alday:2015eya,Alday:2016njk,Caron-Huot:2017vep}, which studies the spectrum of CFTs in the limit of large spin. Indeed, as shown in \cite{Lemos:2017vnx}, defect CFTs exhibit similar universal behavior at large \textit{transverse spin}, i.e.~the quantum number associated to rotations around the defect. Note that this quantum number does not exist in the case of boundaries. Other related analytical approaches to defects include Mellin space \cite{Rastelli:2017ecj,Goncalves:2018fwx} and ``alpha space'' \cite{Hogervorst:2017kbj}. We should also mention that, as opposed to local operators in the presence of a defect, one can also study correlation functions of the defects themselves. Works in this direction include \cite{Gadde:2016fbj,Fukuda:2017cup,Kobayashi:2018okw}.

In this article, we will consider operators in a $4d$ $\calN=4$ CFT that are constrained to live on a  supersymmetric line defect. A similar setup without supersymmetry is the monodromy defect of the $3d$ Ising model \cite{Gaiotto:2013nva} (see also \cite{Mazac:2016qev,Qiao:2017xif,Mazac:2018mdx} for recent analytic progress on the $1d$ bootstrap).
Even though the theory living on the defect is a nonlocal CFT,
 as signaled by the absence of a stress tensor, it is possible to write a conformal block expansion and a corresponding crossing equation. This setup also has the added advantage that the non-positivity caveat can be overcome, and the techniques of \cite{Rattazzi:2008pe} can be applied. 

The bootstrap program for supersymmetric defects was initiated in \cite{Liendo:2016ymz}, where a detailed analysis of $\text{OSP}(4^*|4)$ preserving defects was presented, which includes boundaries, interfaces, and line defects. The results of \cite{Liendo:2016ymz} imply that the crossing equations of half-BPS operators of all these configurations are related by a web of analytic continuations. In this work we build on those results and implement the bootstrap for the case of a line defect. Hence, we will work with a $1d$ superconformal theory with $\text{OSP}(4^*|4)$ symmetry.

Although our analysis is mostly based on symmetry without referring to explicit Lagrangian constructions, this setup
corresponds to line defects in $\Nm=4$ SYM, and there is therefore literature that study this system from the gauge theory point of view. Results include exact formulas \cite{Erickson:2000af,Drukker:2000rr,Pestun:2007rz} for the Wilson loop (which is conformally related to the line), perturbative calculations at weak coupling \cite{Cooke:2017qgm}, holographic calculations at strong coupling \cite{Giombi:2017cqn}, and integrability-based studies  \cite{Drukker:2012de,Correa:2012hh}. The bootstrap approach of this paper complements these works.

The structure of the paper is as follows. In section \ref{sec: preliminaries} we discuss the preliminaries which include the relevant Ward identities and the superconformal blocks to be used in the crossing equations. Section \ref{sec: gauge theory} reviews some results for line defects in gauge theories which will helps us understand several of our bootstrap results. The crossing equations are presented in section \ref{sec: crossing equations}, they are analyzed numerically in section \ref{sec: numerical results}, and analytically in section \ref{sec: analytical results}. We conclude with a discussion of future directions and open problems.

%% file: sections/preliminaries.tex

\section{Preliminaries}
\label{sec: preliminaries}

Let us begin by summarizing the symmetries preserved by the half-BPS line defect. We will consider a straight line in four dimensions. The bosonic subgroup of the four-dimensional conformal group preserved by this defect is SO(2,1)
$\times$ SO(3), where the SO(2,1) factor is the $1d$ conformal group on the line and the SO(3) represents rotations orthogonal to the defect; in the supersymmetric setup we consider here there is also a $\text{SP}(4)_\RR$ R-symmetry. The bosonic generators together with the 16 fermionic generators left unbroken form the superalgebra $\text{OSP}(4^*|4)$.
The representations of this superalgebra are labeled by the conformal dimension ${\Delta}$, the SO(3) spin $s$ (this variable was dubbed ``transverse spin'' in \cite{Lemos:2017vnx}), and the $\text{SP}(4)_\RR$ Dynkin labels $[a,b]$. In this work we will be particularly interested in certain half-BPS multiplets of the $\text{OSP}(4^*|4)$ algebra which we denote by $\mathcal{B}_k$, where $k$ labels the $[0,k]$ $\text{SP}(4)_\RR$ irrep of the superconformal primary. 

In the presence of defects there is a rich interaction between bulk and defect quantities. As described in the introduction, in this work we will only study operators constrained to the defect, making our theory effectively one-dimensional. 
Among the defect operators a special role is played by the \textit{displacement operator}, which measures deformations orthogonal to the defect. For a line defect in $4d$, this operator has protected dimension $\Delta_{\Dm}=2$, and in the class of supersymmetric theories we are interested in it sits in a $\mathcal{B}_1$ multiplet. The structure of this multiplet is as follows
\beq \label{B1content}
\mathcal{B}_1: \qquad [0,1]_{\Delta=1}^{s=0}\to [1,0]_{\Delta=\frac{3}{2}}^{s=\frac{1}{2}} \to [0,0]_{\Delta=2}^{s=1}\, ,
\eeq
where the highest weight is a scalar with $\Delta=1$ in the $[0,1]$ representation of $\text{SP}(4)_\RR$.\footnote{This representation corresponds to the fundamental of $\text{SO}(5)$. In gauge theories this quantum number is associated to the five scalars that do not couple to the line, see section \ref{sec: gauge theory}.} We will usually call this multiplet the displacement, although technically the displacement operator is just the term $[0,0]_{\Delta=2}^{s=1}$. The remaining components of this supermultiplet correspond to the  R-symmetry and supersymmetry that are broken due to the presence of the defect.

In the bootstrap analysis of subsequent sections we will consider the four-point function of $\mathcal{B}_1$ multiplets, but also mixed correlators with $\mathcal{B}_2$ multiplets. The structure of the latter is given by
\beq 
\mathcal{B}_2: \qquad [0,2]_{\Delta=2}^{s=0}\to [1,1]_{\Delta=\frac{5}{2}}^{s=\frac{1}{2}}
\to [0,1]_{\Delta=3}^{s=1}\, \oplus\, [2,0]_{\Delta=3}^{s=0}
\to [1,0]_{\Delta=\frac{7}{2}}^{s=\frac{1}{2}}
\to [0,0]_{\Delta=4}^{s=0}\, .
\eeq
Even though we will only consider half-BPS multiplets as external operators, more general multiplets can be exchanged in the OPE, the representations relevant for our analysis are summarized in table~\ref{fig: relevant representations}.

\subsection{Superconformal blocks}

A particularly useful superspace for the study of correlation functions of $\mathcal{B}_k$ multiplets was introduced in \cite{Liendo:2016ymz}. The superspace coordinate on the defect reads
\begin{equation}
X=
\begin{pmatrix}
x\,\epsilon^{ab} & \theta^{a\beta} \\
\theta^{b\alpha}  & y^{(\alpha\beta)}
\end{pmatrix}
\end{equation}
where $a,b=1,2$ are the transverse spin indices, $\alpha,\beta=1,2$, the $\theta^{a\alpha}$ are fermionic and $\epsilon^{ab}$ is the antisymmetric tensor. Let $\calD_k$ be the operators sitting in the short multiplets $\mathcal{B}_k$. Generically we have to deal with operator multiplicities, but let us ignore that for a moment. We will return to that issue in section~\ref{subsec: Topological structure constants}. The two-point functions of the $\calD_k$ operators take the form 
\begin{equation}\label{twoPOINT}
\langle \calD_k(1)\calD_l(2)\rangle\,=\,\delta_{k,l}(12)^k\,,\quad \text{ where }\quad (ij)\equiv \frac{1}{(\text{spf}(X_{12}))^2}= \frac{y^2_{12}}{x_{12}^2}+\text{ferm}\,.
\end{equation}
where $y^2_{12}\equiv\det y_{12}$.
It follows from superconformal symmetry that the  four-point function of 
$\mathcal{B}$-type multiplets can be written as
\beq
\label{eq: correlation function in K and A}
\langle  \calD_{m_1}(1)\calD_{m_2}(2)\calD_{m_3}(3)\calD_{m_4}(4)\rangle \,=\,
K_{\{m_1,m_2,m_3,m_4\}}\,
\calA_{\{m_1,m_2,m_3,m_4\}}(\chi,\zeta_1,\zeta_2)\,,
\eeq
where $m_1+\cdots+m_4$ is even due to R-symmetry and the prefactor reads
\begin{equation}
K_{\{m_1,m_2,m_3,m_4\}}\,=\,
(12)^{\frac{1}{2}(m_1+m_2)}
(34)^{\frac{1}{2}(m_3+m_4)}
\left(\frac{(14)}{(24)}\right)^{\frac{1}{2}(m_1-m_2)}
\left(\frac{(13)}{(14)}\right)^{\frac{1}{2}(m_3-m_4)}\,.
\end{equation}
The quantities $(\chi,\zeta_1,\zeta_2)$ are the eigenvalues of the supermatrix
\begin{equation}
\mathcal{Z}=X_{12}^{}X_{13}^{-1}X_{34}^{}X_{24}^{-1}\,,
\qquad 
X_{ij}:=X_{i}-X_{j}\,.
\end{equation}
Notice that the expression \eqref{eq: correlation function in K and A} implies
 that the correlation functions of all superconformal descendants can be recovered from the one of the corresponding primaries in this case. 
 If the fermionic variables are set to zero by a superconformal transformation, 
 the cross-ratios take the familiar form
\beq
\label{eq: definition of the chi and xi variables}
\chi=\frac{x_{12}x_{34}}{x_{13}x_{24}}\,,\qquad 
\zeta_1\zeta_2=\frac{y_{12}^2y_{34}^2}{y_{13}^2y_{24}^2}\,,
\qquad
(1-\zeta_1)(1-\zeta_2)=\frac{y_{14}^2y_{23}^2}{y_{13}^2y_{24}^2}\,.
\eeq
Notice that since we are in one dimension there is only one spatial cross-ratio and $\zeta_1,\zeta_2$ are defined up to permutations so that $\calA$ has to be symmetric with respect to  the exchange $\zeta_1\leftrightarrow \zeta_2$.

The dependence of  $\calA$ on $\zeta_1,\zeta_2$ is further  restricted by the fact that the 
correlator \eqref{eq: correlation function in K and A} has to be a polynomial in the $y^{\alpha\beta}_i$ coordinates. 
This translates to a condition on the $\zeta_1,\zeta_2$ dependence of $\mathcal{A}$ that singles out a number of linearly independent terms which is equal to the number of $\text{SP}(4)_\RR$ singlets in the tensor product $[0,m_1]\otimes [0,m_2]\otimes [0,m_3]\otimes [0,m_4]$; examples are give in \eqref{eq: direct R-symmetry structures},
\eqref{eq: mixed Rsym structures 1, -1}.
 It is convenient for later to define the shorthand combinations
\beq
\label{eq: definition of sX}
\sX\equiv \frac{\chi^2}{\zeta_1\zeta_2}\,,\qquad \sXt\equiv\frac{(1-\chi)^2}{(1-\zeta_1)(1-\zeta_2)}\,.
\eeq

\paragraph{The Ward identities.} Superconformal symmetry puts strong constraints on the form of correlation functions, these constraints are captured by the superconformal Ward identities. In our setup, the Ward identities take a compact form and can be obtained from the analytic continuations described in \cite{Liendo:2016ymz} (see also \cite{Dolan:2004mu,Doobary:2015gia,Liendo:2015cgi,Lemos:2016xke} for Ward identities in higher spacetime dimensions), in our coordinates they read
\beq
\label{eq: Ward identities}
\left(\frac{\partial \calA}{\partial\zeta_1}+\frac{1}{2}\frac{\partial \calA}{\partial\chi}\right)_{\big| \zeta_1=\chi}\,=\,
\left(\frac{\partial \calA}{\partial\zeta_2}+\frac{1}{2}\frac{\partial \calA}{\partial\chi}\right)_{\big| \zeta_2=\chi}\,=\,0\,,
\eeq
where $\calA\equiv \calA_{\{m_1,m_2,m_3,m_4\}}$. Let us start with the simplest case of identical $\calD_1$ external operators. The solution to \eqref{eq: Ward identities} can be written in an elegant form:
\beq
\label{eq:WI_solution}
\calA_{\{1,1,1,1\}}(\chi,\zeta_1,\zeta_2)=F\sX + \mathbb{D} f(\chi)\,,
\eeq
where $F$ is a constant and the differential operator $\mathbb{D}$ is defined as
\beq
\mathbb{D} = \left(2\chi^{-1}-\zeta_1^{-1}-\zeta_2^{-1}\right)-\chi ^2 \left(\zeta_1^{-1}-\chi^{-1}\right) \left(\zeta_2^{-1}-\chi^{-1}\right) \frac{\partial }{\partial \chi }\,.
\eeq
The different solutions to these equations correspond to different superblocks associated to the $\text{OSP}(4^*|4)$ multiplets being exchanged in the OPE. Below we list all the relevant solutions.
\begin{itemize}
\item $\mathcal{I}$. The simplest solution represents the contribution of the identity operator $\mathcal{I}$:
\beq\label{blockID}
F_{\mathcal{I}}=1\,,\qquad f_{\mathcal{I}}(\chi)=\chi\,.
\eeq
\item $\mathcal{B}_2$. This solution can be interpreted as the contribution of a $\mathcal{B}_2$ half-BPS multiplet:
\beq\label{blockB2}
F_{\mathcal{B}_2}=1\,,\qquad f_{\mathcal{B}_2}(\chi)=\chi\left(1-\, _2F_1\left(1,2;4;\chi \right)\right)
\,.
\eeq
\item $\mathcal{L}^{\Delta}_{[0,0]}$. The final solution has no constant term and can be identified as the superblock of a generic long block whose primary is neutral under $\text{SP}(4)_\RR$:
\beq\label{blockLong}
F_{\mathcal{L}^{\Delta}_{[0,0]}}=0\,,\qquad f_{\mathcal{L}^{\Delta}_{[0,0]}}=\frac{1}{1-\Delta}\left[\chi^{\Delta +1} \, _2F_1\left(\Delta +1,\Delta +2;2 (\Delta +2);\chi\right)\right]\,.
\eeq
\end{itemize}
In general, superconformal blocks are given by a finite sum of bosonic blocks, although not obvious from the expressions listed above, this is indeed case. For example, using the $\mathbb{D}$ operator we obtain 
\beq
\mathcal{G}_{\mathcal{I}}(\chi,\zeta_1,\zeta_2) = 
F_{\mathcal{I}}\sX +\mathbb{D} f_{\mathcal{I}}(\chi,\zeta_1,\zeta_2) = 1\, ,
\eeq
as expected for the identity contribution. More illuminating is the expansion of the $\mathcal{B}_2$ short block:
\beq
\begin{split}
\mathcal{G}_{\mathcal{B}_2}(\chi,\zeta_1,\zeta_2)&= F_{\mathcal{B}_2} \sX +\mathbb{D} f_{\mathcal{B}_2}(\chi)\,,\\
&= \BS_{[0,2]}\CB_2(\chi )+\frac{1}{10} \BS_{[2,0]} \CB_{3}(\chi )+\frac{3}{350} \BS_{[0,0]} \CB_4(\chi)\,.
\end{split}
\eeq
Here the $\CB_{h}(\chi)$ corresponds to the one-dimensional bosonic block in \eqref{eq: bosonic blocks} with the external dimensions set to $\Delta_{12}=\Delta_{34}=0$. The terms $\BS_{[p,q]}$ are polynomials in $\zeta_{i}^{-1}$ given in \eqref{eq: direct R-symmetry structures}. They are appropriately normalized eigenfunctions of the $\text{SP}(4)_\RR$ quadratic Casimir \eqref{eq: quadratic Casimir for SP4} and capture the R-symmetry structures associated with the $[p,q]$ irrep of $\text{SP}(4)_\RR$.\footnote{In appendix \ref{eq: superconformal blocks} there is an extra label ``$0,0$'' in the superblocks $\mathcal{G}$ and the R-symmetry polynomials $\BS$, this label can be ignored in sections \ref{sec: preliminaries} and \ref{sec: gauge theory} but it will play a role later when we discuss mixed correlators.} The corresponding expansion for the long block $\mathcal{L}^{\Delta}_{[0,0]}$ reads
\begin{align}\label{LongblockExpanded}
&\mathcal{G}_{\mathcal{L}_{[0,0]}^\Delta}(\chi,\zeta_1,\zeta_2)\,=\,\BS_{[0,0]}\CB_{\Delta}(\chi )-\frac{\Delta }{\Delta -1}  \BS_{[2,0]}\CB_{\Delta +1}(\chi )\nonumber\\
&+\left(\frac{\Delta +1 }{\Delta -1}\BS_{[0,2]}+\frac{3 \Delta  (\Delta +1) (\Delta +3)}{10 (\Delta -1) (2 \Delta +1) (2 \Delta +5)}\BS_{[0,0]}\right) \CB_{\Delta +2}(\chi )
\\&-\frac{((\Delta +1) (\Delta +2) (\Delta +3) }{4 (\Delta -1) (2 \Delta +3) (2 \Delta +5)}\BS_{[2,0]} \CB_{\Delta +3}(\chi )+\frac{(\Delta +1) (\Delta +2) (\Delta +3)^2 (\Delta +4) }{16 (\Delta -1) (2 \Delta +3) (2 \Delta +5)^2 (2 \Delta +7)}\BS_{[0,0]}\CB_{\Delta +4}(\chi )\,.\nonumber
\end{align}

In later sections we will consider more general correlators involving different half-BPS multiplet as external operators. In particular, we will study the full mixed system given by 
\beq
\label{eq:all_correlators}
\calA_{\{1,2,1,2\}}(\chi,\zeta_1,\zeta_2)\,, \qquad \calA_{\{1,2,2,1\}}(\chi,\zeta_1,\zeta_2)\,, \qquad \calA_{\{2,2,2,2\}}(\chi,\zeta_1,\zeta_2)\, .
\eeq
The Ward identities for these cases can be solved similarly as we did for $ \calA_{\{1,1,1,1\}}$, although for the mixed system it is convenient to use a different parameterization for the correlators. We have collected all the solutions and the corresponding expansions in bosonic blocks in appendix \ref{eq: superconformal blocks}, together with the explicit map between the two parameterizations (see \eqref{eq:map_mixed}).

\subsection{The OPE selection rules}
\label{subsec: selection rules}

The solutions to the Ward identities together with the knowledge of the $\text{SP}(4)_\RR$ tensor products give us the OPE selection rules,
which state which supermultiplets are allowed to appear in a particular OPE. The full set of selection rules relevant for our system reads
\beq
\label{eq: D with D OPE}
\begin{split}
\mathcal{B}_1\,\times\,\mathcal{B}_1\,=\,&\mathcal{I}+\mathcal{B}_2+\mathcal{C}_{[2,0]}+\sum_{\Delta\geq 1}\mathcal{L}^{\Delta}_{[0,0]}\,,\\
\mathcal{B}_1\, \times \,\mathcal{B}_2\,=\,& \mathcal{B}_1+\mathcal{B}_3+\mathcal{C}_{[2,1]}+\sum_{\Delta\geq 2} \mathcal{L}^{\Delta}_{[0,1]}\,,\\
\mathcal{B}_2\, \times \,\mathcal{B}_2\,=\,&
\mathcal{I}+\mathcal{B}_2+\mathcal{B}_4+\mathcal{C}_{[2,0]}+\mathcal{C}_{[4,0]}+\mathcal{C}_{[2,2]}\\&+\sum_{\Delta\geq 1}\mathcal{L}^{\Delta}_{[0,0]}+\sum_{\Delta\geq 3}\left(\mathcal{L}^{\Delta}_{[2,0]}+\mathcal{L}^{\Delta}_{[0,2]}\right)\,.
\end{split}
\eeq
The operators $\mathcal{C}_{[a,b]}$ are semi-short multiplets that can be obtained by putting the longs $\mathcal{L}^{\Delta}_{[a,b]}$ at their unitarity bound. In fact, we have normalized the superblocks (see appendix \ref{eq: superconformal blocks}) so that
\beq
\label{eq: long block at the unitarity bound}
\lim_{\Delta\rightarrow 1+a+b}\Big(\Delta-(1+a+b)\Big)\mathcal{G}_{\mathcal{L}^{\Delta}_{[a,b]}}=\mathcal{G}_{\mathcal{C}_{[a+2,b]}}\,.
\eeq
Notice also that \eqref{eq: D with D OPE} only contains multiplets with zero SO(3) spin. As an aside (based on an analysis of the OPEs of some higher $\mathcal{B}_k$) we conjecture that the OPE relation can be generalized as 
\beq
\nn
\label{eq: short and long OPE}
\begin{split}
\mathcal{B}_k\, \times \,\mathcal{B}_l\,=&\, 
\sum_{m=|k-l|, \text{ step } 2}^{k+l}\mathcal{B}_{m}+\sum_{i=0}^{\min(k-1,l-1)} \sum_{j=0}^i \mathcal{C}_{\big[2i-2j+2,2j+|k-l|\big]}\\&+\sum_{i=0}^{\min(k-1,l-1)}\sum_{j=0}^i\sum_{\Delta > 2i+|k-l|+1} \mathcal{L}^{\Delta}_{\big[2i-2j,2j+|k-l|\big]}\,,
\end{split}
\eeq 
where $\mathcal{I}\equiv \mathcal{B}_0$. The main properties of the exchanged representations are listed in table~\ref{fig: relevant representations}.
\begin{table}[!ht]
\centering
\begin{tabular}{c|ccc}
Name & $\Delta$ & SO(3) Spin & R-Symmetry\\\hline
$\mathcal{B}_k$ & $k$ & 0 & $[0,k]$\\
$\mathcal{C}_{[a,b]}$ & $a+b$ & 0 & $[a,b]$\\
$\mathcal{L}_{[a,b]}^{\Delta}$ & $\Delta$ & 0 & $[a,b]$\\
\end{tabular}
\caption{Representations of OSP(4$^*|$4) that are relevant for the line defect bootstrap.}
\label{fig: relevant representations}
\end{table}

\subsection{Topological structure constants}
\label{subsec: Topological structure constants}

Correlators of half-BPS operators on the line have a topological sector that does not depend on the positions of the operators \cite{Drukker:2009sf,Giombi:2018qox}, and whose existence follows directly from the superconformal Ward identities \eqref{eq: Ward identities}.\footnote{This subsector is closely related to the exact truncations recently uncovered in superconformal theories \cite{Beem:2013sza,Beem:2014kka,Chester:2014mea,Beem:2016cbd}.}  The restriction to this subsector in the four-point functions \eqref{eq: correlation function in K and A} is equivalent to setting $\chi=\zeta_1=\zeta_2$. In this limit only the exchange of the identity operator and $\mathcal{B}_k$-type multiplets survives in the OPE decomposition.
In this section, we summarize some information regarding the structure constants (or three-point couplings) of the topological sector relevant for the mixed correlator system to be studied below. Our basic assumption is that the displacement operator $\calD_1$ is \textit{unique}, given this assumption we make the following definitions:
\begin{enumerate}
\item We define $\calD_2$ as the operator in the multiplet $\mathcal{B}_2$ that appears in the OPE $\calD_1\times \calD_1$.
\item We define $\calD_3$ as the operator in the multiplet $\mathcal{B}_3$ that appears in the OPE $\calD_1\times \calD_2$.
\item We define $\calD_4$ as the operator in the multiplet $\mathcal{B}_4$ that appears in the OPE $\calD_2\times \calD_2$.
\end{enumerate}
Since there can be several operators that sit in $\mathcal{B}_2$ multiplets, we denote by $\calD_2'$ the contribution of the remaining operators\footnote{Note that there can be many operators in $\mathcal{B}_2$ multiplets, but only one linear combination will appear in the OPE as $\calD_2'$. } that appears in the OPE $\calD_2\times \calD_2$. Schematically we have
\beq
\calD_2\times \calD_2=\mathcal{I}+\calD_2+\calD_2'+\calD_4+\cdots\, .
\eeq
Having defined the operators $\calD_m$ we introduce the following notation for their structure constants
\beq
C_{\calD_m \calD_n\calD_r}\equiv C_{m,n,r}\,,
\eeq
where we shall write $2'$ for the $\calD_2'$ operator; the index $0$ stands for $\calD_0\equiv \mathcal{I}$. 
In addition to  \textbf{cyclicity}, one-dimensional structure constants also exhibit \textbf{time-reversal symmetry} \cite{Qiao:2017xif}
\beq
C_{i,j,k}=C_{k,j,i}^*\,,\qquad  
C_{i,j,k}=C_{j,k,i}=C_{k,i,j}\,.
\eeq
In particular, it follows that $C_{i,i,j}$ are real for all $i$ and $j$.
Choosing appropriate normalizations it is possible to set $C_{1,1,0}=C_{2,2,0}=1$, and also $C_{1,1,2} \geq 0$. Considering the correlator $\vac{\calD_1\calD_1\calD_1\calD_3}$ restricted to the topological sector it follows that $C_{1,2,3}$ is real. Then, the independent real OPE structures entering the mixed correlator system we are interested in are 
\beq
\label{eq: inequality C112 vs C222}
C_{1,1,2}\geq 0\,,\quad C_{1,2,3}\,,\quad C_{2,2,2}\,,\quad C_{2,2,2'},\quad C_{2,2,4}\,.
\eeq
In addition, from the correlator $\vac{\calD_1\calD_1\calD_2\calD_2}$ restricted to the topological sector the following condition can be obtained
\beq
\label{eq: mixed equations minibootstrap}
1+C_{1,1,2}C_{2,2,2}=C_{1,1,2}^2+C_{1,2,3}^2\,,\quad  \Longrightarrow\quad C_{2,2,2}\geq C_{1,1,2}-C_{1,1,2}^{-1}\,.
\eeq
There are no simple conditions on the OPE coefficients $C_{2,2,2'}$ or $C_{2,2,4}$ that do not involve OPE coefficients of higher half-BPS operators. 

%% file: sections/gauge_theory.tex

\section{Line defects in gauge theory}
\label{sec: gauge theory}
As discussed in the introduction, the bootstrap philosophy aims at solving theories using only symmetry requirements without relying on explicit Lagrangian formulations, nevertheless, part of our motivation is the understanding of concrete line defects in gauge theory. In this section we collect some results regarding Wilson lines in $\Nm=4$ SYM, that will be relevant for our subsequent analysis.

The defect is defined by 
\beq
\calW_{\rep}=\text{tr}_{\rep}\,\text{Pexp}\int_\gamma dt\left[i\dot{x}^\mu A_\mu +|\dot{x}|\,\theta^I\Phi^I\right]\,,
\eeq
where the path $\gamma$ is a straight line and the scalar part is given by $\theta^I\Phi^I=\Phi^6$. 
With this choice, the configuration preserves an $\text{SO}(5)_\RR \sim \text{SP}(4)_\RR$ R-symmetry since we can freely rotate the five scalars $\Phi^a$, and also the $\text{SO}(2,1)\times \text{SO}(3)$ which is the $1d$ conformal algebra together with rotations orthogonal to the line. In addition to the bosonic generators the defect also preserves 16 supercharges that form the $1d$ $\text{OSP}(4^*|4)$ superconformal algebra. Our bootstrap setup then describes this particular class of line defects. Gauge invariant correlation function on the line are then given by
\beq
\left\langle\left\langle\calO_1(x_1)\cdots \calO_n(x_n)\right\rangle\right\rangle = \frac{\left\langle \text{Tr}_{\rep}\left[\calO_1(x_1)\text{Pexp}(\cdots) \cdots \calO_n(x_n)\text{Pexp}(\cdots)\right]\right\rangle}{\langle \calW_{\rep}\rangle}\,.
\eeq
Note that in this paper we will ignore the double-bracket notation and denote correlators using single brackets.
In gauge theory, the displacement operator contains the elementary excitations of the Wilson line and sits in the half-BPS $\calD_1$ multiplet described in the previous section. 
Its bosonic content is the five scalars $\Phi^a$ not coupled to the Wilson line with $\Delta=1$ and the three components of the field strength $F_{t\mu}\equiv i F_{t\mu}+D_{\mu}\Phi^6$ along the directions $\mu = 1, 2, 3$ transverse to the line with $\Delta=2$ \footnote{Like before, the actual ``displacement operators" are the three $F_{t\mu}$ that measure the change of the Wilson loop under deformations orthogonal to the contour, the remaining elements of $\calD_1$ are their supersymmetric partners.}.

\subsection{Localization results}

Thanks to localization techniques \cite{Pestun:2007rz}, it is possible to calculate the CFT data for the topological sector of a half-BPS circular Wilson loop. Because the line is conformally related to the circle, the localization results are also valid for this geometry, as long as the correlators are properly normalized. Most of the results of this subsection were already obtained in \cite{Giombi:2018qox}.

For a gauge group $G$ and a representation $\rep$ of $G$, the vacuum expectation value (setting the radius of the circle to one) reads \cite{Pestun:2007rz}
\beq\label{GaussianPestun}
\vac{\calW_{\rep}(\lambda)}=\frac{\int_{\mathfrak{g}}[da]e^{-\frac{8\pi^2 h_G}{\lambda}(a,a)} \text{tr}_{\rep}e^{2\pi a} }{\int_{\mathfrak{g}}[da]e^{-\frac{8\pi^2 h_G }{\lambda}(a,a)}}\,,
\eeq
where $\mathfrak{g}$ is the Lie algebra of $G$, $\lambda=g_{YM}^2h_G$ with $g_{YM}$ the Yang-Mills coupling, $a\in \mathfrak{g}$ and $h_G$ the dual Coxeter number\footnote{We remind that $h_{\SU(N)}=N$.} of $G$. This formula can be generalized to a $\tfrac{1}{8}$-BPS Wilson loop whose contour is not a line but a generic curve that is entirely contained in an $S^2$ \cite{Drukker:2007yx, Giombi:2009ds}. For this Wilson loop one obtains the identity
\beq
\vac{\calW_{\rep}^{1/8}(\lambda ;A)}=\vac{\calW_{\rep}\left(\lambda \frac{A(4\pi-A)}{4\pi^2}\right)}\,,
\eeq
where $A$ is the area in $S^2$ enclosed by the curve. 

Let us denote the correlation functions in the topological sector by a tilde over the operators. The $n$-point function of $\tilde{\calD}_1$ is given by \cite{Giombi:2017cqn}
\beq
\label{eq: tildeD1n}
\vac{\tilde{\calD}_1^n}_{\text{non-norm.}}\,=\,\left[\partial^n_A \vac{\calW_{\rep}^{1/8}(\lambda;A)}\right]_{\big| A=2\pi}\,,\quad \Rightarrow \quad \vac{\tilde{\calD}_1^n}\,=\,\frac{\vac{\tilde{\calD}_1^n}_{\text{non-norm.}}}{\vac{\tilde{\calD}_1^2}_{\text{non-norm.}}^{\frac{n}{2}}}\,.
\eeq
Using the OPE relations \eqref{eq: D with D OPE} and the orthonormalization of the operators, the four-point function is then
\beq
\vac{\tilde{\calD}_1^4}=\vac{(1+C_{1,1,2}\tilde{\calD}_2)^2}=1+C_{1,1,2}^2\,.
\eeq
Now, plugging \eqref{eq: tildeD1n} in the above we find
\beq
\label{eq: localization formula for C112}
C_{1,1,2}^2=-1+3\frac{\calW_{\rep}(\lambda)\calW_{\rep}^{\prime\prime}(\lambda)}{(\calW_{\rep}^{\prime}(\lambda))^2}\,,
\eeq
where, by definition, we take the root so that $C_{1,1,2}$ is positive\footnote{ 
For $g_{YM}=0$ this OPE coefficient takes the value  
$C_{112}^2=2-\tfrac{1}{2} \tfrac{\text{Cas}_2(\text{Adj})}{\text{Cas}_2(\mathcal{R})}$, where $\text{Cas}_2$ denotes the quadratic Casimir of $\mathfrak{g}$. See figure~\ref{fig:xiRank} for more details.
}.

\begin{figure}[htbp!]
             \begin{center}       
              \includegraphics[scale=0.7]{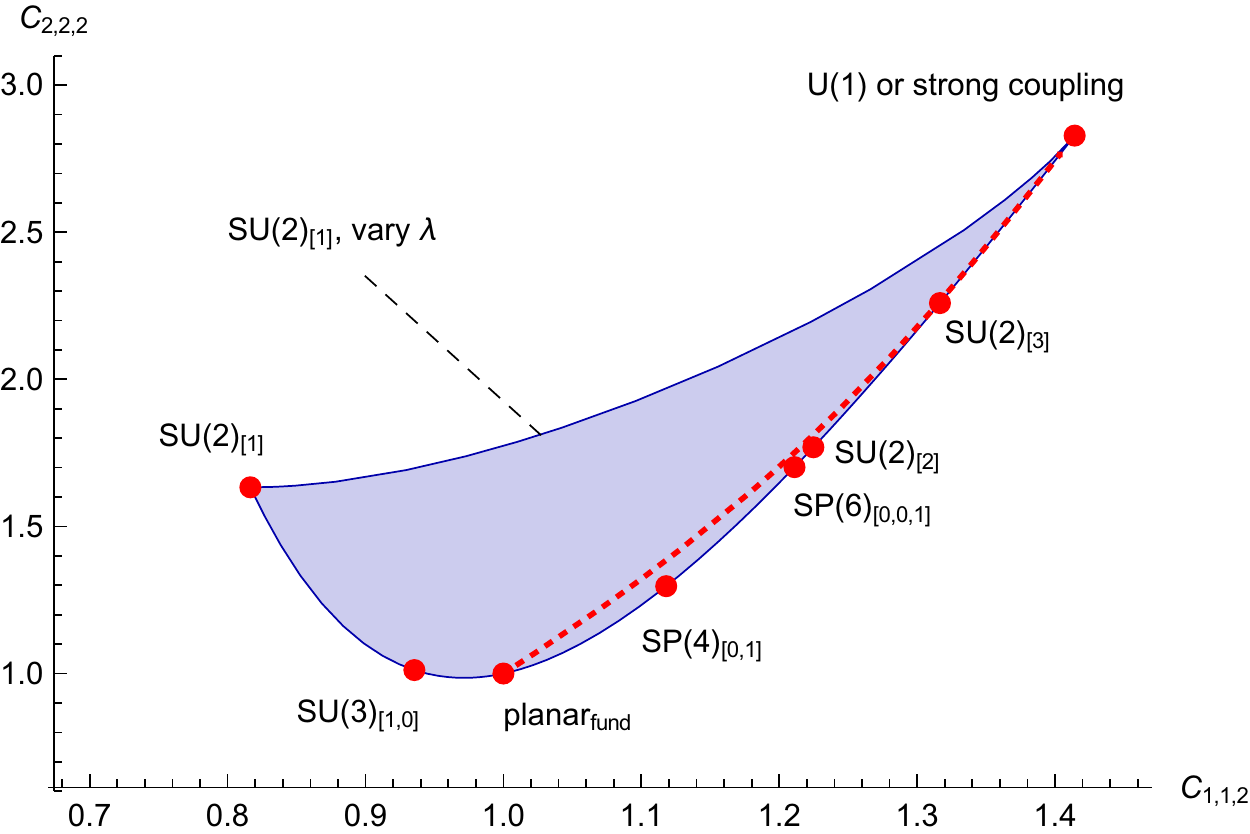}
              \caption{The allowed region for  $C_{1,1,2}$, $C_{2,2,2}$ from localization for classical groups $G$. Extremal points corresponding to free theories are marked by red points and the planar theory in the fundamental representation is marked by a dotted red curve. The U(1) theory at $(\sqrt{2},2\sqrt{2})$ (for any value of the coupling) has the same OPE coefficients as the strong coupling limit of any other case that we looked at. The notation for the theories is $G_{\rep}$, where the representation $\rep$ is given by its Dynkin labels. 
}
              \label{fig:LocalizationResults}
            \end{center}
\end{figure}

We can also study the correlators with $\tilde{\calD}_2$ by using the relation $\tilde{\calD}_2=\frac{1}{C_{1,1,2}}(\tilde{\calD}_1^2-1)$, which stems from the OPE relation \eqref{eq: short and long OPE} and the discussion of the second part of section~\ref{subsec: selection rules}. Then, we find the following relation between $C_{2,2,2}$ and $C_{1,1,2}$
\beq
\label{eq: localization formula for C222}
\begin{split}
C_{2,2,2}=\vac{\tilde{\calD}_2^3}=\frac{\calW_{\rep}'\left(\lambda\right) C_{1,1,2}}{\left(\calW_{\rep}'\left(\lambda\right)^2-3 \calW_{\rep}\left(\lambda\right) \calW_{\rep}''\left(\lambda\right)\right)^2}\Big[&(15 \calW_{\rep}\left(\lambda\right)^2 \calW_{\rep}^{(3)}\left(\lambda\right)+2 \calW_{\rep}'\left(\lambda\right)^3\\&
-9 \calW_{\rep}\left(\lambda\right) \calW_{\rep}'\left(\lambda\right) \calW_{\rep}''\left(\lambda\right)\Big]\,.
\end{split}
\eeq
By directly integrating the Gaussian integrals (and for the antisymmetric representations cross-checking with the results of \cite{Fiol:2013hna}) we have computed $C_{1,1,2}$ and $C_{2,2,2}$ for $G=\SU(N)$, $\text{SO}(N)$, $\text{SP}(2N)$ and a variety of representations; the results are shown in figure~\ref{fig:LocalizationResults}. In section \ref{sec: numerical results} this region will be contrasted with the allowed regions coming from the bootstrap and some analytic solutions to crossing (see figure~\ref{fig:GeneralOverviewC112C222}).

We note that the boundary of figure~\ref{fig:LocalizationResults} can be obtained from two simple formulas. First, the upper bound is provided by the $G=\SU(2)$ in the fundamental representation ($\SU(2)_{[1]}$)
\beq
C_{1,1,2}^2\,=\,2-\frac{3072}{(\lambda +48)^2}\,,\qquad C_{2,2,2}^2\,=\,\frac{8 (\lambda  (\lambda  (\lambda +144)+2304)+12288)^2}{(\lambda  (\lambda +96)+768)^3}\,,
\eeq
as $\lambda$ varies from zero to infinity. The lower bound curve, starting from the leftmost point $\SU(2)_{[1]}$ and extending to the planar theory in the fundamental representation is given by the free $\SU(N)$ theories in the fundamental representation 
\beq
\label{eq: free theory C112 and C222}
C_{1,1,2}^2\,=\,\frac{N^2-2}{N^2-1}\,,\qquad C_{2,2,2}^2\,=\,\frac{\left(N^4-4 N^2+8\right)^2}{\left(N^2-2\right)^3 \left(N^2-1\right)}\,.
\eeq
Finally, the remaining piece of the lower bound curve is also given by \eqref{eq: free theory C112 and C222}, but this time amusingly for purely imaginary values of $N$, i.e. we set $N=i x $ and vary $x$ over all the reals.
It turns out that various other free theories, such as $\SU(2)$ with higher spins and $\text{SP}(2k)$ in the fundamental representation, sit on that curve for appropriate imaginary values of $N$. It is interesting to notice that for any fixed $G$ and $\mathcal{R}$, all $C_{1,1,2}$ and $C_{2,2,2}$ tend to the same value once $g_{\text{YM}}^2\rightarrow \infty$. This is an experimental observation that should have a proof starting from the expression \eqref{GaussianPestun}.  Alternatively, 
since Wilson lines are S-dual to t' Hooft lines one might imagine showing this fact in perturbation theory in the ``magnetic picture".

\subsection{Operator multiplicities in gauge theories}

In section~\ref{subsec: Topological structure constants} we reviewed the symmetries of the OPE coefficients in the topological sector of any $1d$ CFT with $\text{OSP}(4^*|4)$ symmetry. We mentioned that we generically have to deal with operator multiplicities and now we want to shed some light on the origin of these multiplicities, if the $1d$ CFT is obtained from a $4d$ $\calN=4$ SYM with a given gauge group $G$ and representation $\rep$. We begin by assuming that if the model under investigation is not a product of two or more decoupled theories, then the displacement operator $\calD_1$ sitting in the  multiplet $\mathcal{B}_1$ is \textit{unique}. For the multiplets $\mathcal{B}_{\ell\geq 2}$, however, we have to deal with multiplicities. To understand that, we first remind that the five scalars (those not coupled to the line) $\Phi^a$ lie the $[0,1]$ representation of $\text{SP}(4)_R$ and that $[0,k]$, which is the lowest $\Delta$ piece of $\mathcal{B}_k$, is the $k$-fold symmetric traceless tensor product of $[0,1]$. Thus, to make an operator sitting in $\mathcal{B}_2$, we can for example consider the operator
\beq
\calO\,=\,\text{Tr}_{\rep}(--\Phi^{(a}_{\bullet k}(x)\Phi^{b)}_{k\bullet}(x)--)\,,
\eeq
where $--$ denotes the Wilson line, $(a,b)$ stands for the  traceless symmetrization of the indices and the $\bullet$ are gauge groups indices that are contracted to the line. However, another operator that also sits in this representation is given by $\calW_\rep \times \text{Tr}_{\rep}(\Phi^{(a}(x)\Phi^{b)}(x))$, i.e. a color singlet that is just placed on the line.

In general, for a gauge group $G$, the number of $\mathcal{B}_{\ell}$ multiplets on a Wilson line in the representation $\rep$ is given by the number of singlets in the tensor product
\begin{equation}
\# \mathcal{B}_{\ell}\,=\,\Big[\,\rep\,\otimes\, \rep^* \,\otimes\,\left(\text{Adj}\right)^{\otimes_{\text{sym}}\ell}\Big]_{\text{G-invariant}}\,. 
\end{equation}
The first remark is that for $\ell=1$ and $\rep$ not the trivial representation, there is only one singlet in this tensor product corresponding to the displacement operator.
In the example of gauge group $\SU(2)$ one has
$\left(\text{Adj}\right)^{\otimes_{\text{sym}}\ell}\,=\,
(2\ell+1)\oplus (2\ell+1-4)\oplus  (2\ell+1-8) \oplus \dots$\,,
where $(s)$ denotes the $s$-dimensional representation.
For Wilson lines in the fundamental representation,  there is only one $\mathcal{B}_{\ell}$ for each $\ell$. Some other examples for $\SU(2)$ are shown in table~\ref{fig: table of multiplicity for SU2}.
\begin{table}[!ht]
\centering
\begin{tabular}{c|ccccccc}
 $n\backslash\ell$& 0 & 1 & 2 & 3 & 4 & 5 & 6 \\\hline
1 & 1 & 0 & 1 & 0 & 1  & 0 & 1\\
2 & 1 & 1 & 1 & 1 & 1  & 1 & 1\\
3 & 1 & 2 & 1 & 2 & 1  & 2 & 1\\
4 & 1 & 2 & 2 & 2 & 2  & 2 & 2\\
5 & 1 & 2 & 2 & 2 & 3  & 2 & 3\\
6 & 1 & 2 & 2 & 2 & 3  & 3 & 3
\end{tabular}
\caption{Number of $\mathcal{B}_{\ell}$ for the representation of dimension $n$ for the group $\SU(2)$. }
\label{fig: table of multiplicity for SU2}
\end{table}
We see that the number of operators sitting in $\mathcal{B}_{2}$ is generically higher than one, even for gauge group $\SU(2)$. Nevertheless, through the OPE relation \eqref{eq: D with D OPE}, we see that a certain linear combination of these operators is special, since it is the one that appears on the RHS of $\calD_1\times \calD_1$. In section \ref{subsec: Topological structure constants} we \textit{defined} this linear combination as the operator $\calD_2$ whose correlation functions we study.

\subsection{Strong coupling}
\label{sec: Strong couplingy}

Complementary to the localization results, there was a recent study of this system at strong coupling using $AdS_2$ Witten diagrams \cite{Giombi:2017cqn}. In this section we will review these results and rewrite them in the language of section \ref{sec: preliminaries}. 
At strong coupling in the planar limit, the correlator $\Am$ has the following expansion
\beq 
\label{A0A1Stong}
\calA_{\{1,1,1,1\}}(\chi,\zeta_1,\zeta_2)= \calA^{(0)}(\chi,\zeta_1,\zeta_2) + \frac{1}{\sqrt{\lambda}}\calA^{(1)}(\chi,\zeta_1,\zeta_2) + \ldots  \, ,
\eeq
where the $\calA^{(0)} $ term corresponds to the strong coupling limit described by an abelian U(1) theory, and the $\calA^{(1)} $ term is captured by leading tree-level connected Witten diagrams.

From section \ref{sec: preliminaries} we know that the non-trivial information of this correlator is captured by the constant $F$ and the function $f(\chi)$. In the strong coupling limit these are given by
\beq
\label{eq:strongC_zeroth}
F^{(0)}=3\, , \qquad f^{(0)}(\chi) = \frac{\chi  (2 \chi -1)}{\chi -1}\, .
\eeq
In order to compare with the results of \cite{Giombi:2017cqn} let us define 
\beq 
\calA_{\{1,1,1,1\}}^{abcd}(\chi) = G_S(\chi)\delta^{ab}\delta^{cd} + G_T(\chi)(\delta^{ac}\delta^{bd}+\delta^{bc}\delta^{ad}-\tfrac{2}{5}\delta^{ab}\delta^{cd})+ G_A(\chi)(\delta^{ac}\delta^{bd}-\delta^{bc}\delta^{ad})\,,
\eeq
where the Kronecker deltas capture the different SO(5)$\sim$ SP(4) channels. In our harmonic coordinates the different channels are captured by the $\zeta_i$ variables. The dictionary is as follows
\beq 
\calA_{\{1,1,1,1\}}(\chi,\zeta_1,\zeta_2) \to \calA_{\{1,1,1,1\}}^{abcd}(\chi) \,,
\eeq
where we use
\begin{align}
\label{eq:dictionary}
\nn
\BS_{[0,0]} & \to \delta^{ab}\delta^{cd} \,,
\\
\BS_{[2,0]}  & \to -(\delta^{ac}\delta^{bd}-\delta^{bc}\delta^{ad})\,,
\\ 
\nn
\BS_{[0,2]} & \to \tfrac{1}{2}(\delta^{ac}\delta^{bd}+\delta^{bc}\delta^{ad}-\tfrac{2}{5}\delta^{ab}\delta^{cd}) \,.
\end{align}
Using this dictionary and the $\mathbb{D}$ operator, equation \eqref{eq:strongC_zeroth} implies
\beq 
G^{(0)}_S(\chi) = 1 + \frac{2}{5}G^{(0)}_T(\chi)\, , \quad 
G^{(0)}_T(\chi) = \frac{1}{2}\left(\chi^2 + \frac{\chi^2}{(1-\chi)^2}\right)\, , \quad 
G^{(0)}_A(\chi) = \frac{1}{2}\left(\chi^2 - \frac{\chi^2}{(1-\chi)^2}\right)\, ,
\eeq
which is the correct leading behavior at strong coupling.
As stated above, the corrections to these expressions were calculated in \cite{Giombi:2017cqn}, in our language their result can be written as
\beq 
\label{eq:strongCfirst}
F^{(1)}=-3\, , \qquad f^{(1)}(\chi) = r(\chi)\log(\chi)
- \frac{\chi^2}{(1-\chi)^2}r(1-\chi)\log(1-\chi) + q(\chi)\, ,
\eeq
where 
\beq 
\label{eq:r_and_q}
r(\chi) = \frac{\chi^3(2-\chi)}{(\chi -1)^2}\, , \qquad 
q(\chi) = \frac{\chi(1-2 \chi)}{(\chi -1)}\, .
\eeq
The functions $G^{(1)}_S(\chi)$, $G^{(1)}_T(\chi)$ and $G^{(1)}_A(\chi)$ can be extracted like before, using the $\mathbb{D}$ operator and the dictionary \eqref{eq:dictionary}:
\beq
\label{eq: G1_STA}
\begin{split}
G^{(1)}_S(\chi)  \,=\,&
-\frac{\left(2 \chi ^4-5 \chi ^3-5 \chi +10\right) \log \left(1-\chi \right)}{5 \chi }-\frac{2 \left(\chi ^4-4 \chi ^3+9 \chi ^2-10 \chi
   +5\right)}{5 (\chi -1)^2}
\\ &   
   +\frac{\left(2 \chi ^4-11 \chi ^3+21 \chi ^2-20 \chi +10\right) \chi ^2 \log \left(\chi\right)}{5 (\chi -1)^3}\,,
\\
G^{(1)}_T(\chi) \,=\,& -\chi ^3 \log \left(1-\chi\right)-\frac{\left(2 \chi ^2-3 \chi +3\right) \chi ^2}{2 (\chi -1)^2}+\frac{\left(\chi ^2-3 \chi +3\right)
   \chi ^4 \log \left(\chi\right)}{(\chi -1)^3}\,,
\\
G^{(1)}_A(\chi) \,=\,& \frac{\left(-2 \chi ^3+5 \chi ^2-3 \chi +2\right) \chi }{2 (\chi -1)^2}+\frac{\left(\chi ^3-4 \chi ^2+6 \chi -4\right) \chi ^3 \log \left(\chi\right)}{(\chi -1)^3}
\\ 
  & -\left(\chi ^3-\chi ^2-1\right) \log \left(1-\chi\right)\,,
\end{split}
\eeq
where $\chi\in [0,1]$. These were the results presented in equation (4.19) of \cite{Giombi:2017cqn}. Thanks to the Ward identities, it is possible to rewrite the somehow involved formulas \eqref{eq: G1_STA} in the more compact form (\ref{eq:strongCfirst})--(\ref{eq:r_and_q}). The strong coupling behavior and its first order correction will be important when we interpret the numerical results of section \ref{sec: numerical results}. 
Moreover, due to the simplicity of the functions $r(\chi)$ and $q(\chi)$ (together with $F^{(1)}=-3$) it is natural to ask whether this result can be re-derived using only bootstrap reasoning, bypassing the Witten diagram computation. This will be one of the subjects of section \ref{sec: analytical results}.

\paragraph{Remarks on the  strong coupling $1d$ $\text{CFT}$.} 
A clarification about the $\text{CFT}$ at leading order in $\tfrac{1}{\sqrt{\lambda}}$ is in order, since it is somewhat different 
from the standard supergravity/large $N$ CFTs whose leading behavior is the one of a generalized free theory.
In this case the one dimensional CFT is defined via a two dimensional theory on $\text{AdS}_2$ with $\tfrac{1}{\sqrt{\lambda}}$ as coupling constant, see  \cite{Giombi:2017cqn}. 
The boundary values of the fundamental fields of the bulk theory transform as 
 the displacement supermultiplet $\mathcal{D}_1$.  When the bulk theory is free, i.e.~at leading order for $\sqrt{\lambda}\rightarrow \infty$,  the spectrum consists of composite operators made of a single displacement supermultiplet, and  correlation functions are the one obtained by Wick contractions using the super-propagator \eqref{twoPOINT}.
 Some examples are
 \begin{align}\label{corrSTRONG}
 \langle \mathcal{D}_1(1) \mathcal{D}_1(2) \mathcal{D}_k(3)\mathcal{D}_k(4)\rangle&=
 (12)(34)^k\,+k\left( (13)(24)+(14)(23)\right)(34)^{k-1}\,,
 \\
 \langle \mathcal{D}_2(1) \mathcal{D}_2(2) \mathcal{D}_2(3)\mathcal{D}_2(4)\rangle&=
 (12)^2(34)^2\,+ (13)^2(24)^2+(14)^2(23)^2\\
\nonumber
 &+4\left((12)(23)(34)(41)+(13)(32)(24)(41)+(14)(43)(32)(21)\right)\,,
 \end{align}
 where $\mathcal{D}_k\sim\mathcal{D}_1^k$.
 This is of course different from what is usually referred to as \textit{generalized free theory}. 
 Since at leading order all the operators are words made of components of the displacement supermultiplet, which is of type 
 $\mathcal{B}_1$, their partition function is given by
 \begin{equation}\label{Zpartition}
 \mathcal{Z}_{\text{strong}}
 =\text{P.E.}\left(\chi_{\mathcal{B}_1}(q,z;x,y)\right)\,,
 \qquad 
 \text{P.E.}(f(t)):=\exp\left(\sum_{n=1}^\infty f(t^n)\right)\,.
 \end{equation}
Above $\text{P.E.}$ is the plethystic exponential and $\chi_{\mathcal{B}_1}(q,z;x,y)$ is the character of the $\mathcal{B}_1$ multiplet, see \eqref{B1content}, with $q,z,(x,y)$ character variables for scaling weight, transverse spin and $\text{SP}(4)_\RR$
respectively. 
By expanding \eqref{Zpartition} in the $q$ variable one notices that it can be written as 
 \begin{equation}\label{ZpartitionDEC}
 \mathcal{Z}_{\text{strong}}=1+\mathcal{Z}^{\text{Short}}_{\text{strong}}+\mathcal{Z}^{\text{Long}}_{\text{strong}}\,,
 \qquad
\mathcal{Z}^{\text{Short}}_{\text{strong}}=\sum_{k=1}^{\infty}\,\chi_{\mathcal{B}_k}(q,z;x,y)\,,
 \end{equation}
 Notice that there are no multiplicities for the half-BPS operators $\mathcal{B}_k$ and no long multiplets at the unitarity bound.
 It will be useful for a later discussion to collect here the content of long operators of low dimensions in the representation $[0,0]$ and with vanishing transverse spin: 
  \begin{equation}\label{longs000}
\mathcal{Z}^{\text{Long}}_{\text{strong}}\Big{|}_{[0,0],s=0}\,=\,\chi_{\mathcal{L}^{\Delta=2}_{[0,0]}}
+2\,\chi_{\mathcal{L}^{\Delta=4}_{[0,0]}}
+3\, \chi_{\mathcal{L}^{\Delta=6}_{[0,0]}}+\dots
 \end{equation}

%% file: sections/crossing.tex

\section{Crossing equations}
\label{sec: crossing equations}

\begin{figure}[htbp!]
             \begin{center}       
              \includegraphics[scale=0.35]{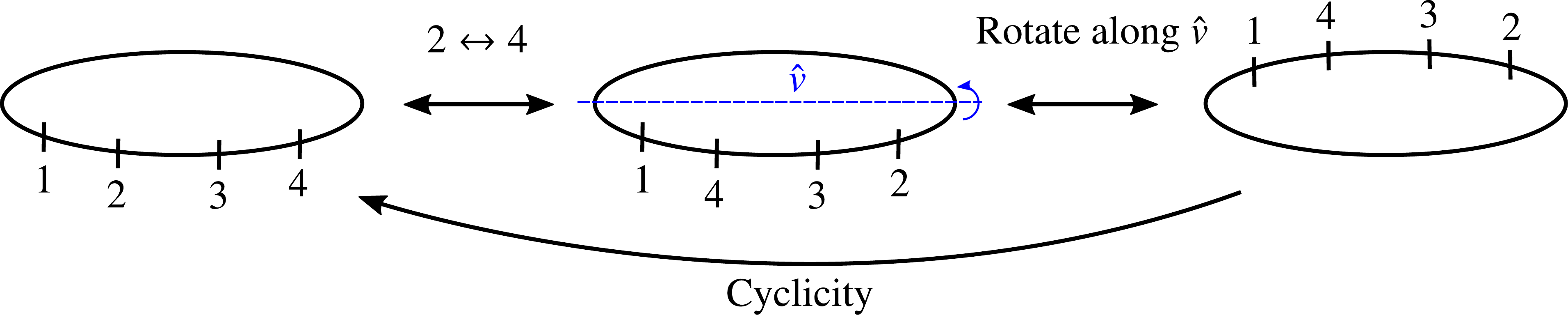}
              \caption{Exchanging the points $2$ and $4$ is a symmetry of the system.}
              \label{fig:flip24}
            \end{center}
\end{figure}

We now present the crossing equations to be studied numerically in section \ref{sec: numerical results}.
On a line, after identifying the endpoints at infinity we are allowed to exchange the points $2$ and $4$ as illustrated in figure~\ref{fig:flip24}.
This operation acts on the four-point invariants as
$ \chi\leftrightarrow 1-\chi$ and $\zeta_i\leftrightarrow 1-\zeta_i$. Using this and the prefactor convention of \eqref{eq: correlation function in K and A}, the crossing relation 
\begin{equation}
\langle  \calD_{m_1}(1)\calD_{m_2}(2)\calD_{m_3}(3)\calD_{m_4}(4)\rangle \,=\,
\langle  \calD_{m_1}(1)\calD_{m_4}(4)\calD_{m_3}(3)\calD_{m_2}(2)\rangle
\end{equation}
implies the following identity
\beq
\label{eq: nice crossing equations for calA}
\sXt^{\frac{m_2+m_3}{2}}\calA_{\{m_1,m_2,m_3,m_4\}}(\chi,\zeta_i)\,=\,
\sX^{\frac{m_3+m_4}{2}}\calA_{\{m_1,m_4,m_3,m_2\}}(1-\chi,1-\zeta_i)\,,
\eeq
where we remind of \eqref{eq: definition of sX} for the definitions of $\sX$ and $\sXt$.
This relation implies in fact multiple equations, one for each R-symmetry factor, however, due to superconformal symmetry not all of them are independent. This is a general feature of the half-BPS bootstrap, see for example \cite{Beem:2013qxa,Chester:2014fya,Beem:2015aoa,Beem:2014zpa,Lemos:2016xke,Beem:2016wfs,Chang:2017xmr,Chang:2017cdx}

\subsection{The single correlator \texorpdfstring{$\calD_1$}{D1}} 

If we just consider the four-point function of the displacement multiplet $\calD_1$, then \eqref{eq: nice crossing equations for calA} implies
\beq
\label{eq: nice crossing equations for calA1111}
\sXt\, \calA_{\{1,1,1,1\}}(\chi,\zeta_i)=\sX\,\calA_{\{1,1,1,1\}}(1-\chi,1-\zeta_i)\,,
\eeq
and the expansion of  $\calA_{\{1,1,1,1\}}(\chi,\zeta_i)$ in terms of superblocks reads
\beq
\label{eq:single_block_exp}
\calA_{\{1,1,1,1\}}(\chi,\zeta_i)=\sum_{\calO\in \calD_1\times \calD_1} C_{1,1,\calO}^2\, \block^{0,0}_{\calO}(\chi,\zeta_i)\, .
\eeq
The upper indices were added to distinguish the different channels (see also \eqref{eq: expansion of A in superblocks}). In order to study the mixed correlator system below it will be convenient to change the parametrization of the single correlator with respect to section \ref{sec: preliminaries}. The new basis is explained in appendix \ref{eq: superconformal blocks}, where we defined functions $f^{a,b}_{\calO}$ and constants $F^{a,b}_{\calO}$ that are related to the blocks $\block^{a,b}$ according to \eqref{eq: extracting f and F} and \eqref{eq: irreducible functions parametrization}. 
With this new parametrization the three crossing equations in \eqref{eq: nice crossing equations for calA1111} (one for each of R-symmetry structures $\BS^{0,0}_{[0,0]}$, $\BS^{0,0}_{[2,0]}$ and $\BS^{0,0}_{[0,2]}$, see \eqref{eq: direct R-symmetry structures}) are satisfied iff the following single equation holds 
\beq
\label{eq: direct crossing 1}
\begin{split}
\left[\chi\,f_{1,\mathcal{I}}^{0,0}\right]_s+C_{1,1,2}^2\left[\chi\,f_{1,\mathcal{B}_2}^{0,0}\right]_s+C_{1,1,\mathcal{C}_{[2,0]}}^2\left[\chi\,f_{1,\mathcal{C}_{[2,0]}}^{0,0}\right]_s
+\sum_{\substack{X=\mathcal{L}^{\Delta}_{[0,0]}\\\Delta>1}} C_{1,1,X}^2\left[\chi\,f_{1,X}^{0,0}\right]_s=&0\,.
\end{split}
\eeq
The translation between the two parametrizations for the single correlator is given in \eqref{eq:map_mixed}.
Notice that only the functions $f_{1,\calO}^{0,0}$ from \eqref{eq: irreducible functions parametrization} appear in \eqref{eq: direct crossing 1}. The other two, $f_{2,\calO}^{0,0}$ and $f_{3,\calO}^{0,0}$, will make their appearance only when we consider the $\calD_2$ multiplet. Furthermore, the constants $F^{0,0}_{\calO}$ do not appear in \eqref{eq: direct crossing 1}. They also will make their appearance later in the topological sector relation \eqref{eq: mini-bootstrap equation} that comes from analyzing the mixed crossing system. 
Finally, in writing \eqref{eq: direct crossing 1}, we have made use of one of the following useful shorthands:
\beq
\label{eq: def of fsa}
[f]_s\equiv f(\chi)+f(1-\chi)\,,\qquad [f]_a\equiv f(\chi)-f(1-\chi)\,.
\eeq
The analysis of the single correlator crossing \eqref{eq: direct crossing 1} already gives several numerical constraints which we present in section~\ref{sec: numerical results}.

\subsection{The full mixed system}
\label{subsec: The full mixed system} 

 Let us now consider the mixed system of four-point functions of $\calD_1$ and $\calD_2$ operators. The crossing equations \eqref{eq: nice crossing equations for calA} imply in addition to \eqref{eq: nice crossing equations for calA1111} the following equations
\beq
\label{eq: full not irreducible crossing equations}
\begin{split}
\sXt^2\calA_{\{2,2,2,2\}}(\chi,\zeta_i)&=\sX^2\calA_{\{2,2,2,2\}}(1-\chi,1-\zeta_i)\,,\\
\sXt^{\frac{3}{2}}\calA_{\{1,2,1,2\}}(\chi,\zeta_i)&=\sX^{\frac{3}{2}}\calA_{\{1,2,1,2\}}(1-\chi,1-\zeta_i)\,,\\
\sXt^2 \calA_{\{1,2,2,1\}}(\chi,\zeta_i)&=\sX^{\frac{3}{2}}\calA_{\{1,1,2,2\}}(1-\chi,1-\zeta_i)\,.
\end{split}
\eeq
The final equation involving $\calA_{\{1,1,2,2\}}$ is actually not a crossing equation since it becomes trivial when using the cyclicity of the correlation functions. However, it is useful since one can decompose the two sides in different sets of blocks and obtain a non-trivial relation. In addition, there is a crossing equation for $\calA_{\{2,1,2,1\}}$ which is simply the complex conjugate of the second equation in \eqref{eq: full not irreducible crossing equations} due to the time-reversal symmetry. 

According to the discussion on the reality and cyclicity conditions of the structure constants in section~\ref{subsec: Topological structure constants}, and using the blocks that we present in detail in appendix~\ref{eq: superconformal blocks}, we can expand all these functions as 
\beq
\label{eq: expansion of A in superblocks}
\begin{split}
\calA_{\{1,1,1,1\}}(\chi,\zeta_i)&=\sum_{\calO\in \calD_1\times \calD_1}C_{1,1,\calO}^2\, \block^{0,0}_{\calO}(\chi,\zeta_i)\,,\\
\calA_{\{2,2,2,2\}}(\chi,\zeta_i)&=\sum_{\calO\in \calD_2\times \calD_2} C_{2,2,\calO}^2\,\block^{0,0}_{\calO}(\chi,\zeta_i)\,,\\
\calA_{\{1,1,2,2\}}(\chi,\zeta_i)&=\sum_{\calO\in \calD_1\times \calD_1}C_{1,1,\calO}C_{2,2,\calO}\block^{0,0}_{\calO}(\chi,\zeta_i)\,,\\
\calA_{\{1,2,1,2\}}(\chi,\zeta_i)&=\sum_{\tilde{\calO}\in \calD_1\times \calD_2}(C_{1,2,\tilde{\calO}})^2\,\block^{1,1}_{\tilde{\calO}}(\chi,\zeta_i)\,,\\
\calA_{\{1,2,2,1\}}(\chi,\zeta_i)&=\sum_{\tilde{\calO}\in \calD_1\times \calD_2}|C_{1,2,\tilde{\calO}}|^2\block^{1,-1}_{\tilde{\calO}}(\chi,\zeta_i)\,.
\end{split}
\eeq
We remind that $C_{1,1,\calO}$ and $C_{2,2,\calO}$ are real while $C_{1,2,\tilde{\calO}}$ is complex and that we need to supplement the complex crossing equations (the second one in \eqref{eq: full not irreducible crossing equations}) by their complex conjugate.

In appendix~\ref{appendix: Comments on the derivation of the crossing equations}, we explain in more detail how using the blocks \eqref{eq: irreducible functions parametrization}, the crossing equations  \eqref{eq: nice crossing equations for calA1111} and \eqref{eq: full not irreducible crossing equations} can be written in the compact form
\beq
\label{eq: final crossing equations 1}
\sum_{\calO\in \calD_2\times \calD_2} \left(\begin{array}{cc}C_{11\calO} & C_{22\calO}\end{array}\right)V_{\calO}\left(\begin{array}{c}C_{11\calO}\\C_{22\calO}\end{array}\right)\\+ \sum_{\tilde{\calO}\in \calD_1\times \calD_2} \left(\begin{array}{cc}\text{Re}\,C_{12\tilde{\calO}} & \text{Im}\,C_{12\tilde{\calO}}\end{array}\right)\tilde{V}_{\tilde{\calO}}\left(\begin{array}{c}\text{Re}\,C_{12\tilde{\calO}}\\\text{Im}\,C_{12\tilde{\calO}}\end{array}\right) =0\,.
\eeq
In the above, the $V_{\calO}$ and the $\tilde{V}_{\tilde{\calO}}$ are vectors with eight components, each of which is a $2\times 2$ matrix. Specifically, we find for $V_{\calO}$ and $\tilde{V}_{\tilde{\calO}}$ the expressions
\beq
\label{eq: define V and tilde V}
V_{\calO}=\left(\begin{array}{c}
\theta(\calO)\left(\begin{array}{cc}[\chi f_{1,\calO}^{0,0}(\chi)]_s& 0\\ 0 & 0\end{array}\right)\\
\left(\begin{array}{cc} 0 & 0\\ 0 & [ f_{1,\calO}^{0,0}(\chi)]_a\end{array}\right)\\
\left(\begin{array}{cc} 0 & 0\\ 0 & [f_{2,\calO}^{0,0}(\chi)]_s\end{array}\right)\\
\left(\begin{array}{cc} 0 & 0\\ 0 & \left[f_{3,\calO}^{0,0}(\chi)\right]_a\end{array}\right)\\
\left(\begin{array}{cc}0& 0\\ 0 & 0\end{array}\right)\\
\left(\begin{array}{cc}0& 0\\ 0 & 0\end{array}\right)\\
\frac{\theta(\calO)}{2}\left(\begin{array}{cc}0& [ \chi f_{1,\calO}^{0,0}(\chi)]_s\\ {}[ \chi f_{1,\calO}^{0,0}(\chi)]_s & 0\end{array}\right)\\
\frac{\theta(\calO)}{2}\left(\begin{array}{cc}0& [ \chi f_{1,\calO}^{0,0}(\chi)]_a\\ {}[ \chi f_{1,\calO}^{0,0}(\chi)]_a & 0\end{array}\right)
\end{array}\right)\,,
\quad
 \tilde{V}_{\tilde{\calO}}=\left(\begin{array}{c}
\left(\begin{array}{cc} 0& 0\\ 0 & 0\end{array}\right)\\
\left(\begin{array}{cc} 0& 0\\ 0 & 0\end{array}\right)\\
\left(\begin{array}{cc} 0& 0\\ 0 & 0\end{array}\right)\\
\left(\begin{array}{cc} 0& 0\\ 0 & 0\end{array}\right)\\
\left(\begin{array}{cc} [f^{1,1}_{\tilde{\calO}}]_s& 0\\ 0 & -[f^{1,1}_{\tilde{\calO}}]_s\end{array}\right)\\
\left(\begin{array}{cc} 0 & [f^{1,1}_{\tilde{\calO}}]_s\\ {}[f^{1,1}_{\tilde{\calO}}]_s & 0\end{array}\right)\\
\left(\begin{array}{cc} [f^{1,-1}_{\tilde{\calO}}]_s& 0\\ 0 & [f^{1,-1}_{\tilde{\calO}}]_s\end{array}\right)\\
\left(\begin{array}{cc} -[f^{1,-1}_{\tilde{\calO}}]_a& 0\\ 0 & -[f^{1,-1}_{\tilde{\calO}}]_a\end{array}\right)
\end{array}\right)\,,
\eeq
where\footnote{The function $\theta(\calO)$ is included due to the fact that only operators in $\calD_1\times \calD_1$ contribute to the $\calA_{\{1,1,2,2\}}$ function.}
\beq
\theta(\calO)=1 \text{ if } \calO\in \calD_1\times \calD_1 \text{ and is zero otherwise}\,.
\eeq
To \eqref{eq: final crossing equations 1}, we have to supplement the topological sector relation 
\beq
\label{eq: mini-bootstrap equation}
C_{1,2,3}^2=1+C_{1,1,2}C_{2,2,2}-C_{1,1,2}^2\,,
\eeq
which is the only crossing equation in which the constants $F^{a,b}_{\calO}$ make an apparition. We remind that these constants are equal to one if $\calO$ is a short operator $\mathcal{B}_k$, and zero otherwise. 

The first line of \eqref{eq: final crossing equations 1} contains the crossing equations of the $\calD_1$ system \eqref{eq: direct crossing 1}. However, the mixed system of equations \eqref{eq: final crossing equations 1} is not the final word.  We must rewrite them a bit in order to take into account several facts: $C_{1,2,1}=C_{1,1,2}$,  $C_{1,2,3}$ is real,  $C_{1,1,X}=0$ for many $X$, and the multiplicity of $\mathcal{B}_2$ is in general greater than zero (said otherwise,  $C_{2,2,2}$ and $C_{2,2,2'}$ enter the equations differently). Moreover, we can eliminate $C_{1,2,3}^2$ by using \eqref{eq: mini-bootstrap equation}. 
Putting it all together, we get the following expression for the crossing equations of the mixed system
\beq
\label{eq: final crossing equations 2}
\begin{split}
0=&\left(\begin{array}{cc}1 & 1\end{array}\right)V_{\mathcal{I}}\left(\begin{array}{c}1\\1\end{array}\right)+(\tilde{V}_{\mathcal{B}_3})_{11}\\&+\left(\begin{array}{cc}C_{1,1,2} & C_{2,2,2}\end{array}\right)\left[V_{
\mathcal{B}_2}+\left(\begin{array}{cc}(\tilde{V}_{\mathcal{B}_1})_{11} & 0\\0 & 0\end{array}\right)+\left(\begin{array}{cc}-(\tilde{V}_{\mathcal{B}_3})_{11} & \frac{(\tilde{V}_{\mathcal{B}_3})_{11}}{2}\\\frac{(\tilde{V}_{\mathcal{B}_3})_{11}}{2} & 0\end{array}\right)\right]\left(\begin{array}{c}C_{1,1,2}\\C_{2,2,2}\end{array}\right)\\
&+(V_{\mathcal{B}_2})_{22}C_{2,2,2'}^2+(V_{\mathcal{B}_4})_{22}C_{2,2,4}^2\\
&+\sum_{X=\mathcal{L}_{[0,0]}^{\Delta}}\left(\begin{array}{cc}C_{1,1,X} & C_{2,2,X}\end{array}\right)V_{X}\left(\begin{array}{c}C_{1,1,X}\\C_{2,2,X}\end{array}\right)+\sum_{X=\mathcal{L}_{[2,0]}^{\Delta}}C_{2,2,X}^2(V_X)_{22}+\sum_{X=\mathcal{L}_{[0,2]}^{\Delta}}C_{2,2,X}^2(V_X)_{22}\\
&+\sum_{X=\mathcal{L}_{[0,1]}^{\Delta}}\left(\begin{array}{cc}\text{Re}\,C_{1,2,X} & \text{Im}\,C_{1,2,X}\end{array}\right)\tilde{V}_{X}\left(\begin{array}{c}\text{Re}\,C_{1,2,X}\\\text{Im}\,C_{1,2,X}\end{array}\right)\,,
\end{split}
\eeq
where $(V)_{ab}$ is the $ab$-component of the corresponding $2\times 2$ matrix in \eqref{eq: define V and tilde V}. 

We must remark that even though we have eliminated $C_{1,2,3}^2$ using \eqref{eq: mini-bootstrap equation}, we cannot completely forget about it when implementing the numerical bootstrap as we shall note in section~\ref{sec: numerical results}. 

%% file: sections/numerical_results.tex

\section{Numerical results}
\label{sec: numerical results}

In this section we apply the methods of the numerical conformal bootstrap to the crossing equations \eqref{eq: final crossing equations 2}, and obtain bounds on the conformal dimensions of the long operators $\mathcal{L}^{\Delta}_{[a,b]}$ (in section~\ref{subsec: Dimension bounds}) as well as on the OPE coefficients of the theory (in section~\ref{subsec: OPE bounds}). In each case, the presentation of the numerical results is preceded by a short review of the way that semi-definite programming is applied to the problem at hand. The actual numerics are then performed by using \textsf{Mathematica} as a front end to the semi-definite program solver \textsf{SDPB} \cite{Simmons-Duffin:2015qma}.\footnote{Other front end options are the \textsf{Python} package \textsf{PyCFTBoot} \cite{Behan:2016dtz} or the \textsf{Sage} package \textsf{cboot} \cite{Ohtsuki:2016} (see also \cite{Paulos:2014vya} for an alternative to \textsf{SDPB}).}  For more details on the numerical implementation, see for example \cite{Rattazzi:2008pe,Poland:2011ey,Kos:2014bka} for original literature, and \cite{Rychkov:2016iqz,Simmons-Duffin:2016gjk} for introductory lectures.

\subsection{Dimension bounds}
\label{subsec: Dimension bounds}

Both the $\calD_1$ crossing equation \eqref{eq: direct crossing 1} as well as the full system \eqref{eq: final crossing equations 2} can be put schematically into the form
\beq
\label{eq: schematically dimension bounds 1}
0=\textbf{P}_{id}+\sum_{X \text{ short}}C_X^2\textbf{P}_{X}+\sum_{Y \text{ long }}C_Y^2\textbf{P}_{Y}\,,
\eeq
where the $\textbf{P}_{\calO}$ can be sets of $2\times 2$ matrices in the mixed case. Importantly, the above equation separates into a part that does not depend on the OPE coefficients (because those involving the identity have been normalized to one) and on a part that does. We can then search via SDPB for a functional $\alpha$ that satisfies
\beq
\label{eq: schematically dimension bounds 2}
\alpha(\textbf{P}_{id})=1\,,\qquad \alpha(\textbf{P}_{X})\geq 0\  \forall X\text{ short }\,,\qquad  \alpha(\textbf{P}_{Y})\geq 0\  \forall Y \text{ with } \Delta_Y\geq \text{Bound}
\eeq
In the above $\geq 0$ means semi-definite positive for the blocks involving $2\times 2$ matrices. 
The bounds for the conformal dimensions of the long operators $Y=\mathcal{L}^{\Delta}_{[a,b]}$ can be different for different $[a,b]$. Thus, in the full mixed system \eqref{eq: final crossing equations 2}  we have to deal with four a priori different bounds: $\Delta_{[0,0]}$, $\Delta_{[0,2]}$, $\Delta_{[2,0]}$ and $\Delta_{[0,1]}$, of which only $\Delta_{[0,0]}$ is relevant in the analysis of the $\calD_1$ crossing equation \eqref{eq: direct crossing 1}. 
If a linear functional $\alpha$ can be found such that \eqref{eq: schematically dimension bounds 2} holds, then clearly \eqref{eq: schematically dimension bounds 1} cannot be true and the corresponding bound structure is forbidden. The space of functionals that we consider is given by
\beq
\alpha(f)=\sum_{n=0}^{\Lambda} \alpha_n\frac{\partial^n}{\partial \chi^n} f_{\big| \chi=1/2}\,,
\eeq
and the numerics improve as we increase the number of derivatives $\Lambda$.  

\begin{figure}[htbp!]
             \begin{center}       
              \includegraphics[scale=0.55]{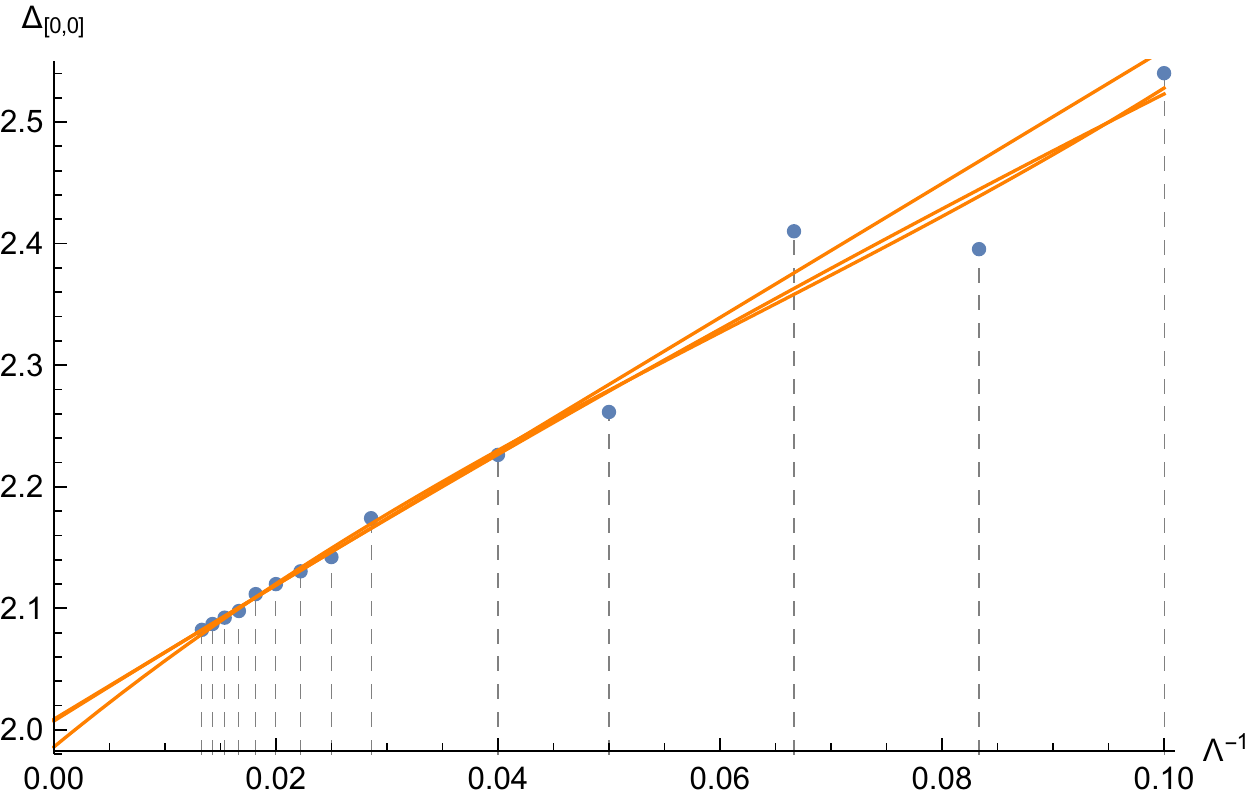}\qquad               \includegraphics[scale=0.55]{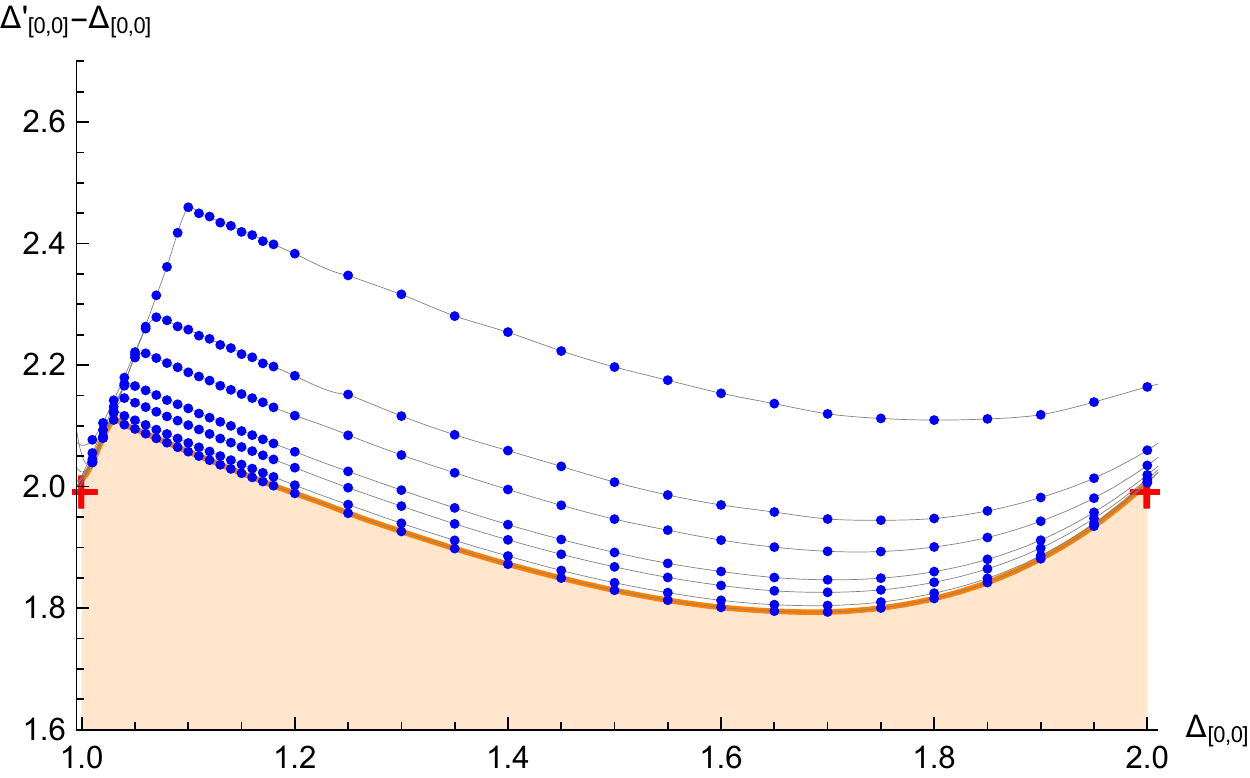}
              \caption{Left: Upper bounds on $\Delta_{[0,0]}$ as a function of $\Lambda^{-1}$. Several fits (linear while ignoring the first 4 points, quadratic and cubic) were done are are plotted in orange. Extrapolated to $\Lambda\rightarrow \infty$, they lead to $\Delta_{[0,0]}\leq 2.009, 2.007, 1.986$ respectively. Right: Bounds on the difference  $\Delta'_{[0,0]}-\Delta_{[0,0]}$ between the conformal dimensions of the first two longs for a given first long with dimension $\Delta_{[0,0]}$. The plot was done for $\Lambda=20, 30, \ldots,  80$ and only the allowed region for $\Lambda=80$ was shaded. The left red dot denotes the analytic solution \eqref{eq: most general analytic solution} for $\xi=-1$, while the right one corresponds to $\xi=1$. For the other values of $\xi$ we have $\Delta_{[0,0]}=1$ and $\Delta'_{[0,0]}=2$, which is too low to be interesting. }
              \label{fig:DDBootstrapDelta0}
            \end{center}
\end{figure}

\paragraph{The $\calD_1$ four-point function. } Let us first analyze \eqref{eq: direct crossing 1} using the above discussion. Letting the sum over longs be restricted to operators with $\Delta\geq \Delta_{[0,0]}$, we obtain the bounds of the left side of figure~\ref{fig:DDBootstrapDelta0}. We remind that in our conventions, the semi-short $\mathcal{C}_{[2,0]}$ can be thought of as a long at the unitarity bound $\Delta=1$. Thus, having $\Delta_{[0,0]}>1$ implies that the $\mathcal{C}_{[2,0]}$ multiplet is absent.
For $\Lambda\rightarrow \infty$, the bounds of figure~\ref{fig:DDBootstrapDelta0} seem to extrapolate to 
$\Delta_{[0,0]}\lesssim 2$.  This could potentially be rigorously proven \`a la \cite{Mazac:2016qev}. 
In addition, we can consider the case of a double gap, in which we allow one long operator with dimension $\Delta_{[0,0]}$ and then require that the other longs have dimensions $\Delta\geq \Delta'_{[0,0]}$. We have plotted the allowed region in the two gaps for various $\Lambda$ on the right hand  side of figure~\ref{fig:DDBootstrapDelta0}. 
The kink in the allowed region is related to the absence of a (strictly positive) lower  bound  for the OPE coefficient $C_{1,1,2}^2$ as it is clear by looking at figure~\ref{fig:DDBootstrapOPEshort}.
It is expected that this kink will disappear  as $\Lambda\rightarrow \infty$.

\paragraph{The full mixed system. } 

\begin{figure}[htbp!]
             \begin{center}       
              \includegraphics[scale=0.55]{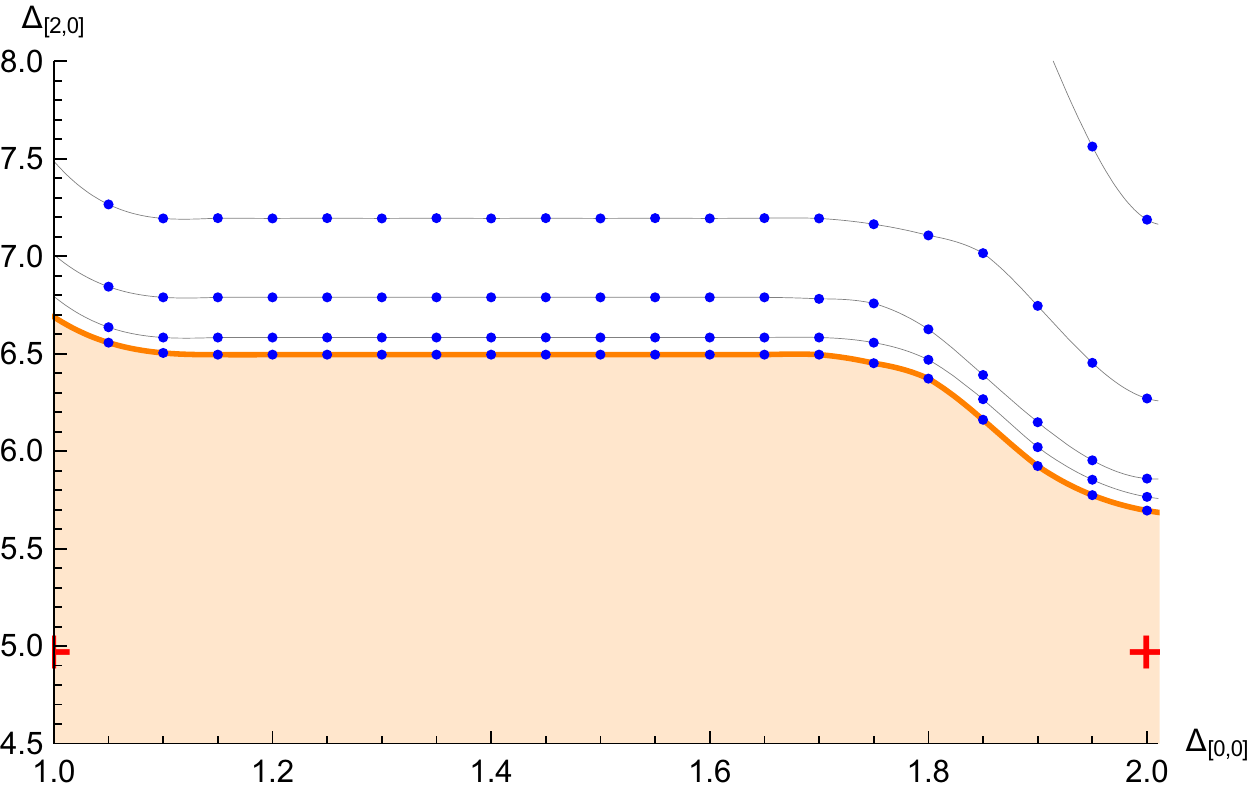}\quad \quad                \includegraphics[scale=0.55]{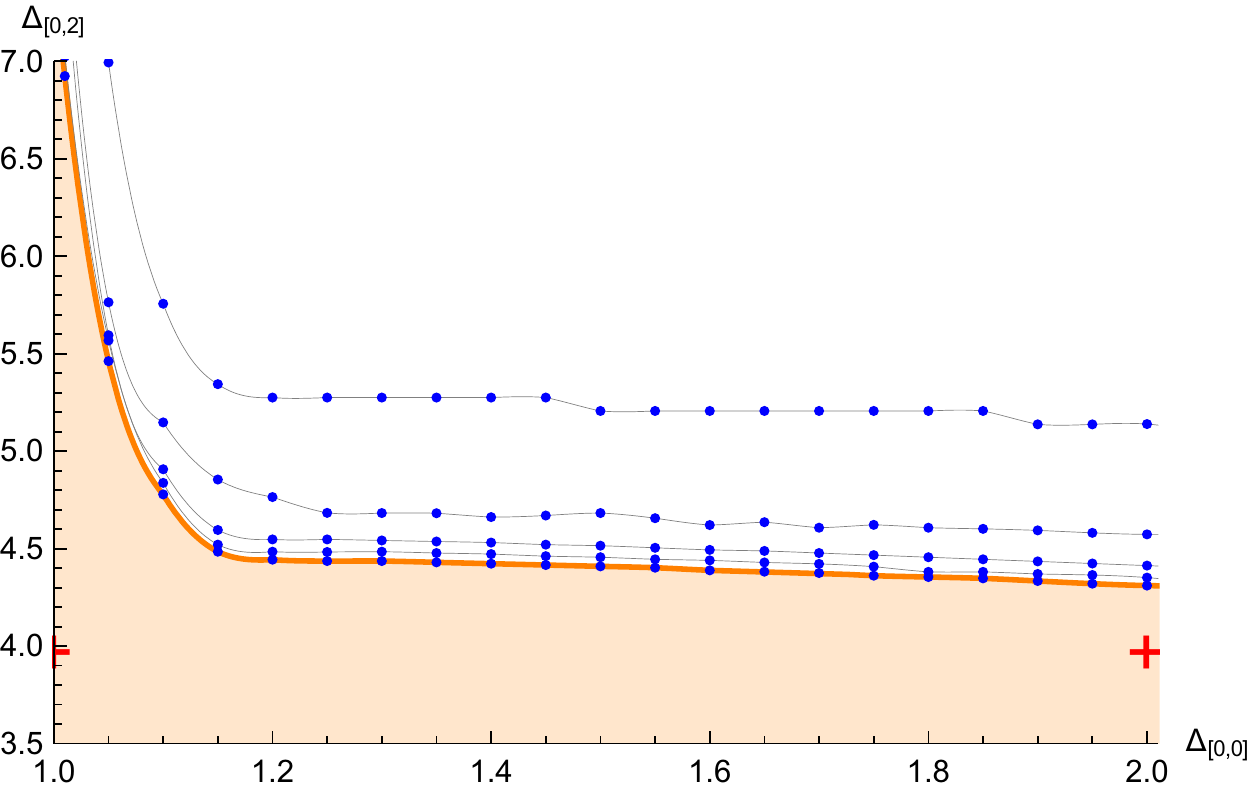}\quad \quad \includegraphics[scale=0.55]{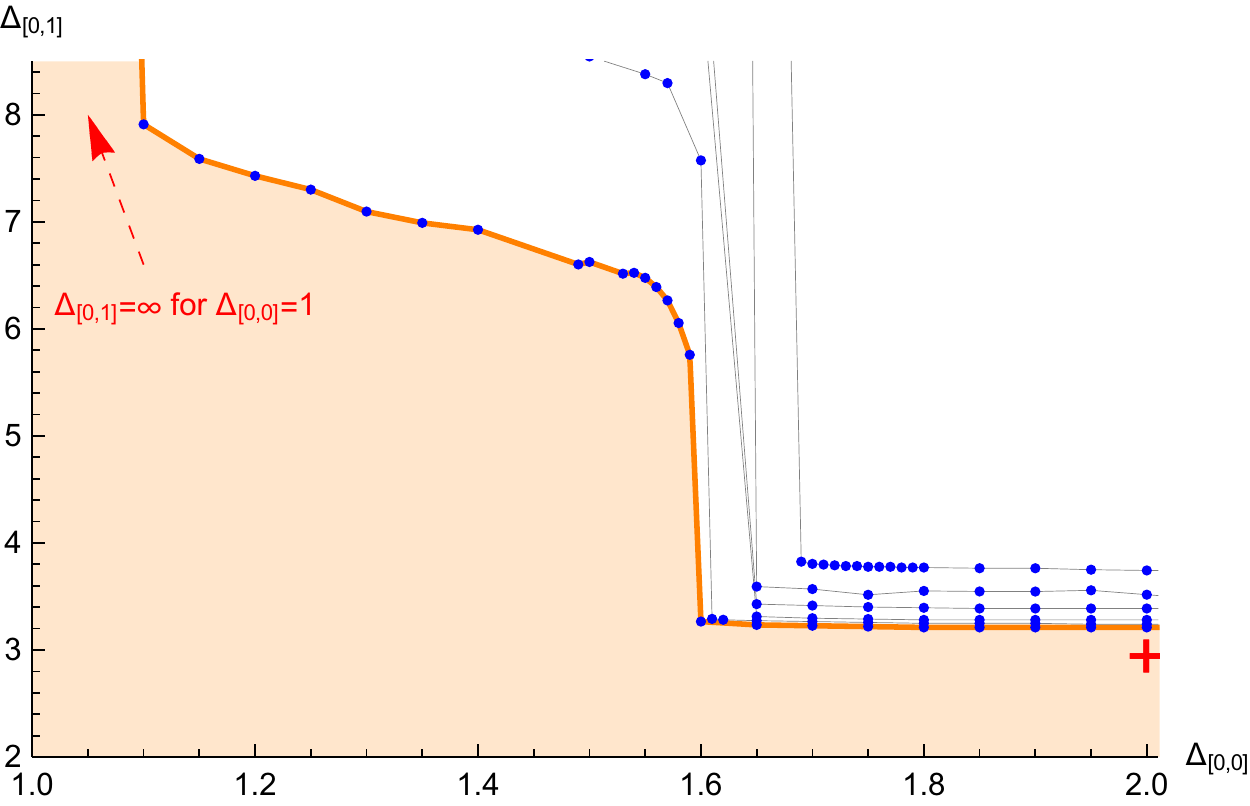}
                        \caption{
                        Upper bounds for $\Delta_{[2,0]}$, $\Delta_{[0,2]}$, $\Delta_{[0,1]}$ 
                        as a function of $\Delta_{[0,0]}$ for $\Lambda=10,20,30,35, 40$.
                         Gap structures coming from the analytic solutions \eqref{eq: most general analytic solution} for special values of the parameters are shown with red crosses.  Among them there is an analytic solution for which $\Delta_{[0,0]}=1$ and $\Delta_{[0,1]}=\infty$, which explains why the bound on  $\Delta_{[0,1]}$ diverges for small $\Delta_{[0,0]}$. 
                        It seems plausible that for $\Delta_{[0,0]}=2$, the bounds would converge to the  strong coupling  values 
                       $\Delta_{[2,0]}=5$, $\Delta_{[2,0]}=4$, $\Delta_{[0,1]}=3$ for infinite $\Lambda$.}
             \label{fig:Delta20And02vsDelta00Full}
            \end{center}
\end{figure}

In the full system the maximal bound for the gap $\Delta_{[0,0]}$ does not change. We can plot the upper bounds of the other gaps as a function of $\Delta_{[0,0]}$ in figure~\ref{fig:Delta20And02vsDelta00Full}.  

It is suggestive that the bottom plot of figure~\ref{fig:Delta20And02vsDelta00Full} shows a drop in the upper bound for $\Delta_{[0,1]}$ around $\Delta_{[0,0]}=1.6$ for $\Lambda=40$. For a similar value of $\Delta_{[0,0]}$ and for the same precision, the LHS of figure~\ref{fig:C112C222BoundsVaryDelta} shows the sudden appearance of an upper bound for the OPE coefficient $C_{2,2,2}$. It is likely that the two phenomena are related, similarly to what happens in the $3d$ Ising model, where the appearance of a kink can be traced back to the vanishing of a certain OPE coefficient 
\cite{El-Showk:2014dwa}. 

\subsection{OPE bounds}
\label{subsec: OPE bounds}

In order to obtain bounds on the OPE coefficients of an operator $X$, we rewrite \eqref{eq: schematically dimension bounds 1} as\footnote{The ``rest" in \eqref{eq: schematically OPE bounds} is made out of long and short operators and takes into account the unitarity bounds on the spectrum of long operators. }
\beq
\label{eq: schematically OPE bounds}
0=\textbf{P}_{id}+\left\{\begin{array}{c}C_X^2\textbf{P}_X\\ \text{ or } \\ (a_X\ b_X)\textbf{P}_X\left(\begin{array}{c}a_X\\b_X\end{array}\right)\end{array}\right\}+\sum_{Y \text{ rest }}C_Y^2\textbf{P}_{Y}\,,
\eeq
depending on whether $C_X$ appears alone or is mixed like $C_{1,1,2}$ and $C_{2,2,2}$ in \eqref{eq: final crossing equations 2}. In the latter case, $\textbf{P}_X$ is a $2\times 2$ matrix and we set $a_X=C_X \cos(\theta)$ and $b_X=C_X \sin(\theta)$, where $\theta$ is an angle over whose values we have to sweep, see \cite{Kos:2016ysd}. We then act on \eqref{eq: schematically OPE bounds} with the functional $\alpha$ and require
\beq
\label{eq: schematically OPE bounds functional}
\begin{split}
&\alpha(\textbf{P}_{id})\text{ is maximized}\,,\qquad \alpha(\textbf{P}_Y)\geq 0\ \text{ for all } Y \in \text{ Rest}\,,\\
&\alpha(\textbf{P}_X)=\pm 1 \quad \text{ or } \quad  \alpha\left((\cos(\theta)\ \sin(\theta))\textbf{P}_X\left(\begin{array}{c}\cos(\theta)\\\sin(\theta)\end{array}\right)\right)=\pm 1 \,.
\end{split}
\eeq
Depending on the normalization condition (the last condition in \eqref{eq: schematically OPE bounds functional}), we get the bounds
\beq
C_X^2\leq -\alpha(\textbf{P}_{id})\ (\text{for }+)\qquad \text{or}\qquad  C_X^2\geq \alpha(\textbf{P}_{id})\ (\text{for }-)\,.
\eeq 
We remark that in order to get a positive lower bound for $C_X$ it is necessary that $X$ be an isolated operator in the spectrum \cite{Poland:2011ey}.

\begin{figure}[htbp!]
             \begin{center}       
              \includegraphics[scale=0.52]{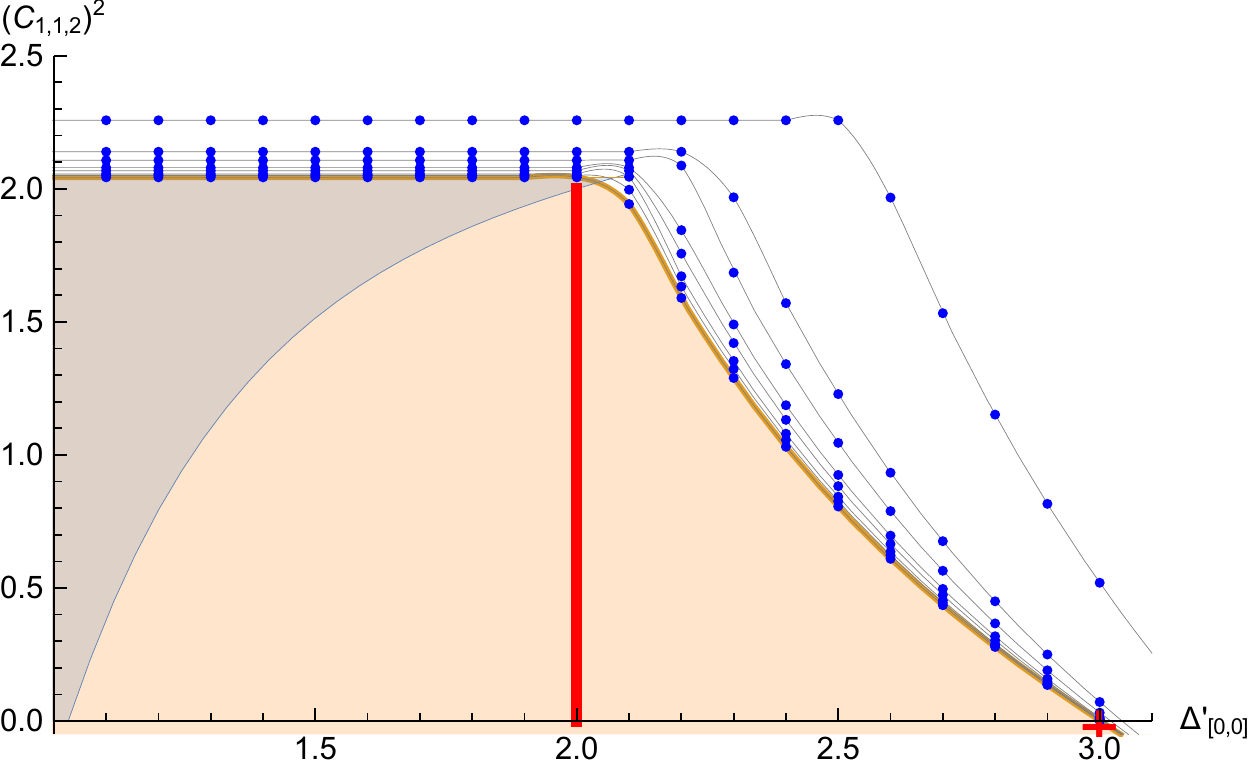}\qquad  \qquad  \includegraphics[scale=0.52]{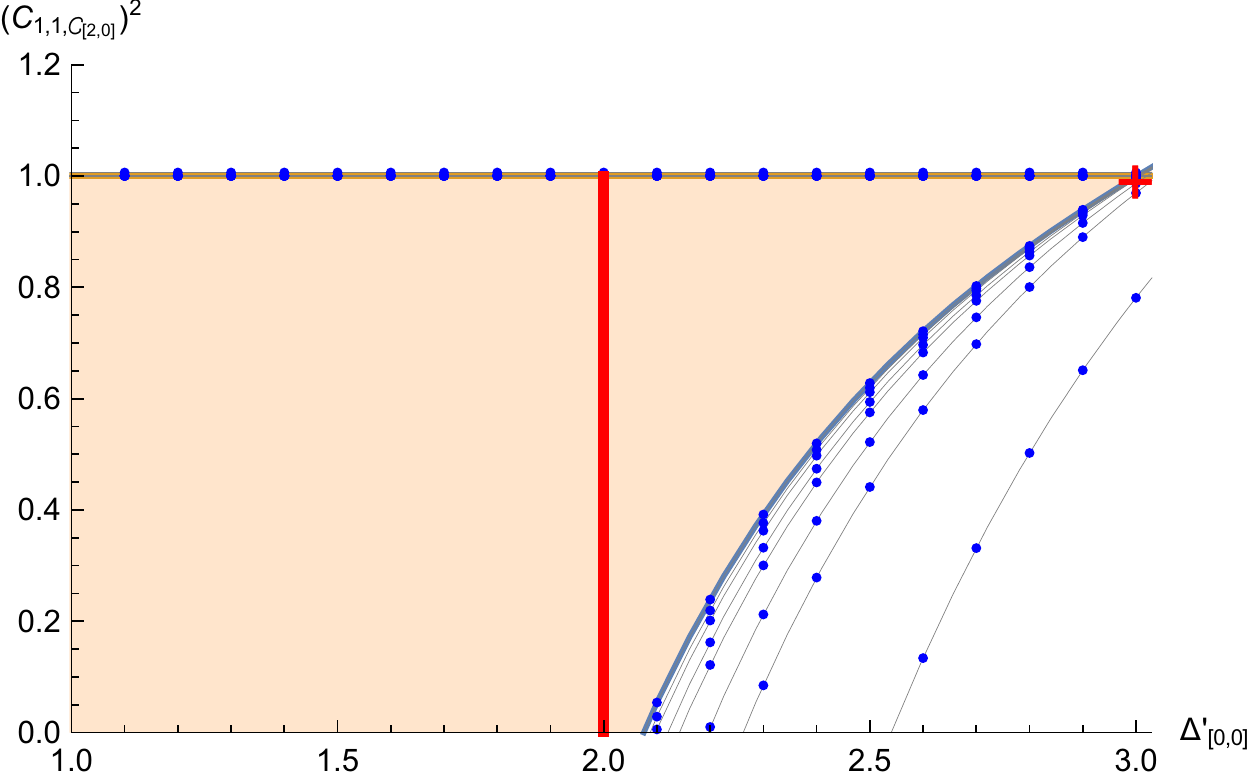}
                        \caption{Left: the bounds on $C_{1,1,2}^2$ as a function of $\Delta'_{[0,0]}$ if the semi-short $\mathcal{C}_{[2,0]}$ is present. Right: the bounds on $C_{1,1,\mathcal{C}_{[2,0]}}^2$ as a function of $\Delta'_{[0,0]}$.  The numerics are done for $\Lambda=10,20,\ldots, 80$ and the allowed regions for $\Lambda=80$ are shaded in orange. Analytic solutions from \eqref{eq: most general analytic solution} are marked in red. For the purpose of comparison, we overlay in light blue on the left the allowed region of figure~\ref{fig:DDBootstrapOPEshort}. One must keep in mind that if $\mathcal{C}_{[2,0]}$ decouples, we can identify $\Delta_{[0,0]}'$ with $\Delta_{[0,0]}$ here, since we consider a single gap in the long spectrum. }
             \label{Fig:OPEMinMaxD1D1WithLongAtBound12}
            \end{center}
\end{figure}

\paragraph{The $\calD_1$ four-point function. } Let us start by considering the case in which the semi-short $\mathcal{C}_{[2,0]}$ is present. This implies setting $\Delta_{[0,0]}=1$ in which case the maximal value of the second gap $\Delta'_{[0,0]}$ seems to go to $\Delta'_{[0,0]}=3$, from the extrapolation from figure~\ref{fig:Delta20And02vsDelta00Full}. For the OPE coefficients, we find the result of Figure~\ref{Fig:OPEMinMaxD1D1WithLongAtBound12}. The position of the ``kink" on the left plot is the position at which the lower bound appears in the right plot. On both plots, there is a line of analytic solutions for $\Delta_{[0,0]}'=2$ (corresponding to $-1<\xi<1$ in  \eqref{eq: most general analytic solution}) and a point (corresponding to $\xi=-1$) for $\Delta_{[0,0]}'=3$.
Furthermore, we can ask for the allowed region in the OPE coefficients of the operator $\calD_2$ and the semi-short $\mathcal{C}_{[2,0]}$ for a given value of $\Delta'_{[0,0]}$. The results are shown in figure~\ref{fig:DDBootstrapOPElong2D} below. 

\begin{figure}[htbp!]
             \begin{center}       
              \includegraphics[scale=0.8]{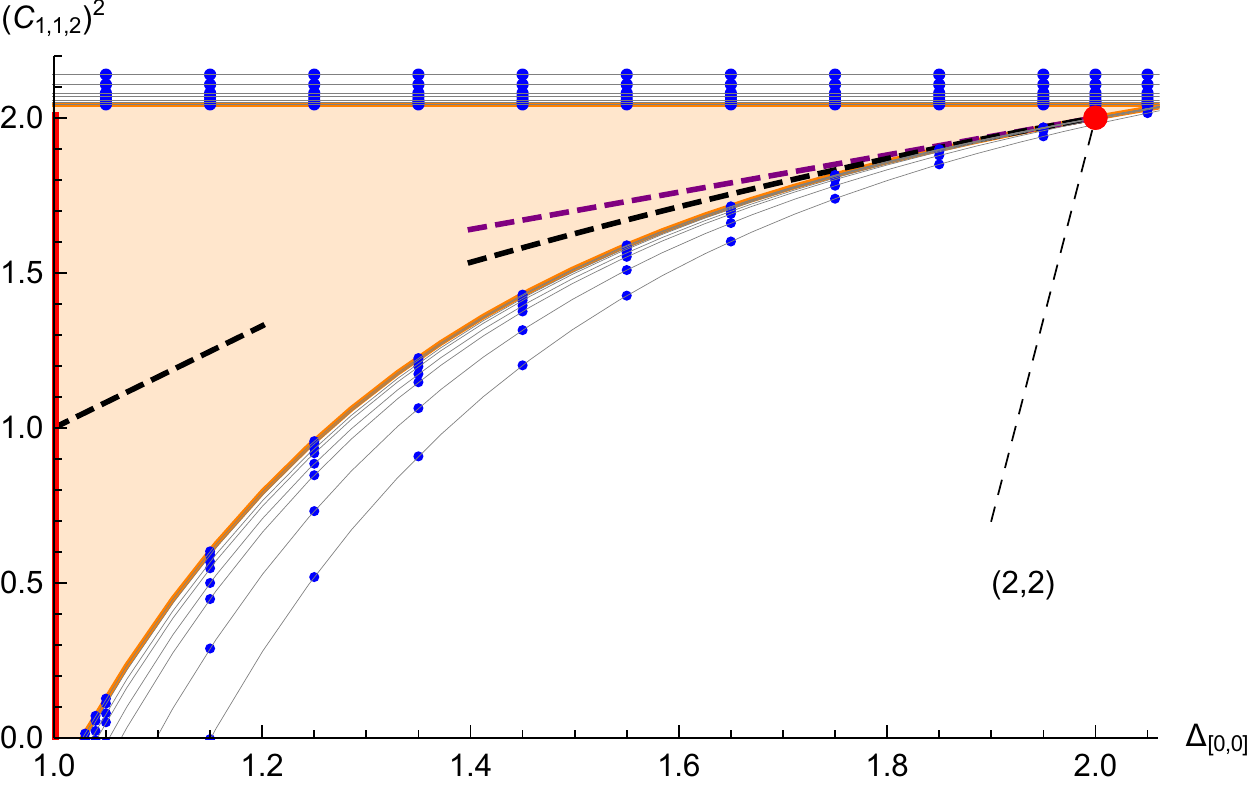}
              \caption{Upper/Lower Bounds on the OPE coefficient $C_{1,1,2}^2$ for $\Lambda=10, 20, \ldots, 80$. The allowed region for $\Lambda=80$ is shaded. The point $(\Delta_{[0,0]}=2, C_{1,1,2}^2=2)$ represented by a bigger red dot is occupied by the solution \eqref{eq: most general analytic solution} with $\xi=1$. The thick red line refers to the solutions with $-1\leq \xi<1$. For a given $\Lambda$, the value of $\Delta_{[0,0]}$ for which a non-trivial lower bound on $C_{1,1,2}$ appears is the value of $\Delta_{[0,0]}$ for which there is a kink on the RHS of figure~\ref{fig:DDBootstrapDelta0}. 
              The black dashed line  starting from the point $(1,1)$ represents the behaviour of Wilson lines in planar $\mathcal{N}=4$ SYM for which $\Delta_{[0,0]}=1+\tfrac{\lambda}{4\pi^2}+\dots$ as first computed in \cite{Alday:2007he} and  $C_{112}^2=1+\tfrac{\lambda}{24}+\dots$ as follows from localization.
              The (upper) purple and (lower) black dashed curves starting from the point $(2,2)$  are the first and second order perturbative approximation of the lower bound curve given  by
             \eqref {aB2analytic}, compare to \eqref{numericslopSTRONG}. 
              }
              \label{fig:DDBootstrapOPEshort}
            \end{center}
\end{figure}

In the theories that are not free, it is expected that the semi-short $\mathcal{C}_{[2,0]}$ would be absent. In our framework, this implies setting $\Delta_{[0,0]}>1$. Computing the upper and lower bounds on the OPE $C_{1,1,2}^2$ in this case leads to the bounds of figure~\ref{fig:DDBootstrapOPEshort}.
As an aside, we note that since a long at the unitarity bound becomes a semi-short \eqref{eq: long block at the unitarity bound}, the analytic solutions \eqref{eq: most general analytic solution} with\footnote{For these values of $\xi$ the analytic solutions contain a semi-short, see the block expansion \eqref{eq: most general analytic solution Block Expansion}. } $\xi\in [-1,1)$ will appear in figure~\ref{fig:DDBootstrapOPEshort} for $\Delta_{[0,0]}=1$. This is the reasoning behind the red line in figure~\ref{fig:DDBootstrapOPEshort}. 

We note furthermore, that we can compute the slope of the lower bound in $C_{1,1,2}$ around $\Delta_{[0,0]}=2$. Specifically, the lower bound of Figure~\ref{fig:DDBootstrapOPEshort} at $\Delta_{[0,0]}=2$ gives $C_{1,1,2}\geq 1.9998$ with the tangent vector $(1,0.6063)$ at that point.
In fact, we can compute even more terms and write (by Taylor-expanding the interpolation of the $\Lambda=80$ result in \textsf{Mathematica}),
\beq\label{numericslopSTRONG}
\text{lower bound of }C_{1,1,2}^2(\Delta_{[0,0]})=1.9998+0.6063 (\Delta_{[0,0]}-2)-0.3801(\Delta_{[0,0]}-2)^2+\mathcal{O}(\Delta_{[0,0]}-2)^3\,,
\eeq
where we would like to remark that we have significantly less control over the second order term.

On the other hand, analyzing the lower bound around $\Delta_{[0,0]}=1$ is difficult. From the way that the intersection of the lower bound with the $\Delta_{[0,0]}$ axis moves to the left as $\Lambda$ increases, it seems natural to expect that at $\Lambda=\infty$ the only way to have $C_{1,1,2}=0$ is to also have $\Delta_{[0,0]}=1$. For these values we have an analytic solution, namely \eqref{eq: most general analytic solution} with $\xi=-1$. It would be very interesting to know the value of the slope of the lower bound at that point for $\Lambda=\infty$, but the numerics do not seem to be able to give us a conclusive answer. 

\begin{figure}[htbp!]
             \begin{center}       
              \includegraphics[scale=0.8]{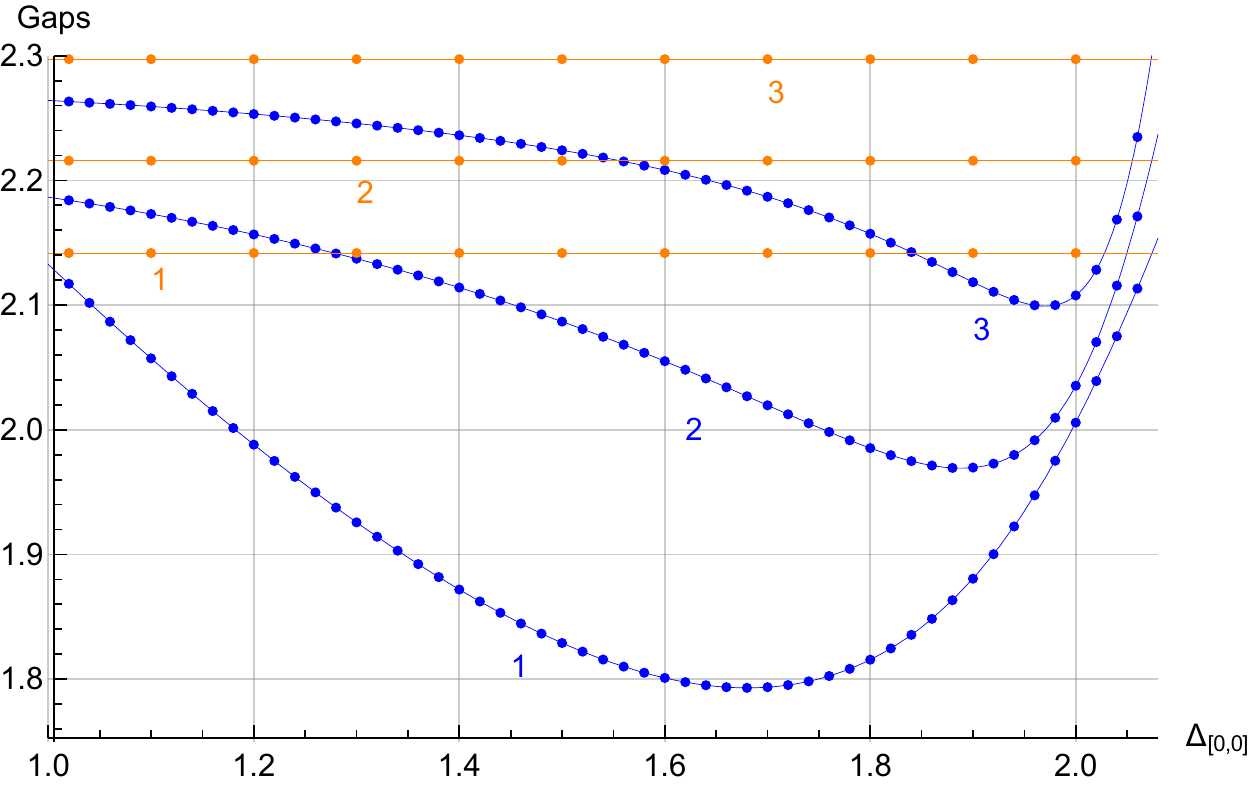}
              \caption{Plot of the difference $\Delta^{(i)}_{[0,0]}-\Delta^{(i-1)}_{[0,0]}$ (for $i=1,2,3$) between the conformal dimensions of the lowest lying longs as a function of the gap $\Delta^{(0)}_{[0,0]}\equiv \Delta_{[0,0]}$. We have depicted in blue the spectrum of for the lower bound and in orange the spectrum of the upper bound with the numerics done for $\Lambda=80$. Observe that the upper bound spectrum is independent of $\Delta_{[0,0]}$ and that the two become identical for the maximal value of $\Delta_{[0,0]}$ allowed for $\Lambda=80$. }
              \label{fig:C112Spectrum}
            \end{center}
\end{figure}

It is interesting to take a more careful look at the spectrum $\{\Delta_{[0,0]}^{(0)},\Delta_{[0,0]}^{(1)}, \ldots\}$ of longs operators of the theories that extremize the $C_{1,1,2}^2$ bounds of figure~\ref{fig:DDBootstrapOPEshort}. One can extract this spectrum by computing the zeroes of the extremizing functional \cite{ElShowk:2012hu}, the results are presented in figure~\ref{fig:C112Spectrum} where we plot the differences $\Delta^{(i)}_{[0,0]}-\Delta^{(i-1)}_{[0,0]}$ (for $i=1,2,3$) between the conformal dimensions of the lowest-lying longs as a function of the gap $\Delta^{(0)}_{[0,0]}\equiv \Delta_{[0,0]}$.
We see that for $\Delta_{[0,0]}=1$ and for $\Delta_{[0,0]}=2$ the gaps of the extremizing solutions are roughly equal to 2. This is also the case for the analytic solutions with $\xi=-1$ (for $\Delta_{[0,0]}=1$) and $\xi=1$ (for $\Delta_{[0,0]}=2$), see the block decompositions \eqref{eq: most general analytic solution Block Expansion}. On the other hand, the analytic solutions with $-1<\xi<1$ have gaps of 1 between the conformal dimensions of the long operators. This suggests that if we want to perform a conformal perturbation analysis that starts from the point $\Delta_{[0,0]}=1, C_{1,1,2}^2=0$ and follows the lower bound curve of figure~\ref{fig:DDBootstrapOPEshort}, we would need to start from a spectrum for which the long operators have gaps of 2. 

Interestingly, we also see in figure~\ref{fig:C112Spectrum} that the spectra of the lower bound (in blue) and of the upper bound (in orange) agree for the maximal possible value of $\Delta_{[0,0]}$ for our $\Lambda$. Hence, this suggests that the spectrum of the theory at the right tip of the allowed ``triangle'' in figure~\ref{fig:DDBootstrapOPEshort} should be unique, at least as far as the single $\calD_1$ correlator is concerned. This is compatible with the bound of the RHS of figure~\ref{fig:DDBootstrapDelta0}, though that latter one only provides an upper bound on the gaps and not a lower one.

Lastly, we can obtain upper bounds on the OPE coefficient of the first long operator in the spectrum, namely $\mathcal{L}_{[0,0]}^{\Delta_{[0,0]}}$,  for a given value of $\Delta_{[0,0]}$ and of $C_{1,1,2}^2$. The results are depicted on the left side of figure~\ref{fig:DDBootstrapOPElong2D}. 

\begin{figure}[htbp!]
             \begin{center}       
              \includegraphics[scale=0.67]{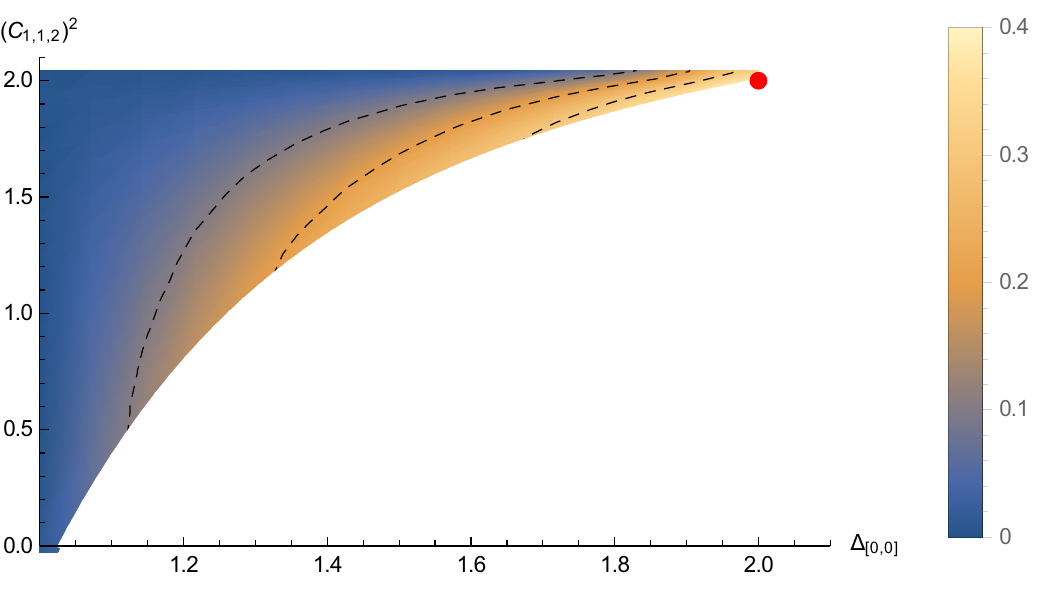}\quad \quad \includegraphics[scale=0.5]{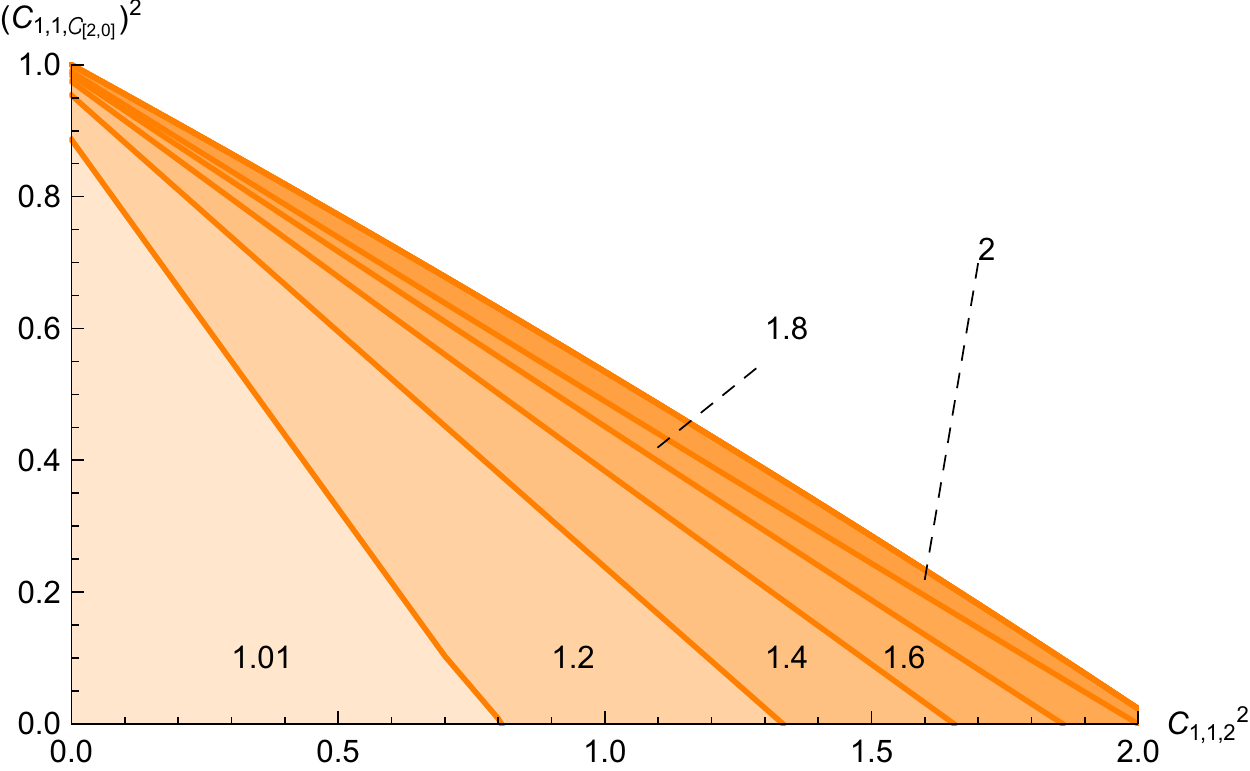}
              \caption{Left: Upper bounds on the OPE coefficient $\big(C_{1,1,\mathcal{L}_{[0,0]}^{\Delta_{[0,0]}}}\big)^2$ for given $\Delta_{[0,0]}$ and of $C_{1,1,2}^2$. The numerics were done for $\Lambda=80$. The upper bound at the analytic solution for $\xi=1$, represented by the red dot, is $\big(C_{1,1,\mathcal{L}_{[0,0]}^{2}}\big)^2\leq 0.4$. To our precision, the numerics exactly saturate the bound. The dashed lines show the levels $0.1$, $0.2$ and $0.3$. Right: Allowed region for the OPE coefficients $C_{1,1,2}^2$ and $C_{1,1,\mathcal{C}_{[2,0]}}^2$ for a given value of $\Delta'_{[0,0]}$. Again, the numerics were done for $\Lambda=80$. The upper bound does not change as $\Delta'_{[0,0]}$ is varied and we denote the value $\Delta'_{[0,0]}$ in black close to the lower bound. Note the region for a given $\Delta'_{[0,0]}$ contains the regions for larger values of $\Delta'_{[0,0]}$.}
              \label{fig:DDBootstrapOPElong2D}
            \end{center}
\end{figure}

\begin{figure}[htbp!]
             \begin{center}       
              \includegraphics[scale=0.8]{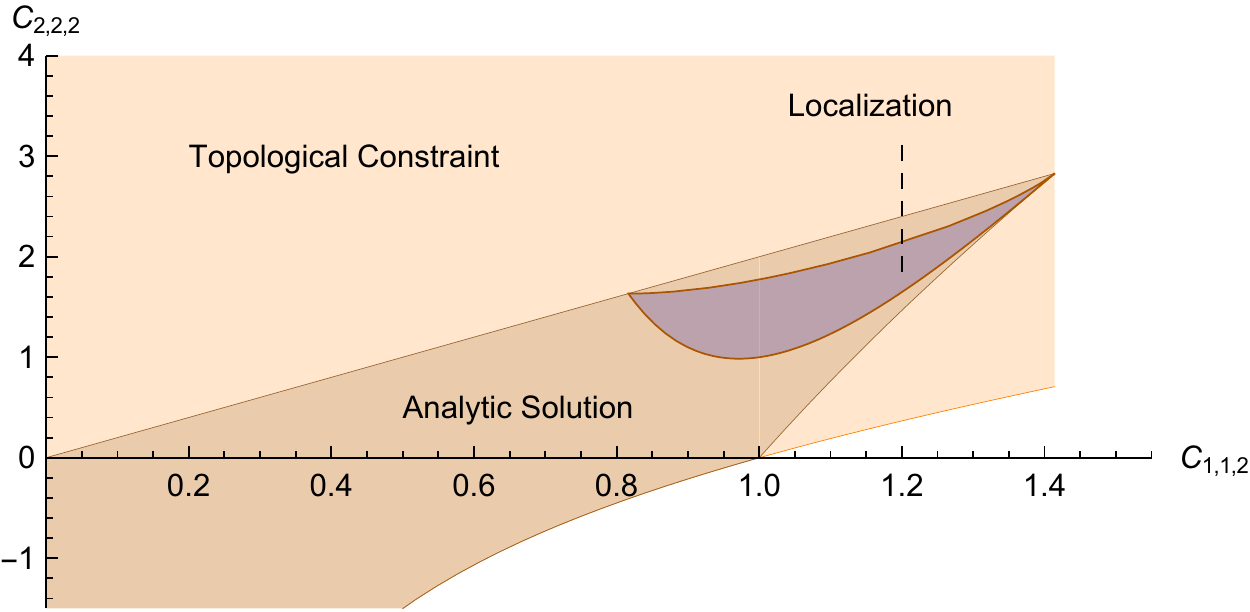}
              \caption{Allowed region for  $C_{1,1,2}$, $C_{2,2,2}$ from the topological contraint (which coincides with the numerical analysis for small values of the gaps above the unitarity bounds). There is a general upper bound on $C_{1,1,2}$, namely $C_{1,1,2}\leq \sqrt{2}$, but there is no upper or lower bound on $C_{2,2,2}$. On top are the analytical solutions \eqref{eq: most general analytic solution}, and the localization region from formulas \eqref{eq: localization formula for C112}, \eqref{eq: localization formula for C222}. 
}
              \label{fig:GeneralOverviewC112C222}
            \end{center}
\end{figure}

\paragraph{The full mixed system. } In the full system of the crossing equation \eqref{eq: final crossing equations 2}, we can set $C_{1,1,2}=r \cos(\theta)$ and $C_{2,2,2}=r\sin(\theta)$ and search for bounds on $r$ as a function of $\theta$. The general situation is illustrated in figure~\ref{fig:GeneralOverviewC112C222}. There we show the areas allowed by the topological sector relation, the region covered by the analytic solutions \eqref{eq: most general analytic solution Block Expansion}, and also the most general localization region. 

\begin{figure}[htbp!]
             \begin{center}                  
              \includegraphics[scale=0.55]{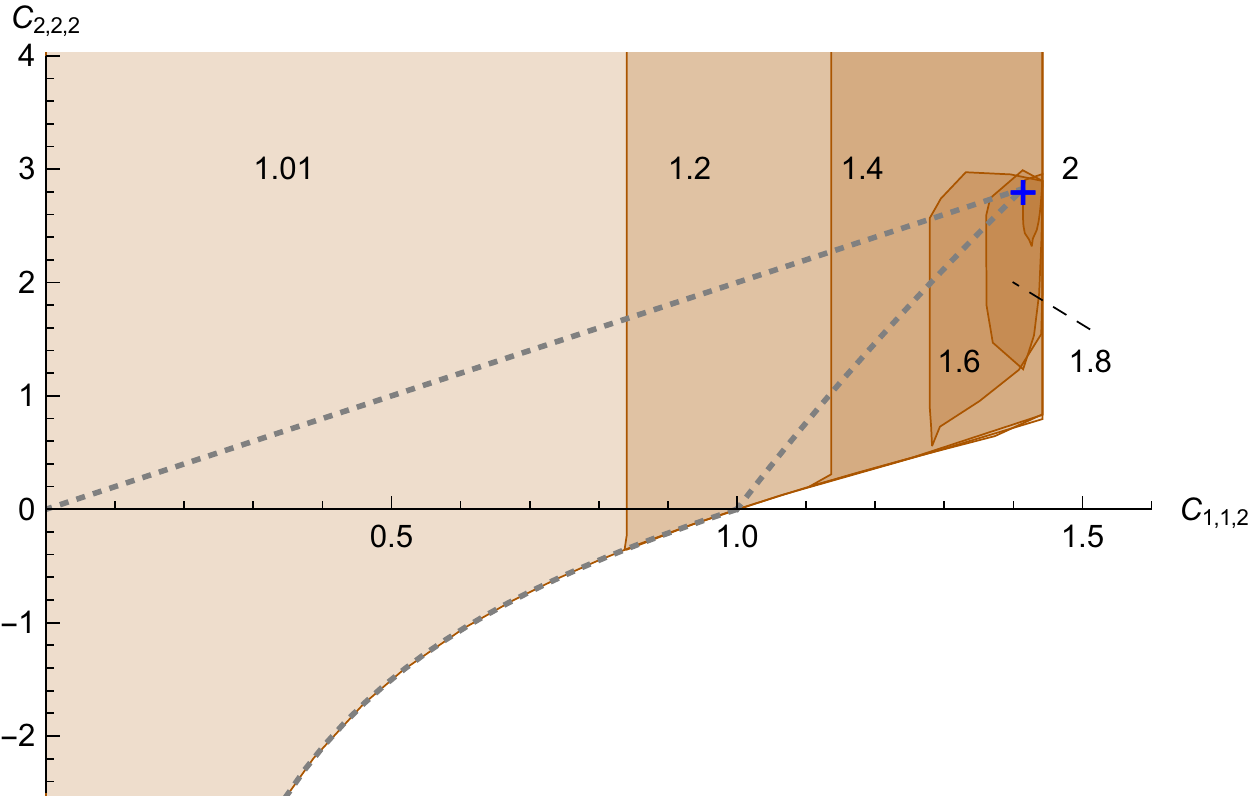}\quad \quad \includegraphics[scale=0.55]{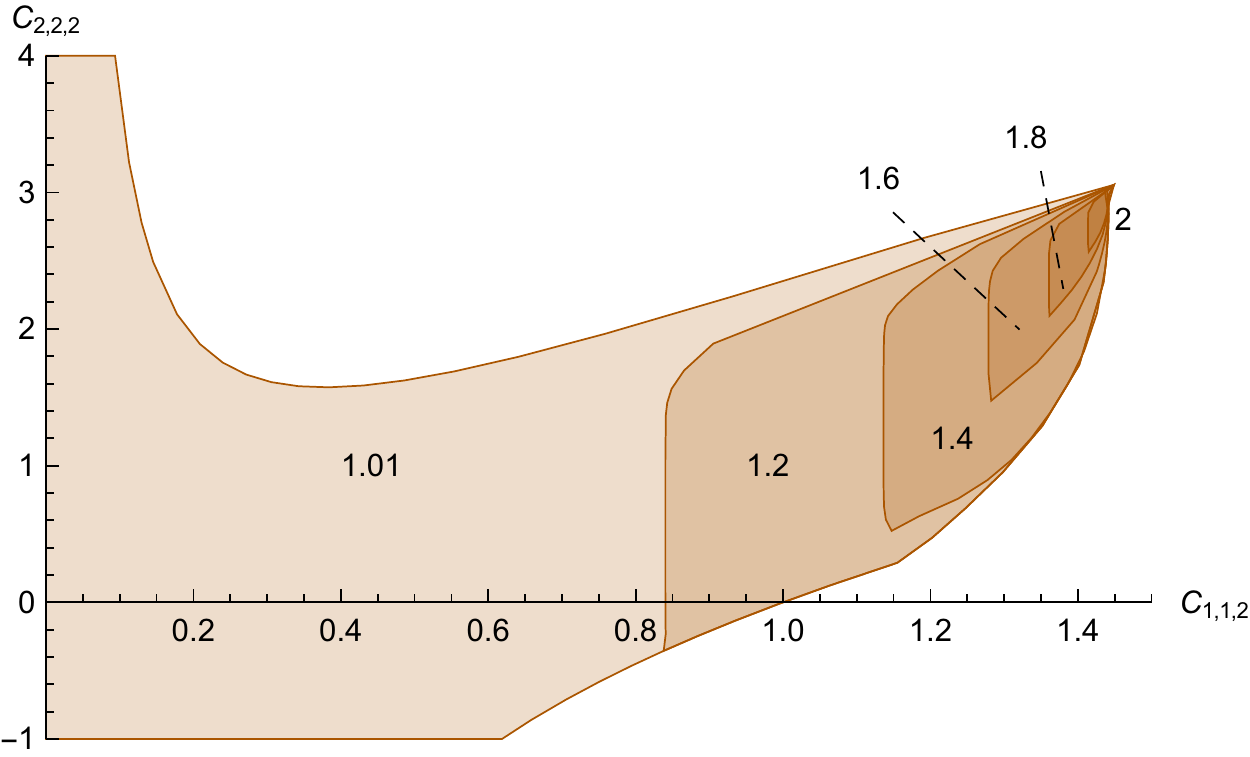}
                        \caption{
                        Left: bounds for $\Lambda=40$ and for the gaps $\Delta_{[0,0]}=1.01, 1.2, 1.4, 1.6, 1.8, 2$, while keeping the other gaps just above their respective unitarity bounds. The allowed region for given $\Delta_{[0,0]}$ contains the allowed regions for larger values of $\Delta_{[0,0]}$. The small blue cross at $(\sqrt{2},2\sqrt{2})$ is an analytic solution for $\Delta_{[0,0]}=2$. The subregion enclosed by the dotted curve comes from the analytic solutions \eqref{eq: most general analytic solution} with the parameters \eqref{eq: analytic solution reparametrization} subject to \eqref{eq: allowed range xi and eta} and is also shown in figure~\ref{fig:GeneralOverviewC112C222}. Right: Allowed region for $C_{1,1,2}$ and $C_{2,2,2}$ if $C_{2,2,2'}=0$. The numerics were done for $\Lambda=40$ and for $\Delta_{[0,0]}=1.01, 1.2, 1.4, 1.6, 1.8,  2$. The other gaps are just slightly above their unitarity bounds. Note that the allowed region for a given $\Delta_{[0,0]}$ contains the regions for larger values of $\Delta_{[0,0]}$.}
             \label{fig:C112C222BoundsVaryDelta}
            \end{center}
\end{figure}

Turning now to the numerical analysis and assuming the lowest possible values for the gaps just slightly above the unitarity bounds,\footnote{This way, we exclude the presence of the semi-shorts operators.} namely $\Delta_{[0,0]}=1.01$, $\Delta_{[2,0]}=3.01$, $\Delta_{[0,2]}=3.01$ and $\Delta_{[0,1]}=2.01$, we get the results of the LHS of figure~\ref{fig:C112C222BoundsVaryDelta}. 

We can leave the gaps $\Delta_{[2,0]}$, $\Delta_{[0,2]}$ and $\Delta_{[0,1]}$ just above the unitarity bound and  vary the gap $\Delta_{[0,0]}$. Doing so, we get the results of the LHS of figure~\ref{fig:C112C222BoundsVaryDelta}, where we have also overlayed the allowed region for the analytic solutions \eqref{eq: most general analytic solution}. We observe that until $\Delta_{[0,0]}\approx 1.6$, there is no upper bound on $C_{2,2,2}$. The appearance of this upper bound, which transforms the allowed region into an island might be connected to the drop in the upper bound on $\Delta_{[0,1]}$, see figure~\ref{fig:Delta20And02vsDelta00Full}. The fact that, for suitable gaps in the long spectrum, the allowed region for the OPE coefficients $C_{1,1,2}$ and $C_{2,2,2}$ becomes an island can be compared with similar phenomena in \cite{Kos:2016ysd, Dymarsky:2017yzx}. 

It is also interesting to investigate the consequences of the decoupling of the operator $\calD_2'$. This can happen due to the multiplicity of $\mathcal{B}_2$ being equal to one as in the case of the $\SU(2)$ theory with $R$ the fundamental representation, or in the case of the analytic solution of appendix~\ref{app: analytic solutions}, see \eqref{eq: OPE for the modified Wick prescription}. Alternatively, it could be that the multiplicity is higher than one but that the operator $\calD_2'$ still decouples, implying $C_{2,2,2'}=0$. The results for the allowed OPE of $C_{1,1,2}$ and $C_{2,2,2}$ for various values of $\Delta_{[0,0]}$ are shown on the RHS of figure~\ref{fig:C112C222BoundsVaryDelta}. The main difference with the general case is the appearance of an upper bound on $C_{2,2,2}$ even for very small values of the gap $\Delta_{[0,0]}$.

\begin{figure}[htbp!]
             \begin{center}       
              \includegraphics[scale=0.5]{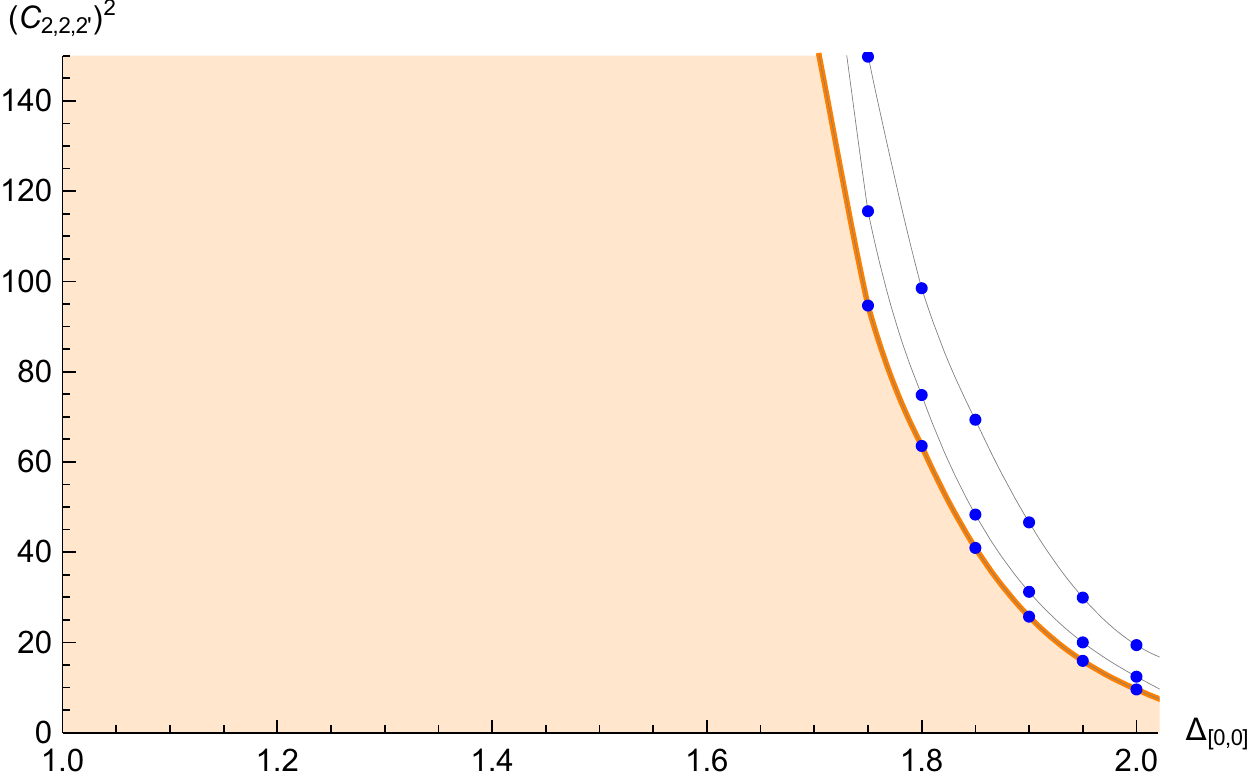}\qquad  \qquad  \includegraphics[scale=0.5]{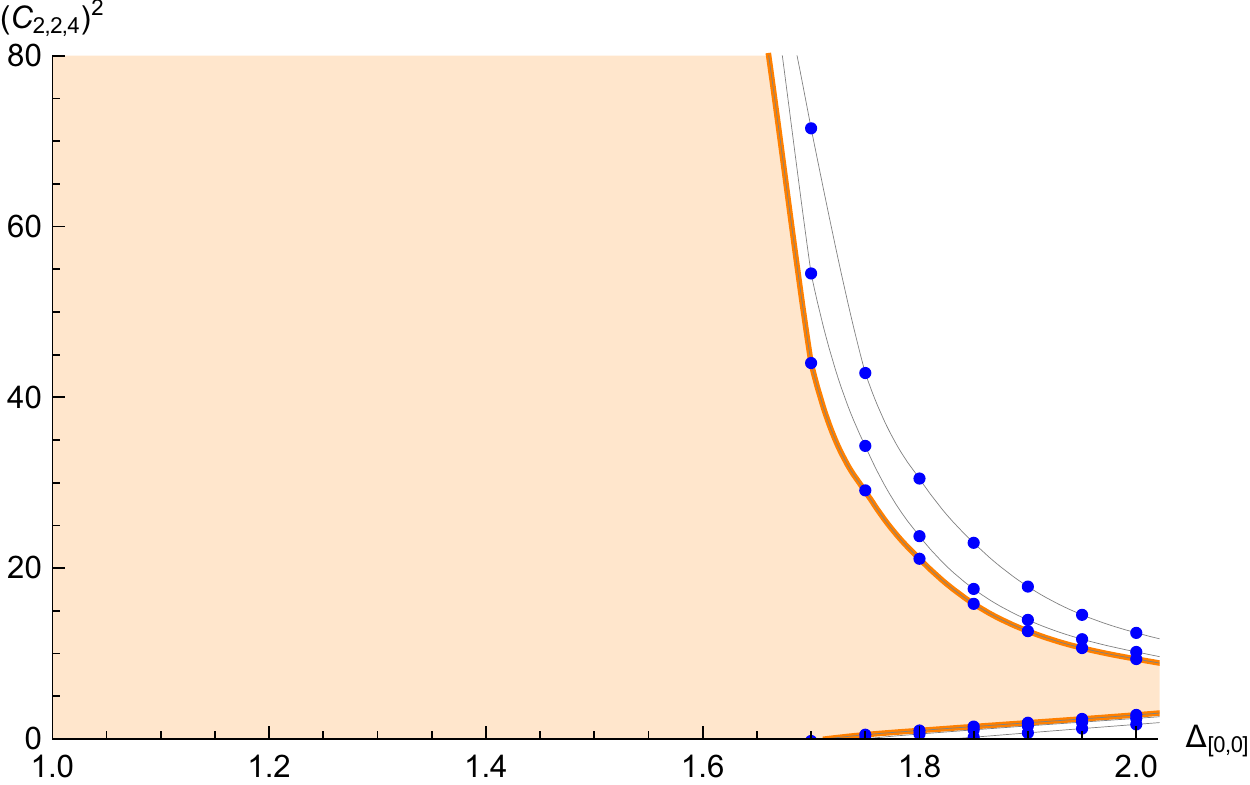}
                        \caption{Left: Upper bounds on the OPE coefficient $C_{2,2,2'}^2$ as a function of $\Delta_{[0,0]}$. We find no lower bounds. One the right: Upper/lower bounds on the OPE coefficient $C_{2,2,4}^2$ as a function of $\Delta_{[0,0]}$. The numerics are done for $\Lambda=20,30, 40$. We see that increasing $\Lambda$ improves the bounds significantly and it is plausible that the bounds will converge to the strong coupling solution  $C_{2,2,2'}^2=0$, $ C_{2,2,4}^2=6$.}
             \label{fig:OPEboundsD2primeD4}
            \end{center}
\end{figure}

We can also obtain upper/lower bounds on the remaining short operators $\calD_2'$ and $\calD_4$. Keeping again the gaps $\Delta_{[2,0]}$, $\Delta_{[0,2]}$ and $\Delta_{[0,1]}$ just above the unitarity bound and varying $\Delta_{[0,0]}$, we find the results of figure~\ref{fig:OPEboundsD2primeD4}. Compared to the other OPE bounds, they are weaker, in particular the one for $C_{2,2,2'}$. No lower bound for $C_{2,2,2'}$ was found, which is consistent with the possibility of setting $C_{2,2,2'}=0$ in the plot of the RHS of figure~\ref{fig:C112C222BoundsVaryDelta}, and yet still obtaining results for all allowed values of $\Delta_{[0,0]}$. 

\begin{figure}[htbp!]
             \begin{center}       
              \includegraphics[scale=1]{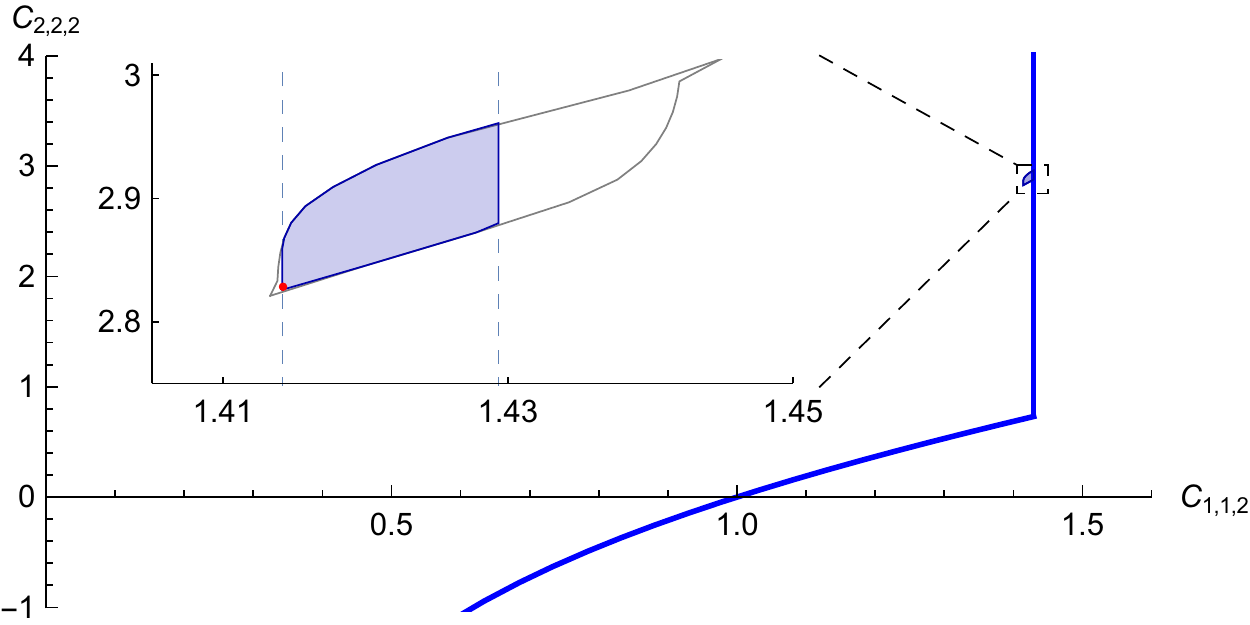}
                        \caption{Allowed region for the OPE coefficients $C_{1,1,2}$ and $C_{2,2,2}$ for the gaps $\Delta_{[0,0]}=2$, $\Delta_{[2,0]}=5$, $\Delta_{[0,2]}=4$ and $\Delta_{[0,1]}=3$. The allowed region is shaded in blue and is the result of intersecting the numerics for the mixed system done for $\Lambda=40$ with the bound $1.414\leq C_{1,1,2}\leq 1.429$ done for $\Lambda=80$, see figure~\ref{fig:DDBootstrapOPEshort}. The red cross corresponds to $(\sqrt{2},2\sqrt{2})$ for which we have an analytic solution. The thick blue lines correspond to the inequality $C_{2,2,2}\geq C_{1,1,2}-C_{1,1,2}^{-1}$ \eqref{eq: inequality C112 vs C222} and to the bound $C_{1,1,2}\leq 1.429$. We zoom in on the allowed region. }
             \label{fig:C112C222BoundsSC}
            \end{center}
\end{figure}

\paragraph{The strong coupling case.} Lastly,
we can impose that the gap structure  is the one of the leading strong coupling solution given in \eqref{corrSTRONG}, namely 
$\Delta_{[0,0]}=2$, $\Delta_{[2,0]}=5$, $\Delta_{[0,2]}=4$ and $\Delta_{[0,1]}=3$ and compute bounds on the OPE coefficients $C_{1,1,2}$ and $C_{2,2,2}$.
This results in an island of allowed values shown in figure~\ref{fig:C112C222BoundsSC}. 
Additionally, we can combine the stronger (for $\Lambda=80$) upper/lower bounds on $C_{1,1,2}$ from the analysis  of the $\calD_1$ four-point function in figure~\ref{fig:DDBootstrapOPEshort}. This excludes about half of the island that the $\Lambda=40$ numerics for the mixed system have given us. 
From figure~\ref{fig:C112C222BoundsSC}, we read that for this gap structure we have in particular the inequalities
\beq
\label{eq: strong coupling gap}
\Delta_{[0,0]}=2\,,\  \Delta_{[2,0]}=5\,,\  \Delta_{[0,2]}=4\,,\  \Delta_{[0,1]}=3\ \Rightarrow \left\{\begin{array}{l}1.414\leq C_{1,1,2}\leq 1.429\\ 2.821\leq C_{2,2,2}\leq 2.961\end{array}\right.\,.
\eeq
The above is suggestive of there being only one possible value of the OPE coefficients $C_{1,1,2}$ and $C_{2,2,2}$ that solves the crossing equations for $\Lambda\rightarrow \infty$. For the OPE coefficients $C_{2,2,2'}$ and $C_{2,2,4}$, the situation is less clear. Just imposing the gap $\Delta_{[0,0]}=2$ for $\Lambda=40$ does not place high enough restrictions on them. However, the situation improves if we also demand that $\Delta_{[2,0]}=5$, $\Delta_{[0,2]}=4$, $\Delta_{[0,1]}=3$ and, since we are still in a regime of $\Lambda$ in which the numerics improve drastically with increased precision, it is possible that for $\Lambda\rightarrow \infty$ the OPE coefficients would be restricted to the values $C_{2,2,2'}=0$ and $C_{2,2,4}=6$ which correspond to the strong coupling solution.

The expectation then is that this is the unique solution to crossing with the maximal gap 
$\Delta_{[0,0]}=2$. Further support of this claim could be produced by increasing the values of $\Lambda$ and extracting the spectrum as in \cite{Simmons-Duffin:2016wlq}.
This behavior is somewhat similar to the one obtained in  \cite{Beem:2013qxa,Beem:2016wfs} 
when considering the four-point function of stress-tensor supermultiplets in $4d$ $\mathcal{N}=4$ SCFT. In that case, the extremal solution to crossing at large central charge seems to coincide with mean field theory, and its first correction with tree level supergravity, 
see conjecture 3 in \cite{Beem:2016wfs}. 

%% file: sections/analytical_results.tex

\section{Analytical results}
\label{sec: analytical results}

In this section we present an analytic study of the  four-point function of the displacent operator $\mathcal{D}_1$ corresponding to the lower bound in figure~\ref{fig:DDBootstrapOPEshort} in the vicinity of the point $(\Delta_{[0,0]},C_{1,1,2}^2)=(2,2)$.\footnote{
Recently, there has been substantial progress in the application of  analytic bootstrap methods to problems of the type addressed here, see e.g.~\cite{Aharony:2018npf} and references therein.
}  The latter is associated  to a very simple four-point function and coincides with the leading strong-coupling solution $\mathcal{A}^{(0)}$, see \eqref{A0A1Stong}, described in section \ref{sec: Strong couplingy}.
Correlators saturating the lower bound of figure~\ref{fig:DDBootstrapOPEshort} are  solutions of crossing for which the number of operators exchanged is minimized. The end points of the lower bound curve, namely $(1,0)$ and  $(2,2)$ in figure~\ref{fig:DDBootstrapOPEshort}, illustrate this point neatly: the (non half-BPS) operators (in the $\mathcal{L}_{[0,0]}^{\Delta}$ representation) being exchanged are given by the sets $\{\Delta\}_{(1,0)}=\{1,3,5,\dots\}$ and $\{\Delta\}_{(2,2)}=\{2,4,6,\dots\}$ respectively. In between these extrema the spectrum of dimensions  starts from a lower bound $1<\Delta_{[0,0]}<2$  and continues with a spacing of roughly  two units, see right-hand side of  figure~\ref{fig:DDBootstrapDelta0} and figure~\ref{fig:C112Spectrum}.

In the following, we derive the first and second order perturbations\footnote{
We thank Fernando Alday for important discussions on this problem and for sharing some unpublished notes with us.
} of  the  $(2,2)$ solution corresponding to the lower bound curve. The first order perturbation  coincides with the string theory result described in section \ref{sec: Strong couplingy}. 
 At second order, the lower bound solution might differ from the second order perturbation in  string theory due to degeneracies of the operators. We comment on this point in the end of the section. The point $(1,0)$ in figure~\ref{fig:DDBootstrapOPEshort} corresponds to a very simple four point function as well, see  \eqref{4pt10}. Unfortunately, the vicinity of this point, for which the gap $\Delta_{[0,0]}$ approaches the unitarity bound, is hard to probe with the numerics as the convergence of the bound to its  $\Lambda\rightarrow \infty$  limit is very slow in this region.
 We present an analysis of the vicinity of this point in appendix \ref{app:10Perturbation}.
It is relatively easy to generalize such expansions to the case of mixed correlators and this is crucial to resolve the issue of degeneracy and make contact with the second order correction in string theory. We postpone this interesting problem to future work.

\subsection{Setup}

As explained in section \ref{sec: preliminaries}, we parametrize the four-point function of displacement operators in terms of a function $f(\chi)$ and a constant $F$. They can be expanded in superconformal blocks as
\begin{equation}\label{blockExpansioninSetup}
f(\chi)=f_{\text{id}}(\chi)
+a_{\mathcal{B}_2}\,f_{\mathcal{B}_2}(\chi)
+a_{\mathcal{C}_{[2,0]}}\,f_{\mathcal{C}_{[2,0]}}(\chi)+
\sum_{\Delta \in \mathsf{S}}\,a_{\Delta}\,f_{\Delta}(\chi)\,,
\qquad
F=1+a_{\mathcal{B}_2}\,,
\end{equation}
where the blocks are given in \eqref{blockID}, \eqref{blockB2}  \eqref{blockLong} and 
\eqref{eq: long block at the unitarity bound}. In order to shorten the notation we use 
$f_{\Delta}:=f_{\mathcal{L}^{(\Delta)}_{[0,0]}}$,
and introduce the notation $a_{\star}$ for OPE coefficients $C_{1,1,\star}^2$. The crossing equation reads
\begin{equation}\label{crossingSec6}
(\chi-1)^2f(\chi)+\chi^2f(1-\chi)=0\,,
\end{equation}
where $\chi\in [0,1]$.
We will consider a perturbation of a given solution to crossing denoted by $\left(f^{(0)}(\chi), F^{(0)}\right)$, with associated CFT data $a_{\star}^{(0)}, \mathsf{S}^{(0)}$ such that the number of operators appearing in the OPE is unchanged and no hidden degeneracy is lifted by the perturbation.
We introduce the notation
\begin{equation}\label{crossingf}
f(\chi)=f^{(0)}(\chi)+\epsilon \,f^{(1)}(\chi)+\dots\quad
a_{\star}=a_{\star}^{(0)}+\epsilon \,a_{\star}^{(1)}+\dots\quad
\mathsf{S}=\{\Delta+\epsilon\,\gamma^{(1)}_{\Delta}+\dots\}_{\Delta\in  \mathsf{S}^{(0)}}\,,
\end{equation}
with $\star \neq \mathcal{C}_{[2,0]}$. We will discuss the special case of including that operator shortly.

The crossing equations \eqref{crossingf} are valid order by order in $\epsilon$.  The conformal block expansion on the other hand mixes CFT data from different orders, which is crucial. 
Expanding the conformal block decomposition in $\epsilon$ gives  at first order
\begin{equation}\label{f1logandnolog}
f^{(1)}(\chi)=f^{(1)}_{\log}(\chi)\log \chi+f^{(1)}_{\log^0}(\chi)\,,
\end{equation}
with the two new functions given by
\begin{align}\label{f1LogandnoLogdef}
f^{(1)}_{\log}(\chi)&=\sum_{\Delta \in \mathsf{S}^{(0)}} \,a^{(0)}_{\Delta}\gamma^{(1)}_{\Delta}
f_{\Delta}(\chi)\,
+\,a_{\mathcal{C}}\,\gamma^{(1)}_{\mathcal{C}}\,f_{\mathcal{C}}(\chi)\,,
\\
\qquad
f^{(1)}_{\log^0}(\chi)&=a^{(1)}_{\mathcal{B}_2}\,f_{\mathcal{B}_2}(\chi)+
\sum_{\Delta \in \mathsf{S}^{(0)}}
\left(
a^{(1)}_{\Delta}f_{\Delta}(\chi)
+
a^{(0)}_{\Delta}\gamma^{(1)}_{\Delta}\,f^{(1)}_{\Delta}(\chi)
\right)
+
\left(
\tilde{a}_{\mathcal{C}}\,f_{\mathcal{C}}(\chi)
+
a_{\mathcal{C}}\gamma^{(1)}_{\mathcal{C}}\,f^{(1)}_{\mathcal{C}}(\chi)
\right)\,,\nonumber
\end{align}
where for notational convenience $\mathcal{C}\equiv \mathcal{C}_{[2,0]}$.
In the above, we have defined
\begin{equation}
f^{(\ell)}_{\Delta}(\chi):=\chi^{\Delta}
\left( \frac{\partial}{\partial \Delta}\right)^{\ell}
\left(\chi^{-\Delta} f_{\Delta}(\chi)\right)\,.
\end{equation}
This function has a regular expansion around $\chi =0$ starting at order $\chi^{\Delta+1}$, so that $f^{(1)}_{\log}(\chi)$  and 
$f^{(1)}_{\log^0}(\chi)$ are both regular at $\chi =0$.
It should be noted that 
the operator of type $\mathcal{B}_2$ cannot have anomalous dimension, but the OPE coefficient $a_{\mathcal{B}_2}$ 
can vary with $\epsilon$.  The contribution of $\mathcal{C}_{[2,0]}$ requires a small discussion.  
Operators of type $\mathcal{C}_{[2,0]}$
appear as a subrepresentation of long operators at the unitarity bound:
 $\mathcal{L}^{\Delta=1}_{[0,0]}=\mathcal{C}_{[2,0]}+\dots$, and 
the order $\epsilon^0$ contribution results from a cancellation 
between a pole in the conformal block with a zero in the OPE coefficient.\footnote{
More explicitly
\begin{align}
f_{1+\epsilon \gamma_1+\epsilon^2 \gamma_2+\dots}(\chi)
&=\frac{1}{\epsilon\, \gamma_1}\,f_{\mathcal{C}_{[2,0]}}(\chi)+\left(\log \chi -\frac{\gamma_2}{\gamma_1^2}\right)
f_{\mathcal{C}_{[2,0]}}(\chi)+\,f^{(1)}_{\mathcal{C}_{[2,0]}}(\chi)+\dots \,.\nonumber
\,
\\
a_{\Delta=1+\epsilon \gamma_1+\epsilon^2 \gamma_2+\dots}&=(\epsilon\, \gamma_1+\epsilon^2 \,\gamma_2+\dots)\,\left(a_{\mathcal{C}_{[2,0]}}
+\epsilon \,\tilde{a}_{\mathcal{C}_{[2,0]}}\right)+\dots\,.\nonumber
\end{align}
It is worth mentioning that this is exactly what happens for Wilson lines in $\mathcal{N}=4$ SYM at weak coupling for the multiplet with highest weight $\phi^{I=6}$. It is indeed cleat that $\langle \phi^a \phi^b \phi^6 \rangle=\mathcal{O}(\epsilon)$.
The multiplet recombination in this case has been analyzed in  \cite{Cooke:2017qgm}.
} 
The first-order correction \eqref{f1logandnolog} should be crossing symmetric, see \eqref{crossingSec6}.
In order to make it manifest we rewrite it as
\begin{equation}\label{f1form}
 f^{(1)}(\chi) = r(\chi)\log(\chi)
- \frac{\chi^2}{(1-\chi)^2}r(1-\chi)\log(1-\chi) + q(\chi)\,,
\end{equation}
where  $q(\chi)$ is crossing symmetric by itself
 and has a regular expansion around $\chi=0$.
Comparing \eqref{f1logandnolog} with \eqref{f1form} we obtain
\begin{equation}\label{f1sFROMrq}
f^{(1)}_{\log}(\chi)=r(\chi)\,,
\qquad
f^{(1)}_{\log^0}(\chi)=
- \frac{\chi^2}{(1-\chi)^2}r(1-\chi)\log(1-\chi) + q(\chi)\,.
\end{equation}
Let us now turn to the description of the solutions to crossing that we are going to perturb.
\paragraph{The free solution.}
We recall that the solution of crossing corresponding to the point $(2,2)$ of figure~\ref{fig:DDBootstrapOPEshort} is 
\begin{equation}
\label{eq: strong coupling free solution}
\langle\mathcal{D}_1\mathcal{D}_1\mathcal{D}_1\mathcal{D}_1  \rangle_{(2,2)}=
(12)(34)+(13)(24)+(14)(23)\,,
\end{equation}
where  $(ij)$ denotes the super-propagator defined in \eqref{twoPOINT}. 
In the parametrization of \eqref{eq:WI_solution} the solution \eqref{eq: strong coupling free solution} corresponds to $f^{(0)}=\tfrac{\chi(2\chi-1)}{\chi-1}$ and $F^{(0)}=3$
 from which one extracts the CFT data:
\begin{equation}
(2,2):\qquad a^{(0)}_{\mathcal{B}_2}=2,\quad a^{(0)}_{\mathcal{C}_{[2,0]}}=0,\quad
 a^{(0)}_{\Delta}=\tfrac{\Gamma(\Delta+3)\Gamma(\Delta+1)(\Delta-1)}{\Gamma(2\Delta+2)},
 \qquad
   \mathsf{S}^{(0)}=\{2,4,6,\dots\}\,.
\end{equation}
The crucial insight for the study of the perturbations of these solutions is to analyze the transformation properties of the various terms in \eqref{f1logandnolog}, \eqref{f1LogandnoLogdef} under the coordinate transformations
\begin{equation}
\chi\,\mapsto\,\frac{\chi}{\chi-1}\,.
\end{equation} 
where $\chi$ is in a neighborhood of zero.
We will use the following identity which are easy to check and generalize to higher values of $\ell$:
\begin{equation}
\label{gdiffeeven}
f^{(\ell)}_{\Delta}(\tfrac{\chi}{\chi-1})+f^{(\ell)}_{\Delta}(\chi)=\begin{cases}
0 & \ell=0\\
-\log(1-\chi)\,f_{\Delta}(\chi) & \ell=1\\
-2\log(1-\chi)\,f^{(1)}_{\Delta}(\chi)+\log(1-\chi)^2\,f_{\Delta}(\chi)  & \ell=2\\
\dots &
\end{cases}
\end{equation}
for $\Delta\in \{2,4,6,\dots\}$ and 
\begin{equation}
\label{gdiffeB2for20}
f_{\mathcal{B}_2}(\tfrac{\chi}{\chi-1})+f_{\mathcal{B}_2}(\chi)=\frac{\chi^2}{\chi-1}\,.
\end{equation}

\subsection{First order perturbation of  \texorpdfstring{$\langle\mathcal{D}_1\mathcal{D}_1\mathcal{D}_1\mathcal{D}_1  \rangle_{(2,2)}$}{D1 4-pt function around (2,2)}}
\label{subsec: first order perturbation}

It is an immediate consequence of \eqref{gdiffeeven}  and \eqref{gdiffeB2for20} that  $f^{(1)}_{\log}(\chi)$ and  $f^{(1)}_{\log^0}(\chi)$ defined in \eqref{f1LogandnoLogdef} satisfy
\begin{equation}\label{f1stransfproperties}
f^{(1)}_{\log}(\tfrac{\chi}{\chi-1})+f^{(1)}_{\log}(\chi)=0\,
\qquad
f^{(1)}_{\log^0}(\tfrac{\chi}{\chi-1})+f^{(1)}_{\log^0}(\chi)+\log(1-\chi)\,f^{(1)}_{\log}(\chi)
=a_{\mathcal{B}_2}^{(1)}\,\frac{\chi^2}{\chi-1}\,.
\end{equation}
Rewriting $f^{(1)}_{\log^0}$ and $f^{(1)}_{\log}$ in terms of $r(\chi)$ and $q(\chi)$, see \eqref{f1sFROMrq}, one notices that the second equation in \eqref{f1stransfproperties} takes the form\footnote{
We use the relation $\log(1-\chi)+\log(1-\chi)^{-1}=0$.
} $A(\chi)+\log(1-\chi) B(\chi)=0$. Assuming that $r(\chi)$ and $q(\chi)$ are rational, which can be justified by looking at the structure of Witten diagrams in the $AdS_2$ dual \cite{Giombi:2017cqn}, one obtains two conditions: $A(\chi)=B(\chi)=0$. 
We have three new equations to be added to the obvious crossing relation for $q(\chi)$ following from the parametrization \eqref{f1form}.
To summarize, we have found the relations
\beq
\begin{split}
\label{randqequations}
r(\tfrac{\chi}{\chi-1})+r(\chi)\,&=\,0\,,
\qquad\quad\qquad
r(\chi)+\chi^2\,r(\tfrac{1}{1-\chi})=\left(\tfrac{\chi}{\chi-1}\right)^2 r(1-\chi)\,,\\
q(\tfrac{\chi}{\chi-1})+q(\chi)\,&=\,a_{\mathcal{B}_2}^{(1)}\,\frac{\chi^2}{\chi-1}\,,
\qquad\quad
q(\chi)+\left(\tfrac{\chi}{\chi-1}\right)^2\,q(1-\chi)\,=\,0\,.
\end{split}
\eeq
The equations \eqref{randqequations} put strong constraints on the functions $r(\chi)$ and $q(\chi)$ but are not powerful enough to specify them uniquely and some sort of boundary conditions need to be imposed. Two  obvious ones follow directly from the   definitions \eqref{f1LogandnoLogdef}, namely
\begin{equation}\label{randqEquation}
r(\chi)=\chi^3 (r_0+\mathcal{O}(\chi))\,,
\qquad
-
\left(\tfrac{\chi}{\chi-1}\right)^2r(1-\chi) \log(1-\chi)
+q(\chi)= -\tfrac{1}{2}\,a_{\mathcal{B}_2}^{(1)}\,\chi^2+\mathcal{O}(\chi)\,.
\end{equation}
Notice that in the second equation there can in principle be cancellations between the contributions from the two factors on the left hand side.
There are additional conditions related to the behaviour of $r(\chi)$ for $\chi$ close to one and these are more subtle. 
They can be translated  to the behavior of $\gamma^{(1)}_{\Delta}$ for large $\Delta$, by the definition of
 $r(\chi)$ in \eqref{f1LogandnoLogdef}, \eqref{f1form}. The intuitive argument is that  by acting with the Casimir operator
 we can increase the order of the pole of $r(\chi)$  for $\chi \sim 1$ at the price of having a more divergent behaviour 
 of  $\gamma^{(1)}_{\Delta}$ at large $\Delta$.
 This follows from the  following relation
\begin{equation}
C_2\, f_{\Delta}(\chi)\,=\,(\Delta+1)(\Delta+2)\, f_{\Delta}(\chi)\,,
\qquad
C_2:=(1-\chi)\partial_{\chi}\chi^2\partial_{\chi}\,.
\end{equation}
The growth of anomalous dimensions is related to local bulk interactions in the $AdS$ dual: the more irrelevant the interaction, the bigger the growth \cite{Heemskerk:2009pn,Fitzpatrick:2010zm}.
Because we are going after the leading correction to the strong coupling behavior, it is natural to choose the solution with the mildest behavior at large $\Delta$. This corresponds to keeping the leading effective vertex in the dual theory. We will therefore 
impose that the behavior at $\chi \sim 1$ is no worse than  $r(\chi) \sim (1-\chi)^{-2}$. 
Under these conditions \eqref{randqEquation} admits a unique solution which of course coincides with the string theory calculation reported above, see \eqref{eq:r_and_q}.
The correction of  CFT data at this order  read\footnote{
It should be noticed that the expression for the anomalous dimensions differs slightly from the one presented in
 \cite{Giombi:2017cqn}.
By looking at the expansion of the superblock \eqref{LongblockExpanded} in bosonic blocks, it is rather clear that the (bosonic) partial wave decomposition we are considering will be degenerate in all R-symmetry channel but the $[0,2]$ one. The latter has a unique representative in each long block. For this reason equation (4.33) in \cite{Giombi:2017cqn} agrees with \eqref{gamma1} while  equation (4.42) there does not. The degeneration between bilinears in the displacement operators is lifted when arranging operators in supermultiplets but there are additional degeneracies which are relevant.
}
  \begin{equation}\label{gamma1}
\gamma^{(1)}_{\Delta}=a^{(1)}_{\mathcal{B}_2}\,\tfrac{\Delta(\Delta+3)}{6}\,,
\qquad
a^{(1)}_{\Delta}=\partial_{\Delta}\left(a_{\Delta}^{(0)}\gamma^{(1)}_{\Delta}\right)\,.
\end{equation}
The results for the anomalous dimensions holds upon assuming that there is no operator degeneracy. This will be further discussed in the end of this section.

\subsection{Second order perturbation of 
\texorpdfstring{$\langle\mathcal{D}_1\mathcal{D}_1\mathcal{D}_1\mathcal{D}_1  \rangle_{(2,2)}$}{D1 4-pt function around (2,2)}}

Let us now consider the second order term in the expansion of the conformal block decomposition
\begin{equation}\label{f2logandnolog}
f^{(2)}(\chi)=f^{(2)}_{\log^2}(\chi)(\log \chi)^2+f^{(2)}_{\log}(\chi)\log \chi+f^{(2)}_{\log^0}(\chi)\,,
\end{equation}
where\footnote{Recall that there are no $\mathcal{C}_{[2,0]}$ appearing in this example.}
\beq
\begin{split}
\label{f2LogandnoLogdef}
f^{(2)}_{\log^2}(\chi)&=\tfrac{1}{2}\sum_{\Delta \in \mathsf{S}^{(0)}} \,a^{(0)}_{\Delta}
\left(\gamma^{(1)}_{\Delta}\right)^2
f_{\Delta}(\chi)\,,\\
\qquad
f^{(2)}_{\log}(\chi)&=
\sum_{\Delta \in \mathsf{S}^{(0)}}
\left(a^{(1)}_{\Delta}\gamma^{(1)}_{\Delta} +a^{(0)}_{\Delta} \gamma^{(2)}_{\Delta}\right)
 f_{\Delta}(\chi)
+
a^{(0)}_{\Delta}\,\left(\gamma^{(1)}_{\Delta}\right)^2\,f^{(1)}_{\Delta}(\chi)
\,,\\
\qquad
f^{(2)}_{\log^0}(\chi)&=
\sum_{\Delta \in \mathsf{S}^{(0)}}
a^{(2)}_{\Delta} f_{\Delta}(\chi)+
\left(a^{(1)}_{\Delta}\gamma^{(1)}_{\Delta} +a^{(0)}_{\Delta}\gamma^{(2)}_{\Delta}\right)
 f^{(1)}_{\Delta}(\chi)
+
\tfrac{1}{2}a^{(0)}_{\Delta}\,\left(\gamma^{(1)}_{\Delta}\right)^2\,f^{(2)}_{\Delta}(\chi)\,.
\end{split}
\eeq
Notice that, without loss of generality corresponding to a redefinition of $\epsilon$,
 we can set $a^{(2)}_{\mathcal{B}_2}=0$.
The contribution of the double logarithm $f^{(2)}_{\log^2}(\chi)$ is expressed in terms of known CFT data\footnote{For this to be the case it is crucial that there is no operator mixing.} and can be resummed in to the rather simple form, see  $a(\chi)$ given in \eqref{abcfomr}.
Next we proceed as in the previous section by considering the transformation properties of $f^{(2)}_{\log^k}(\chi)$ under 
$\chi\mapsto \tfrac{\chi}{\chi-1}$. They follow from   \eqref{gdiffeeven} and the definition  \eqref{f2LogandnoLogdef}:
\beq
\begin{split}
\label{ffSecondord}
f^{(2)}_{\log^2}(\tfrac{\chi}{\chi-1})+f^{(2)}_{\log^2}(\chi)&=0\,,\\
f^{(2)}_{\log^1}(\tfrac{\chi}{\chi-1})+f^{(2)}_{\log^1}(\chi)&=-2\log(1-\chi)\,f^{(2)}_{\log^2}(\chi)\,,\\
f^{(2)}_{\log^0}(\tfrac{\chi}{\chi-1})+f^{(2)}_{\log^0}(\chi)&=-\log(1-\chi)\,f^{(2)}_{\log^1}(\chi)-\log^2(1-\chi)\,f^{(2)}_{\log^2}(\chi)\,.
\end{split}
\eeq
In order to make crossing symmetry manifest we write\footnote{One can also include $\text{Li}_2$ functions in the ansatz, but it turns out that they have to be set to zero in the end.}
\begin{equation}
f^{(2)}(\chi)=\left(a(\chi)\,\log^2 \chi+
b(\chi)\,\log\chi -\left(\tfrac{\chi}{\chi-1}\right)^2( \chi\rightarrow1-\chi)\right)
\,+
c(\chi)\,\log \chi\log (1-\chi)\,,
\end{equation}
where  $c(\chi)$ is crossing symmetric by itself.
The relation between $a(\chi)$, $b(\chi)$, $c(\chi)$ and $f^{(2)}_{\log^k}(\chi)$ is obvious and generalizes  \eqref{f1sFROMrq}.
The relations \eqref{ffSecondord} imply functional relations for $a(\chi)$, $b(\chi)$, $c(\chi)$ by taking the coefficients of the
 $\log^k(1-\chi)$ for $k=0,1,2$.
After imposing  boundary conditions at $\chi=0,1$ for the functions $a(\chi), b(\chi), c(\chi)$ (these boundary conditions follow from similar remarks as in the first order analysis around equation \eqref{randqEquation}), one finds the unique solution to be
\beq
  \begin{split}
  \label{abcfomr}
a(\chi)&=\tfrac{1}{18}\left(\tfrac{\chi}{\chi-1}\right)^3 (2-\chi)(3\chi^2-5 \chi+5)\,a_{\mathcal{B}_2}^{(0)}\,\,,\\
b(\chi)&=\tfrac{1}{36}\tfrac{\chi(\chi-2)}{(\chi-1)^2}(6\chi^2-\chi+1)\,a_{\mathcal{B}_2}^{(0)}\,,\\
c(\chi)&=\tfrac{1}{18}\tfrac{(2\chi-1)}{(\chi-1)^2}(3\chi^4-6\chi^3+3\chi^2+1)\,a_{\mathcal{B}_2}^{(0)}\,.
\end{split}
\eeq
From the above one can extract the CFT data:
\beq
\begin{split}
\label{gamma2}
\gamma^{(2)}_{\Delta}&\,=\,
\tfrac{\Delta(\Delta+3)}{72}
\left(4\Delta-5-\tfrac{2}{\Delta+1}+\tfrac{6}{\Delta+2}+4\,H_{\Delta}\right)\,a_{\mathcal{B}_2}^{(0)}\,\,,
\quad
H_n=\sum_{k=1}^n\tfrac{1}{k}\,,\\
a^{(2)}_{\Delta}\,&=\,
\partial_{\Delta}\Big(a_{\Delta}^{(0)}\gamma_{\Delta}^{(2)}+a_{\Delta}^{(1)}\gamma_{\Delta}^{(1)}\Big)
-\tfrac{1}{2}\partial_{\Delta}^2\Big(a_{\Delta}^{(0)}\left(\gamma_{\Delta}^{(1)}\right)^2\Big)
+a_{\Delta}^{(0)}\,X_{\Delta}\,a_{\mathcal{B}_2}^{(0)}\,,
\end{split}
\eeq
where
  \begin{equation}
X_{\Delta}=\frac{A(u)+(2\Delta+3) B(u)}{72 \Delta (\Delta+1)^2(\Delta+2)^2}+
\frac{u(2-u))}{72}\,\left(\psi^{(1)}(\tfrac{\Delta+1}{2})-\psi^{(1)}(\tfrac{\Delta}{2})\right)\,,
\end{equation}
$\psi^{(n)}$ is the polygamma function of order $n$ and $u=\Delta(\Delta+3)$, $A(u)=-120-36 u+17u^2+5u^3$, $B(u)=8-4u+5u^2+u^3$.

\paragraph{Comparison with the expectations from string theory at second order.}

The results just obtained will most likely differ from the second order perturbation result in string theory. The main reason for the discrepancy is due to degeneracies, as we will show momentarily, see \cite{Alday:2017xua,Aprile:2017bgs} for a recent related discussion.
 We stress once again that with the appropriate modifications to deal with operator mixing, the methods applied in this section can be generalized to this case as well.
 To illustrate the operator mixing let us look at the correlators
\begin{equation}\label{2correlators}
\langle \mathcal{D}_1\mathcal{D}_1\mathcal{D}_1\mathcal{D}_1\rangle_{(2,2)}\,,\qquad
\langle \mathcal{D}_1\mathcal{D}_1\mathcal{D}_2\mathcal{D}_2\rangle_{(2,2)}\,,\qquad
\langle \mathcal{D}_2\mathcal{D}_2\mathcal{D}_2\mathcal{D}_2\rangle_{(2,2)}\,,
\end{equation}  
see \eqref{corrSTRONG} for their explicit expressions and \eqref{eq: expansion of A in superblocks} for their decomposition in  conformal blocks.
Let us start by looking at operators in representation $\mathcal{L}^{\Delta=2}_{[0,0]}$ which are exchanged in all three of the correlators above. We know from \eqref{longs000} that there is no degeneracy of operators with this quantum numbers. Therefore, let us call $\mathcal{O}$ the (normalized) operator sitting in $\mathcal{L}^{\Delta=2}_{[0,0]}$. The fact that there is only one such operator is confirmed a posteriori by using the block decomposition (see \eqref{eq: most general analytic solution Block Expansion}-\eqref{eq: most general analytic solution Block Expansion 3} and \eqref{strongcouplAPPpar}) of the correlators \eqref{2correlators}, which gives the equations:
\beq
\sum_i C_{11\mathcal{O}_i}^2=\tfrac{2}{5}\,,
\quad
\sum_iC_{11\mathcal{O}_i}C_{22\mathcal{O}_i}=\tfrac{4}{5}\,,
\quad
\sum_i C_{22\mathcal{O}_i}^2=\tfrac{8}{5}\,,
\eeq
that have a solution involving just one operator $\mathcal{O}$ with 
$C_{11\mathcal{O}}=\sqrt{\tfrac{2}{5}}$ and $C_{22\mathcal{O}}=2\sqrt{\tfrac{2}{5}}$ (up to an obvious $\mathbb{Z}_2$ ambiguity).
 
Let us turn to operators in the representation $\mathcal{L}^{\Delta=4}_{[0,0]}$. We know from \eqref{longs000} that there are two such operators that we will denote by $\mathcal{O}_{i=1,2}$ with $\langle \mathcal{O}_i\mathcal{O}_j\rangle \sim \delta_{ij}$.
From the known four point functions we know that 
\beq
\begin{split}
&\sum_i\,C_{11\mathcal{O}_i}^2=\tfrac{1}{7}\,,
\quad
\sum_i\,C_{11\mathcal{O}_i}C_{22\mathcal{O}_i}=\tfrac{2}{7}\,,
\quad
\sum_i\,C_{22\mathcal{O}_i}^2=\tfrac{4}{5}\,,\\
&\quad
\Longrightarrow 
\quad
(C_{11\mathcal{O}_1},C_{11\mathcal{O}_2})\sim\,(\sqrt{\tfrac{1}{7}},0)\,,\quad\text{ and }\quad
(C_{22\mathcal{O}_1},C_{22\mathcal{O}_2})\sim\,2(\tfrac{1}{\sqrt{7}},\sqrt{\tfrac{2}{35}})\,,
\end{split}
\eeq
where $\sim$ means up to $O(2)$ rotations. 
The analysis of higher $\Delta$ is similar but requires the knowledge of more correlators.

\paragraph{Comparison with numerics.}
Recall that as  a physical definition of $\epsilon$ we take $a_{\mathcal{B}_2}=2-\epsilon$, which in our notation is equivalent to 
$a^{(1)}_{\mathcal{B}_2}=-1$, $a_{\mathcal{B}_2}^{(n>1)}=0$.
We will now compare the results from the analytic perturbation to the numerical results in the vicinity of the point $(2,2)$
 in figure~\ref{fig:DDBootstrapOPEshort}.
 From the expression for the anomalous dimensions \eqref{gamma1} and  \eqref{gamma2}  for the operator of lowest dimension, i.e.~$\Delta=2$, we can extract 
\begin{equation}\label{aB2analytic}
a_{\mathcal{B}_2}(\Delta_{[0,0]})=2+\frac{3}{5} (\Delta_{[0,0]}-2)-\frac{59}{200}\, (\Delta_{[0,0]}-2)^2+\dots
\end{equation}
Since $\tfrac{3}{5}=0.6$ and $\tfrac{59}{200}=0.295$ this relation is in  good agreement with  \eqref{numericslopSTRONG}.
 Let us comment on the validity of the perturbation at small but finite $\epsilon$.
 At first oder, for any finite $\epsilon$ the unitarity bound will be violated for $\Delta$ large enough, since the anomalous dimensions are negative and are quadratic in $\Delta$, see \eqref{gamma1}.
We can require that $\Delta +\epsilon\,\gamma^{(1)}_{\Delta} \geq 1$ for $\Delta \leq 30$ for example, this gives $ 0 \leq \epsilon \leq 0.17$. In this range $1.72 \leq 2 +\epsilon\,\gamma^{(1)}_{2} \leq 2$. The resulting value for the gap $\Delta_{[0,0]}$ is depicted in figure~\ref{fig:DDBootstrapOPEshort} by the upper purple dashed curve starting from the point $(2,2)$. 
The situation improves quite a bit at second order. In this case for $ 0 \leq \epsilon \leq 0.994$ all OPE coefficients are positive and all the dimensions are above the unitarity bound. In this range one finds that 
$1.69 \leq 2 +\epsilon\,\gamma^{(1)}_{2}+\epsilon^2\,\gamma^{(2)}_{2}  \leq 2$. This rough but reasonable result is drawn as a black dashed curve starting from the point $(2,2)$ in figure~\ref{fig:DDBootstrapOPEshort}.

%% file: sections/conclusions.tex

\section{Conclusions}

Let us now briefly summarize the main points of this article. We implemented the bootstrap for the displacement operator $\calD_1$ and its cousin $\calD_2$ on half-BPS line defects in $4d$ $\calN=4$ SCFTs. Our results include constraining bounds on the conformal dimensions of long operators,
and on the OPE coefficients of short operators. 
While the numerics have not fully converged yet, they suggest a unique solution to crossing, provided that the gap $\Delta_{[0,0]}$ takes its maximal value of 2. This solution can be identified with the strong-coupling behavior of line defects in $\Nm=4$ SYM. Moreover, corrections to the strong-coupling behavior were obtained analytically using only bootstrap reasoning, and they fit nicely with our numerical results.

There are many interesting directions in which to further develop the analysis of this article. 
In this work we focused just on the correlation functions supported on the one-dimensional defect.
The next important step is to impose the additional consistency conditions arising from  coupling this $1d$ CFT to a  four-dimensional  $\mathcal{N}=4$ theory. 
The relevant bootstrap equations are known in this case  \cite{Liendo:2016ymz}, the issue then becomes that one loses positivity and traditional numerical methods cannot be applied. 
Nevertheless, one could use the  alternative  numerical techniques of \cite{Gliozzi:2013ysa}, combined with input from the existence of a topological sector, and an analytical perturbative treatment in the vicinity of some free theories.

Another interesting problem is to study the four-point function of long operators on the defect. The bootstrap problem for long operators has been largely unexplored  due to various technical complications that have to do with the presence of nilpotent invariants in the four point function, see \cite{Cornagliotto:2017dup} for the only example of this type to date. 
Considering these examples has the advantage that one can vary the dimension of the external operator and look for special features in the plots. 
This might shed some light on the interpretation of the drop in the bound on $\Delta_{[0,1]}$ in figure~\ref{fig:Delta20And02vsDelta00Full}.

One could also study line defects in $\calN=2,3$ four-dimensional theories, see for example \cite{Bianchi:2018zpb}, or alternatively in $\calN=4,6,8$ three-dimensional theories, see e.g.~\cite{Bianchi:2017ozk}. While localization results are available for the half-BPS circular Wilson loops even for $\calN=2$ theories (\cite{Pestun:2007rz} provides the fundamental result for circular Wilson loops, see for example \cite{Passerini:2011fe, Mitev:2014yba} for explicit results for some $\calN=2$ SCFTs) in $4d$, there are currently no known results from localization for the loops involving insertions of the displacement supermultiplet. Alternatively, one could also consider  line defects in $\calN=4$ four-dimensional theories supporting less (or no) supersymmetry, see \cite{Polchinski:2011im, Beccaria:2017rbe, Beccaria:2018ocq} for a study of some of these effective $1d$ theories.

One should stress that the bootstrap problem considered in this work  
 is  probably  one of the simplest bootstrap setups on the market, and one could imagine  producing non-trivial solutions to crossing analytically. A particularly interesting solution is the one corresponding to Wilson lines in  planar $\mathcal{N}=4$ SYM, where one could combine bootstrap methods with  integrability  techniques  to determine some of the CFT data.  One could also try to produce relatives of the  SYK model, see e.g.~\cite{Gross:2017vhb}, with  $\text{OSP}(4^*|4)$ symmetry and investigate how they fit in the picture presented in this work.

Finally, as pointed out in \cite{Iliesiu:2018fao}, the study of CFTs at finite temperature shares many similarities with the defect bootstrap program. Hence, we expect that both lines of research will complement each other.

%% file: sections/acknowledgements.tex

\acknowledgments

We have greatly benefited from discussions with F.~Alday, C.~Beem,  M.~Bonini, E.~Casali, L.~Griguolo, A.~Kulanthaivelu, M.~Lemos, J.~Minahan, J.~Penedones, L.~Rastelli, and V.~Schomerus. VM is very grateful to Connor Behan and to David Simmons-Duffin for answering questions regarding the implementation of the SDPB program. 
CM thanks the Galileo Galilei Institute for
Theoretical Physics for the hospitality and the INFN for partial support
during the completion of this work.
 The work of CM is supported in part by grant \#494786 from the Simons Foundation.
The authors gratefully acknowledge the computing time granted on the supercomputer Mogon at Johannes Gutenberg University Mainz (hpc.uni-mainz.de).
We thank the Simons Collaboration on the Non-perturbative Bootstrap
for providing stimulating workshops and conferences.

%% file: sections/appendix.tex

\section{Blockology}
\label{eq: superconformal blocks}

\subsection{The bosonic pieces}

In this section, we shall discuss the $\text{SO}(2,1)$ and $\text{SP}(4)_\RR$ blocks separately before we put them together in the full superblocks in section~\ref{subsec: blocks tables}.

First, we introduce the $1d$ bosonic conformal blocks that we need. They read
\beq
\label{eq: bosonic blocks}
\CB_{\Delta}(\chi)=\chi^{\Delta} \, _2F_1\left(\Delta+\frac{\Delta_2-\Delta_1}{2},\Delta-\frac{\Delta_4-\Delta_3}{2};2 \Delta;\chi\right)\,.
\eeq 
The next ingredient that we need are the $\text{SP}(4)_\RR$ R-symmetry structures.  They can be defined as the eigenfunctions of the quadratic Casimir operator $D_{\varepsilon=1}$ (depending on 3 parameters $a$, $b$ and finally $c=0$ which we omit) of equation (2.3) in \cite{Dolan:2003hv} with $x=\zeta_1$ and $z=\zeta_2$. We can do that since the operator of \cite{Dolan:2003hv} is the quadratic Casimir for the $d=\varepsilon+2=3$ conformal group $\text{SO}(3,2)$, which is $\text{SP}(4)_\RR$ up to reality conditions. Written explicitly, the Casimir operator reads 
\beq
\label{eq: quadratic Casimir for SP4}
\begin{split}
D_{\varepsilon}&\,=\,\zeta_1^2 (1 - \zeta_1) \partial_{\zeta_1}^2+\zeta_2^2 (1 - \zeta_2) \partial_{\zeta_2}^2 - (a + b + 1) \left(\zeta_1^2 \partial_{\zeta_1}+\zeta_2^2 \partial_{\zeta_2}\right)\\&\, - a b (\zeta_1+\zeta_2)+\varepsilon \frac{\zeta_1\zeta_2}{\zeta_1-\zeta_2}\left((1-\zeta_1)\frac{\partial}{\partial \zeta_1}-(1-\zeta_2)\frac{\partial}{\partial \zeta_2}\right)\,,
\end{split}
\eeq
where in our case $a=\frac{k_2-k_1}{2}$ and $b=\frac{k_3-k_4}{2}$ are functions of the R-symmetry labels of the external operators transforming in the of the $[0,k_i]$ representation. 

We first look for the R-symmetry structure in the $\calD_1\times \calD_1$ and $\calD_2\times \calD_2$ OPE channels. 
These R-symmetry structures are polynomial eigenfunctions in $\zeta_i^{-1}$ of the operator $D_{\varepsilon=1}$ with $a=b=c=0$. Up to the polynomial degree that we want, we get the eigenfunctions:
\beq
\label{eq: direct R-symmetry structures}
\begin{split}
&\BS_{[0,0]}^{0,0}=1\,,\qquad \BS_{[2,0]}^{0,0}=1-\frac{1}{\zeta _1}-\frac{1}{\zeta _2}\,,\qquad \BS_{[0,2]}^{0,0}=\frac{3}{10}-\frac{1}{2 \zeta _1}+\frac{1}{\zeta _1 \zeta _2}-\frac{1}{2 \zeta _2}\,,\\
&\BS_{[0,4]}^{0,0}=\frac{5}{126}-\frac{5}{27} \left(\frac{1}{\zeta _2}+\frac{1}{\zeta _1}\right)+\left(\frac{28}{27 }\frac{1}{\zeta _1 \zeta _2}+\frac{1}{6 \zeta _2^2}+\frac{1}{6 \zeta _1^2}\right)-\left(\frac{1}{\zeta _1 \zeta _2^2}+\frac{1}{\zeta _1^2 \zeta _2}\right)+\frac{1}{\zeta _1^2 \zeta _2^2}\,,\\
&\BS_{[4,0]}^{0,0}=-\frac{4}{3} \left(\frac{1}{\zeta _2}+\frac{1}{\zeta _1}\right)+\frac{1}{\zeta _1^2}+\frac{2}{3 }\frac{1}{\zeta _1 \zeta _2}+\frac{1}{\zeta _2^2}+\frac{1}{2}\,,\\
&\BS_{[2,2]}^{0,0}=\frac{5}{7} \left(\frac{1}{\zeta _2}+\frac{1}{\zeta _1}\right)-\frac{1}{2} \left(\frac{4}{\zeta _1 \zeta _2}+\frac{1}{\zeta _2^2}+\frac{1}{\zeta _1^2}\right)+\left(\frac{1}{\zeta _1 \zeta _2^2}+\frac{1}{\zeta _1^2 \zeta _2}\right)-\frac{3}{14}
\end{split}
\eeq
The $D_{\varepsilon=1}$ eigenvalues of $\BS_{[0,0]}^{0,0}$, $\BS_{[2,0]}^{0,0}$, $\BS_{[0,2]}^{0,0}$, $\BS_{[0,4]}^{0,0}$, $\BS_{[4,0]}^{0,0}$, $\BS_{[2,2]}^{0,0}$ are respectively $0, 3, 5, 14 ,8, 10$ and they are normalized such that the coefficient of the term with the highest power of $\zeta_i^{-1}$ is one. These structures are in one to one correspondence with the irreducible representations appearing in the $\text{SP}(4)$ tensor products:
\beq
\label{eq: sp4 tensor product}
\big[0,a\big]\times \big[0,b\big]\,=\,\bigoplus_{i=0}^{\min(a,b)}\bigoplus_{j=0}^i\big[2i-2j,2j+|a-b|\big]\,.
\eeq
Furthermore, in the expansion of the $\block^{1,1}$ superblocks,  we need the eigenfunctions of the operator $D_{\varepsilon=1}$ with $a=\tfrac{1}{2}$, $b=-\tfrac{1}{2}$ and $c=0$. We get the results
\beq
\label{eq: mixed Rsym structures}
\begin{split}
\BS_{[0,1]}^{1,1}&=\frac{1}{\sqrt{\zeta _1} \sqrt{\zeta _2}}\,,\qquad \BS_{[2,1]}^{1,1}=\frac{1}{\zeta _1^{3/2} \sqrt{\zeta _2}}+\frac{1}{\sqrt{\zeta _1} \zeta _2^{3/2}}-\frac{5}{4 \sqrt{\zeta _1} \sqrt{\zeta _2}}\,,\\
\BS_{[0,3]}^{1,1}&=-\frac{2}{3 \zeta _1^{3/2} \sqrt{\zeta _2}}-\frac{2}{3 \sqrt{\zeta _1} \zeta _2^{3/2}}+\frac{1}{\zeta _1^{3/2} \zeta _2^{3/2}}+\frac{10}{21 \sqrt{\zeta _1} \sqrt{\zeta _2}}\,.
\end{split}
\eeq
Finally, for the $\block^{1,-1}$ superblocks, we need instead to use the eigenfunctions for the operator $D_{\varepsilon=1}$ with $a=\tfrac{1}{2}$, $b=\tfrac{1}{2}$ and $c=0$
\beq
\label{eq: mixed Rsym structures 1, -1}
\begin{split}
\BS_{[0,1]}^{1,-1}&=\sqrt{\zeta _2} \sqrt{\zeta _1}-\frac{\sqrt{\zeta _1}}{\sqrt{\zeta _2}}+\frac{1}{\sqrt{\zeta _1} \sqrt{\zeta _2}}-\frac{\sqrt{\zeta _2}}{\sqrt{\zeta _1}}\,,\\
\BS_{[2,1]}^{1,-1}&=-\frac{1}{4} 3 \sqrt{\zeta _2} \sqrt{\zeta _1}+\frac{7 \sqrt{\zeta _1}}{4 \sqrt{\zeta _2}}-\frac{\sqrt{\zeta _1}}{\zeta _2^{3/2}}+\frac{7 \sqrt{\zeta _2}}{4 \sqrt{\zeta _1}}-\frac{11}{4 \sqrt{\zeta _1} \sqrt{\zeta _2}}+\frac{1}{\zeta _1^{3/2} \sqrt{\zeta _2}}-\frac{\sqrt{\zeta _2}}{\zeta _1^{3/2}}+\frac{1}{\sqrt{\zeta _1} \zeta _2^{3/2}}\,,\\
\BS_{[0,3]}^{1,-1}&=\frac{1}{7} \sqrt{\zeta _2} \sqrt{\zeta _1}-\frac{10 \sqrt{\zeta _1}}{21 \sqrt{\zeta _2}}+\frac{\sqrt{\zeta _1}}{3 \zeta _2^{3/2}}-\frac{10 \sqrt{\zeta _2}}{21 \sqrt{\zeta _1}}+\frac{38}{21 \sqrt{\zeta _1} \sqrt{\zeta _2}}\\&-\frac{4}{3 \zeta _1^{3/2} \sqrt{\zeta _2}}+\frac{\sqrt{\zeta _2}}{3 \zeta _1^{3/2}}-\frac{4}{3 \sqrt{\zeta _1} \zeta _2^{3/2}}+\frac{1}{\zeta _1^{3/2} \zeta _2^{3/2}}\,.
\end{split}
\eeq
The eigenvalues of the R-symmetry structures $\BS_{[0,1]}^{m,n}$, $\BS_{[2,1]}^{m,n}$ and $\BS_{[0,3]}^{m,n}$ in \eqref{eq: mixed Rsym structures} and \eqref{eq: mixed Rsym structures 1, -1} are $2$, $6$ and $9$ respectively.

\subsection{Explicit superblocks} 
\label{subsec: blocks tables}

Armed with the bosonic conformal blocks and the R-symmetry structures, we can obtain the full superblocks by making an ansatz of the type
 \beq
 \label{eq: superblock ansatz}
\block^{m,n}_{\calO}(\chi,\zeta_1,\zeta_2)\,=\,\sum_{h=\Delta}^{\Delta+\delta}\sum_R c_{h,R} \CB_{h}(\chi) \BS^{m,n}_R(\zeta_1,\zeta_2)\,,\qquad \{m,n\}=\big\{\{0,0\},\{1,1\}, \{1,-1\} \big\}\,,
\eeq
where the conformal blocks $\CB_{h}(\chi)$ have the correct external dimensions $\Delta_i$ plugged in them (this depends on $m$ and $n$, see \eqref{eq: bosonic blocks}) and the sum over $R$ runs over the appropriate structures for the channel. The prescription of which block indices $m,n$ to use for which function $\calA$ is summarized in \eqref{eq: expansion of A in superblocks}. The constants $c_{h,R}$ are determined by feeding the ansatz \eqref{eq: superblock ansatz} into the superconformal Ward identities with the coefficient of lower conformal dimension bosonic block normalized to one, or to minus one. The correct sign was determined by expanding the analytic solutions in their unitary domain in superblocks using positive coefficients.

The difference between the ``short" $\mathcal{B}_\ell$ , ``semi-short" $\mathcal{C}_{[a,b]}$ and ``long"  $\mathcal{L}^{\Delta}_{[a,b]}$ superblocks is the difference $\delta$ in conformal dimension between the lowest bosonic block and the highest bosonic block appearing in the decomposition of a superblock. Specifically,  we have
\beq
\begin{split}
\text{short}\ :\  &g^{1d}_{\Delta,\tilde{\Delta}}\,, \ldots \,, g^{1d}_{\Delta+2,\tilde{\Delta}} \qquad (\delta=2)\,,\\
\text{semi-short}\ :\  &g^{1d}_{\Delta,\tilde{\Delta}}\,, \ldots \,,   g^{1d}_{\Delta+3,\tilde{\Delta}} \qquad (\delta=3)\,,\\
\text{long}\ :\  &g^{1d}_{\Delta,\tilde{\Delta}}\,, \ldots \,,  g^{1d}_{\Delta+4,\tilde{\Delta}} \qquad (\delta=4)\,.
\end{split}
\eeq
Once the superconformal blocks $\block^{a,b}_{\calO}$ have been determined, we can extract the corresponding functions $f_{\calO,i}^{a,b}$ and constants $F^{a,b}_{\calO}$. First, given a function $\block(\chi,\zeta_1,\zeta_2)$, we define the following functions in $\chi$:
\beq
\textbf{E}[\block]_{m,n}\,\equiv\,\left(\frac{\partial^m\partial^n}{\partial\zeta _1^m\partial\zeta_2^n}\frac{\block(\chi,\zeta_1,\zeta_2)}{\sX^2}\right)_{\big| \zeta_1=\zeta_2=\chi}\,.
\eeq
Then, using these functions, we obtain
\begin{align}
\label{eq: extracting f and F}
&F_{\calO}^{0,0}=\textbf{E}[ \block^{0,0}_{\calO}]_{0,0}= \block^{0,0}_{\calO}(\chi,\chi,\chi)\,,&  &f_{\calO,1}^{0,0}=\frac{1}{2}\textbf{E}[\block^{0,0}_{\calO}]_{0,2}\,,&\nonumber\\
&f_{\calO,2}^{0,0}=2\textbf{E}[ \block^{0,0}_{\calO}]_{0,1}+\left(\frac{1}{2}-\chi\right)\textbf{E}[ \block^{0,0}_{\calO}]_{0,2}\,,&
& f_{\calO,3}^{0,0}=\frac{\textbf{E}[ \block^{0,0}_{\calO}]_{2,2}}{4}-\frac{\textbf{E}[ \block^{0,0}_{\calO}]_{1,2}+ \partial_\chi \textbf{E}[ \block^{0,0}_{\calO}]_{0,2}}{2\chi} \,,&\nonumber\\
&F_{\calO}^{1,1}=\block^{1,1}_{\calO}(\chi,\chi,\chi)\,,& &f_{\calO}^{1,1}=\frac{\chi}{2}\textbf{E}\left[ \frac{1}{\sqrt{\sX}}\block^{1,1}_{\calO}\right]_{0,2}\,,&\\
& F_{\calO}^{1,-1}=\block^{1,-1}_{\calO}(\chi,\chi,\chi)\,. & &f_{\calO}^{1,-1}=\frac{\chi}{2}\textbf{E}\left[ \frac{\sXt}{\sqrt{\sX}}\block^{1,-1}_{\calO}\right]_{0,2}\,,&\nonumber
\end{align}
In the above, it is very important that the blocks satisfy the superconformal Ward identities.
Conversely, given the functions \eqref{eq: extracting f and F}, we can reconstruct the full superblocks by using 
\begin{align}
\label{eq: irreducible functions parametrization}
\block^{0,0}_{\calO}&=\mathbb{D}_1 \left[\chi^2 f_{1;\calO}^{0,0}\right]+\mathbb{D}_2\left[\frac{\chi}{2}\left(f_{2;\calO}^{0,0}-f_{1;\calO}^{0,0}\right)\right]+\mathbb{D}_3 \left[\chi \left(\chi^2 f_{3;\calO}^{0,0}+f_{1;\calO}^{0,0}+\chi \partial_{\chi }f_{1;\calO}^{0,0}\right)\right]+\sX^2\,F^{0,0}_{\calO}\,,\nonumber\\
\block^{1,1}_{\calO}&=\sqrt{\sX}\left[\mathbb{D} \left(\chi^2 f_{\calO}^{1,1}\right)+\sX\,F^{1,1}_{\calO}\right]\,,\\
\block^{1,-1}_{\calO}&=\frac{\sqrt{\sX}}{\sXt}\left[\mathbb{D} \left(\chi^2 f_{\calO}^{1,-1}\right)+\sX\,F^{1,-1}_{\calO}\right]\,,\nonumber
\end{align}
where (the factor $\mathbb{D}_3  \,=\,\frac{\chi  \left(\chi -\zeta _1\right){}^2 \left(\chi -\zeta _2\right){}^2}{\zeta _1^2 \zeta _2^2}$ is a normalization)
\beq
\label{eq: differential operators mathbbD and others}
\begin{split}
\mathbb{D} & \,=\,\left(2\chi^{-1}-\zeta_1^{-1}-\zeta_2^{-1}\right)-\chi ^2 \left(\zeta_1^{-1}-\chi^{-1}\right) \left(\zeta_2^{-1}-\chi^{-1}\right) \frac{\partial }{\partial \chi }\,,\\
\mathbb{D}_1 & \,=\,\chi^2\left(\chi^{-2}-\zeta_1^{-1}\zeta_2^{-1}\right)-\chi^3  \left(\zeta _1^{-1} -\chi^{-1}\right) \left(\zeta _2^{-1} -\chi^{-1}\right)\frac{\partial }{\partial \chi } \,,\\
\mathbb{D}_2 & \,=\, \frac{\chi^4}{\zeta_1\zeta_2}\left(\chi^{-2}-\zeta_1^{-1}\zeta_2^{-1}\right)-\frac{\chi^5}{\zeta_1\zeta_2}  \left(\zeta _1^{-1} -\chi^{-1}\right) \left(\zeta _2^{-1} -\chi^{-1}\right)\frac{\partial }{\partial \chi }\,.
\end{split}
\eeq
Hence, we having the explicit superblocks $\block^{a,b}_{\calO}$ is equivalent to having the functions $f_{\calO,i}^{a,b}$ and constants $F^{a,b}_{\calO}$. The constants $F^{a,b}_{\calO}$ are easy to list, for they are equal to one for the short operators $\calO=\mathcal{B}_{\ell}$ and are zero otherwise. The remaining functions, as well as the explicit superblocks are listed in an auxiliary \textsf{Mathematica} file named ``SuperBlocksResults.nb".

\paragraph{Different notation for the $\calD_1$ system.} It is convenient when discussing the full mixed system to write the blocks $\block^{0,0}_{\calO}$ appearing in the expansion of $\calA_{\{1,1,1,1\}}$ and $\calA_{\{2,2,2,2\}}$ in the same way, namely as in \eqref{eq: irreducible functions parametrization}. Since $\calA_{\{1,1,1,1\}}$ is only quadratic in $\zeta_i^{-1}$, it is possible to also write it in a simpler way as in \eqref{eq:WI_solution}. Writing the superblocks appearing in the decomposition of $\calA_{\{1,1,1,1\}}$ (and only them!) as 
$\block^{0,0}_{\calO}=\sX F_{\calO} + \mathbb{D}f_{\calO}(\chi)$ and equating them to the expression in \eqref{eq: irreducible functions parametrization}, we get
\beq
\label{eq:map_mixed}
F^{0,0}_{\calO}=F_{\calO}\,,\qquad f_{1;\calO}^{0,0}=\frac{f_\calO(\chi )}{\chi ^3}\,,\qquad f_{2;\calO}^{0,0}=\frac{f_\calO(\chi )+2 F_\calO \chi ^2}{\chi ^3}\,,\qquad f_{3;\calO}^{0,0}=\frac{2 f_\calO(\chi )-\chi  f_\calO'(\chi )}{\chi ^5}\,.
\eeq
This dictionary allows one to translate from the \eqref{eq:WI_solution} notation to the \eqref{eq: irreducible functions parametrization} easily as required. 

\section{Comments on the derivation of the crossing equations}
\label{appendix: Comments on the derivation of the crossing equations}

This appendix contains comments relative to the derivation of the  ``irreducible" crossing equation \eqref{eq: final crossing equations 1} starting from \eqref{eq: nice crossing equations for calA1111} and \eqref{eq: full not irreducible crossing equations}. 

The first part involving the irreducible crossing equations for $\calA_{\{1,1,1,1\}}$ is already written in \eqref{eq: direct crossing 1}. Then, the first crossing equation of \eqref{eq: full not irreducible crossing equations} deals with $\calA_{\{2,2,2,2\}}$. It must be decomposed into the six R-symmetry structures of \eqref{eq: direct R-symmetry structures}, giving six equations that are not independent. They are satisfied iff the following three equations are satisfied:
\beq
\label{eq: direct crossing 2}
\begin{split}
\sum_{\calO\in \calD_2\times \calD_2}C_{2,2,\calO}^2\left(\begin{array}{c} \, [f_{1,\calO}^{0,0}]_a \\ \,[f_{2,\calO}^{0,0}]_s\\\, [f_{3,\calO}^{0,0}]_a \end{array}\right)&=0\,,
\end{split}
\eeq
where we refer to \eqref{eq: def of fsa} for the definition of $[f]_{s/a}$. 
Note that all the structure constants in the direct channel are \textit{real}. Hence, all the coefficients appearing in the decomposition are positive. Applying the same procedure to the remaining crossing equations leads to 
\beq
\label{eq: irreducible 1212 and 2121 equations}
\sum_{\tilde{\calO}\in \calD_1\times \calD_2} C_{12\tilde{\calO}}^2 \left[f_{\calO}^{1,1}\right]_s=\sum_{\tilde{\calO}\in \calD_1\times \calD_2} (C_{12\tilde{\calO}}^*)^2 \left[f_{\calO}^{1,1}\right]_s=0\,.
\eeq
for the $\calA_{\{1,2,1,2\}}$ equation (the complex conjugate one is for $\calA_{\{2,1,2,1\}}$) and to 
\beq
\label{eq: irreducible 1122 to 1221 equations}
\sum_{\calO\in \calD_1\times \calD_1} C_{11\calO}C_{22\calO} \left(\begin{array}{c}F_{\calO}^{0,0}\\\chi f^{0,0}_{1,\calO}(\chi)\end{array}\right)+\sum_{\tilde{\calO}\in \calD_1\times \calD_2} |C_{12\tilde{\calO}}|^2 \left(\begin{array}{c}-F_{\tilde{\calO}}^{1,-1}\\f^{1,-1}_{\tilde{\calO}}(1-\chi)\end{array}\right)=0\,,
\eeq
for the one relating $\calA_{\{1,1,2,2\}}$ to $\calA_{\{1,2,2,1\}}$. 
The first line in \eqref{eq: irreducible 1122 to 1221 equations} is the minibootstrap equation \eqref{eq: mixed equations minibootstrap} since $F_{\calO}^{0,0}=1=F_{\tilde{\calO}}^{1,-1}$ if $\calO/\tilde{\calO}$ are short and is zero otherwise. The minibootstrap equation is solved as $C_{1,2,3}^2=1+C_{1,1,2}C_{2,2,2}-C_{1,1,2}^2$, thus eliminating $C_{1,2,3}^2$ out of the game.

We can rewrite the second line of \eqref{eq: irreducible 1122 to 1221 equations} together with \eqref{eq: irreducible 1212 and 2121 equations} as a system of equations in the variables $a_{\tilde{\calO}}\equiv \text{Re}\,C_{12\tilde{\calO}}$ and $b_{\tilde{\calO}}\equiv \text{Im}\,C_{12\tilde{\calO}}$. Then $ C_{12\tilde{\calO}}^2=a_{\tilde{\calO}}^2-b_{\tilde{\calO}}^2+2i a_{\tilde{\calO}}b_{\tilde{\calO}}$ and $(C_{12\tilde{\calO}}^*)^2=a_{\tilde{\calO}}^2-b_{\tilde{\calO}}^2-2i a_{\tilde{\calO}}b_{\tilde{\calO}}$ Thus, taking the real and imaginary part of the two equations of \eqref{eq: irreducible 1212 and 2121 equations}, we get since the blocks are real the equations
\beq
\label{eq: A1212 and A2121 crossing rewritten}
\begin{split}
\sum_{\tilde{\calO}\in \calD_1\times \calD_2} \left(\begin{array}{cc} a_{\tilde{\calO}} & b_{\tilde{\calO}} \end{array} \right) \left(\begin{array}{cc}\left[f_{\tilde{\calO}}^{1,1}\right]_s & 0\\0 & -\left[f_{\tilde{\calO}}^{1,1}\right]_s\end{array}\right)\left(\begin{array}{c} a_{\tilde{\calO}} \\ b_{\tilde{\calO}} \end{array} \right)\,=\,&0\,,\\\sum_{\tilde{\calO}\in \calD_1\times \calD_2} \left(\begin{array}{cc} a_{\tilde{\calO}} & b_{\tilde{\calO}} \end{array} \right) \left(\begin{array}{cc}0 & \left[f_{\tilde{\calO}}^{1,1}\right]_s \\\left[f_{\tilde{\calO}}^{1,1}\right]_s & 0\end{array}\right)\left(\begin{array}{c} a_{\tilde{\calO}} \\ b_{\tilde{\calO}} \end{array} \right)\,=\,&0\,.
\end{split}
\eeq
The remaining ones \eqref{eq: irreducible 1122 to 1221 equations} can also be rewritten in a similar way, where we also use 
\beq
\label{eq: A1221 and A1122 rewritten}
\begin{split}
&0=\sum C_{11\calO}C_{22\calO} \chi f^{0,0}_{1,\calO}(\chi)+ \sum |C_{12\tilde{\calO}}|^2 f^{1,-1}_{\tilde{\calO}}(1-\chi)
\\&\Leftrightarrow \left\{\begin{array}{l}0=\sum C_{11\calO}C_{22\calO}[\chi f^{0,0}_{1,\calO}(\chi)]_s+\sum |C_{12\tilde{\calO}}|^2 [f^{1,-1}_{\tilde{\calO}}(\chi)]_s\\0=\sum C_{11\calO}C_{22\calO}[\chi f^{0,0}_{1,\calO}(\chi)]_a-\sum |C_{12\tilde{\calO}}|^2 [f^{1,-1}_{\tilde{\calO}}(\chi)]_a\end{array}\right.\,,
\end{split}
\eeq
in order to decouple the even from the odd parts of the equation. Combining all the crossing equations into one then directly leads to \eqref{eq: final crossing equations 1} in the main text.

\section{The analytic solutions to the crossing equations}
\label{app: analytic solutions}
It is easy to produce simple four point functions by taking linear combinations of products of super-propagators  defined in \eqref{twoPOINT} and imposing that conformal weights and the relevant permutation symmetry are reproduced correctly.  Below we present such four-point functions together with their conformal block decomposition. 
Knowing these simple solutions is useful when exploring the parameter space of all solutions to crossing. 

\paragraph{The separate analytic solutions. }
 The most general analytic solutions to the crossing equations produced by using  the 
 super-propagators \eqref{twoPOINT} lead to the following $\calA$ functions:
\beq
\label{eq: most general analytic solution}
\begin{split}
\calA_{\{1,1,1,1\}}^{\text{analytic}}&=1+\frac{\sX}{\sXt}+\xi\sX\,,\\
\calA_{\{2,2,2,2\}}^{\text{analytic}}&=1+\left(\frac{\sX}{\sXt}\right)^2 +\xi_1' \sX^2+\xi_2'\frac{ \sX}{\sXt}+\xi_3' \left(\frac{\sX^2}{\sXt}+\sX\right)\,,\\
\calA_{\{1,1,2,2\}}^{\text{analytic}}&=\calA_{\{2,2,1,1\}}^{\text{analytic}}=1+\upsilon _1 \sX+\frac{\upsilon _2 \sX}{\sXt}\,,\\
\calA_{\{1,2,1,2\}}^{\text{analytic}}&=\calA_{\{2,1,2,1\}}^{\text{analytic}}=\sX^{3/2} \left(\upsilon_1'+\frac{\upsilon_2'}{\sX}+\frac{\upsilon_2'}{\sXt}\right)\,,\\
\calA_{\{1,2,2,1\}}^{\text{analytic}}&=\frac{\sqrt{\sX}}{\sXt} \left(\upsilon _2+\upsilon _1 \sX+\frac{\sX}{\sXt}\right)\,,
\end{split}
\eeq
where we have used the shorthands \eqref{eq: definition of sX} and the $\xi$, $\xi'_i$, $\upsilon_i$ and $\upsilon'_i$ are a-priori free parameters that are subject to unitarity and to identifications coming from comparing different block decompositions.

One can expand the solutions in superblocks (we remind that $\block^{0,0}_{\mathcal{I}}=1$). One finds 
\beq
\label{eq: most general analytic solution Block Expansion}
\begin{split}
&\calA_{\{1,1,1,1\}}^{\text{analytic}}=
1+(1+\xi) \block^{0,0}_{\mathcal{B}_2}+\frac{1-\xi}{2} \block^{0,0}_{\mathcal{C}_{[2,0]}} +\sum _{\Delta =2}^{\infty} \frac{\sqrt{\pi } (\Delta -1) \Gamma (\Delta +3)}{2^{2 \Delta +1} \Gamma \left(\Delta +\frac{3}{2}\right)}\frac{1+(-1)^{\Delta } \xi}{2}\block^{0,0}_{\mathcal{L}^{\Delta}_{[0,0]}}\,,\\
& \calA_{\{2,2,2,2\}}^{\text{analytic}}=1 +\left(\xi'_2+\xi'_3\right) \block^{0,0}_{\mathcal{B}_2}+\left(1+\xi'_1+\xi'_3\right)\block^{0,0}_{\mathcal{B}_4}
+\frac{\xi'_2-\xi'_3}{2} \block^{0,0}_{\mathcal{C}_{[2,0]}}
+\left(1-\xi'_1\right)\block^{0,0}_{\mathcal{C}_{[2,2]}} \\
&+\frac{2+2 \xi'_1-\xi'_3}{6}\block^{0,0}_{\mathcal{C}_{[4,0]}}+\sum _{\Delta =2}^{\infty}\frac{ (\Delta -3) (\Delta -2) (\Delta -1) (\Delta +5) (\Delta +6) (5)_{\Delta -2}}{225\times 4^{\Delta-1} \left(\frac{7}{2}\right)_{\Delta -2}}\Bigg[1+(-1)^{\Delta } \xi'_1\\&+\frac{180 \xi'_2}{(\Delta -3) (\Delta -2) (\Delta +5) (\Delta +6)} +\frac{36 \left(\Delta ^2+3 \Delta +6 (-1)^{\Delta } (\Delta  (\Delta +3)-5)-10\right) \xi'_3}{(\Delta -3) (\Delta -2) (\Delta -1) (\Delta +4) (\Delta +5) (\Delta +6)} \Bigg]\block^{0,0}_{\mathcal{L}^{\Delta}_{[0,0]}}\\
&+\sum _{\Delta =4}^{\infty}\frac{(\Delta -3) \Delta  (\Delta +1) (9)_{\Delta -4}}{27\times 4^{\Delta -4} \left(\frac{11}{2}\right)_{\Delta -4}} \Bigg[1+(-1)^{\Delta } \xi'_1+\frac{36 \left(1+(-1)^{\Delta }\right) }{(\Delta +2) (\Delta +4) \left(\Delta ^2-1\right)}\xi'_3\Bigg]\block^{0,0}_{\mathcal{L}^{\Delta}_{[0,2]}}\\
&+\sum _{\Delta =4}^{\infty}\frac{ (\Delta -3) \Delta  (\Delta +5) \left(\Delta ^2+3 \Delta -4\right) (7)_{\Delta -4}}{189\times 4^{\Delta-2} \left(\frac{11}{2}\right)_{\Delta -4}} \Bigg[ 1+ (-1)^{\Delta +1} \xi'_1\\&+\frac{36 \left((-1)^{\Delta }-1\right) \xi'_3}{(\Delta -2) \Delta  (\Delta +3) (\Delta +5)}\Bigg] \block^{0,0}_{\mathcal{L}^{\Delta}_{[2,0]}}\,,
\end{split}
\eeq
for the first two functions. The last remaining function that is expanded in the direct channel blocks is 
\beq
\label{eq: most general analytic solution Block Expansion 2}
\begin{split}
\calA_{\{1,1,2,2\}}^{\text{analytic}}=&1+\left(\upsilon _1+\upsilon _2\right) \block_{\mathcal{B}_2}^{0,0}+\frac{\upsilon _2-\upsilon _1}{2}\block^{0,0}_{\mathcal{C}_{[2,0]}} \\&+\sum _{\Delta =2}^{\infty} \frac{\sqrt{\pi } (\Delta -1) \Gamma (\Delta +3)}{4^{\Delta +1}\Gamma \left(\Delta +\frac{3}{2}\right)} \left((-1)^{\Delta } \upsilon _1+\upsilon _2\right)\block^{0,0}_{\mathcal{L}^{\Delta}_{[0,0]}}\,.
\end{split}
\eeq
Finally, for the mixed correlation functions, one obtains the block decomposition
\beq
\label{eq: most general analytic solution Block Expansion 3}
 \begin{split}
& \calA_{\{1,2,1,2\}}^{\text{analytic}}=\upsilon_2'\,\block^{1,1}_{\mathcal{B}_1}+(\upsilon_1'+\upsilon_2') \block^{1,1}_{\mathcal{B}_3}+\frac{1}{3} (\upsilon_2'-2 \upsilon_1')\block^{1,1}_{\mathcal{C}_{[2,1]}}\\
&+\sum _{\Delta =3}^{\infty} \frac{(\Delta -2) (\Delta +2) (\Delta +3) (4)_{\Delta -3} \left(\upsilon_2'-\frac{1}{6} (-1)^{\Delta -2} (\Delta +1) (\Delta +2) \upsilon_1'\right) }{35\times 2^{2 \Delta -3}  \left(\frac{9}{2}\right)_{\Delta -3}}\block^{1,1}_{\mathcal{L}^{\Delta}_{[0,1]}}\,,\\
& \calA_{\{1,2,2,1\}}^{\text{analytic}}=\upsilon _2\, \block^{1,-1}_{\mathcal{B}_1} +\left(\upsilon _1+1\right) \block^{1,-1}_{\mathcal{B}_3}+\left(\frac{2}{3}-\frac{\upsilon _1}{3}\right)  \block^{1,-1}_{\mathcal{C}_{[2,1]}}\\
&+\sum _{\Delta =3}^{\infty} \frac{(\Delta -2) (\Delta +2) (\Delta +3) (4)_{\Delta -3}\left(\frac{1}{6} (\Delta +1) (\Delta +2)-(-1)^{\Delta } \upsilon _1\right) }{35\times 2^{2 \Delta-3 }  \left(\frac{9}{2}\right)_{\Delta -3}}\block^{1,-1}_{\mathcal{L}^{\Delta}_{[0,1]}}\,.
 \end{split}
 \eeq
 
\paragraph{The solutions taken together. } We can take the solutions \eqref{eq: most general analytic solution} as together describing a mixed $\calD_1$, $\calD_2$ system of correlation functions. In so doing, some of the parameters become identified since the structure constants such as $C_{1,2,3}$ appearing in different channels have to agree. The solution to all the constraints is to reparametrize \eqref{eq: most general analytic solution} through 
\begin{align}
\label{eq: analytic solution reparametrization}
&\upsilon _1= \eta ^2-1\,,& &\upsilon_1'= \eta ^2-\xi -1\,,& &\upsilon _2=\upsilon_2'= 1+\xi \,,&\nonumber\\
&\xi'_1= 1-\omega_3\,,& & \xi'_2= \frac{\left(\eta ^2+\xi \right)^2}{\xi +1}+2\omega_1-3\omega_2-\omega_3+2\,,& & \xi'_3= 3\omega_2+\omega_3-2\,.&
\end{align}
Some OPE coefficients then read
\begin{align}
\label{eq: OPEcoefficients for the analytic solutions}
&C_{1,1,2}=\sqrt{1+\xi }\,,& &C_{1,1,S^{(2)}_{[2,0]}}^2=\frac{1-\xi}{2}&  &C_{1,2,3}=\eta \,,&\nonumber\\
&C_{2,2,2}=\frac{\eta ^2+\xi }{\sqrt{1+\xi}}\,,& &C_{2,2,2'}^2=2\omega_1\,,& & C_{2,2,4}^2=3\omega_2\,,& \\ &C_{2,2,S^{(2)}_{[2,0]}}^2=2+\frac{C_{2,2,2}^2}{2}+\omega_1-3 \omega_2- \omega_3\,, &  & C_{2,2,S^{(4)}_{[4,0]}}^2=\frac{2-\omega_2-\omega_3}{2}& & C_{2,2,S^{(4)}_{[2,2]}}^2=\omega_3\,,&\nonumber
\end{align}
together with $C_{1,2,S^{(3)}_{[2,1]}}^2=1-\frac{\eta^2}{3}$. In the above, we've used the identity \eqref{eq: long block at the unitarity bound} for the semi-short blocks.

It follows that  $\xi \in \{-1,1\}$ due to unitarity and that $\eta, \omega_1, \omega_2$ and $\omega_3$ have to be positive. There are also other positivity conditions due to unitarity. For example, from comparing $\calA^{\text{analytic}}_{\{1,2,1,2\}}$ to $\calA^{\text{analytic}}_{\{1,2,2,1\}}$, we find a constraint on $\eta$. Summarizing: 
\beq
\label{eq: allowed range xi and eta}
-1\leq \xi\leq 1\,,\qquad 0\leq \sqrt{\max(0,3\xi)}\stackrel{!}{\leq} \eta \stackrel{!}{\leq} \sqrt{2+\xi}\leq \sqrt{3}\,.
\eeq
The conditions on the parameters $\omega_i$ are more annoying to state and we omit them since they are not needed.

The \textit{leading order strong coupling solution} given in  \eqref{corrSTRONG} corresponds to 
\eqref{eq: most general analytic solution} with the reparametrization \eqref{eq: analytic solution reparametrization}
and
\begin{equation}
\label{strongcouplAPPpar}
\xi=1\,,
\qquad
\eta^2=3\,,
\qquad
 \omega_1= \omega_3=0\,,
 \qquad
 \omega_2=2\,.
\end{equation}
Notice that for these values of the parameters there are no long at unitarity bound in the conformal block decomposition, i.e.~$\mathcal{C}$-type multiplets appearing in the OPE decomposition, see \eqref{eq: OPEcoefficients for the analytic solutions}.

We remark that the analytic solutions \eqref{eq: most general analytic solution} with the reparametrization \eqref{eq: analytic solution reparametrization} contain the most peculiar unitary solution, namely on with $\calA^{\text{analytic}}_{\{1,2,1,2\}}=0$. This corresponds to $\xi=-1$ and $\eta=0$ and leads to the explosion of the upper bound on $\Delta_{[0,1]}$ for small values of $\Delta_{[0,0]}$, see figure~\ref{Fig:OPEMinMaxD1D1WithLongAtBound12}. For that solution, the value of $C_{2,2,2}$ diverges.

\begin{figure}[htbp!]
             \begin{center}       
              \includegraphics[scale=0.55]{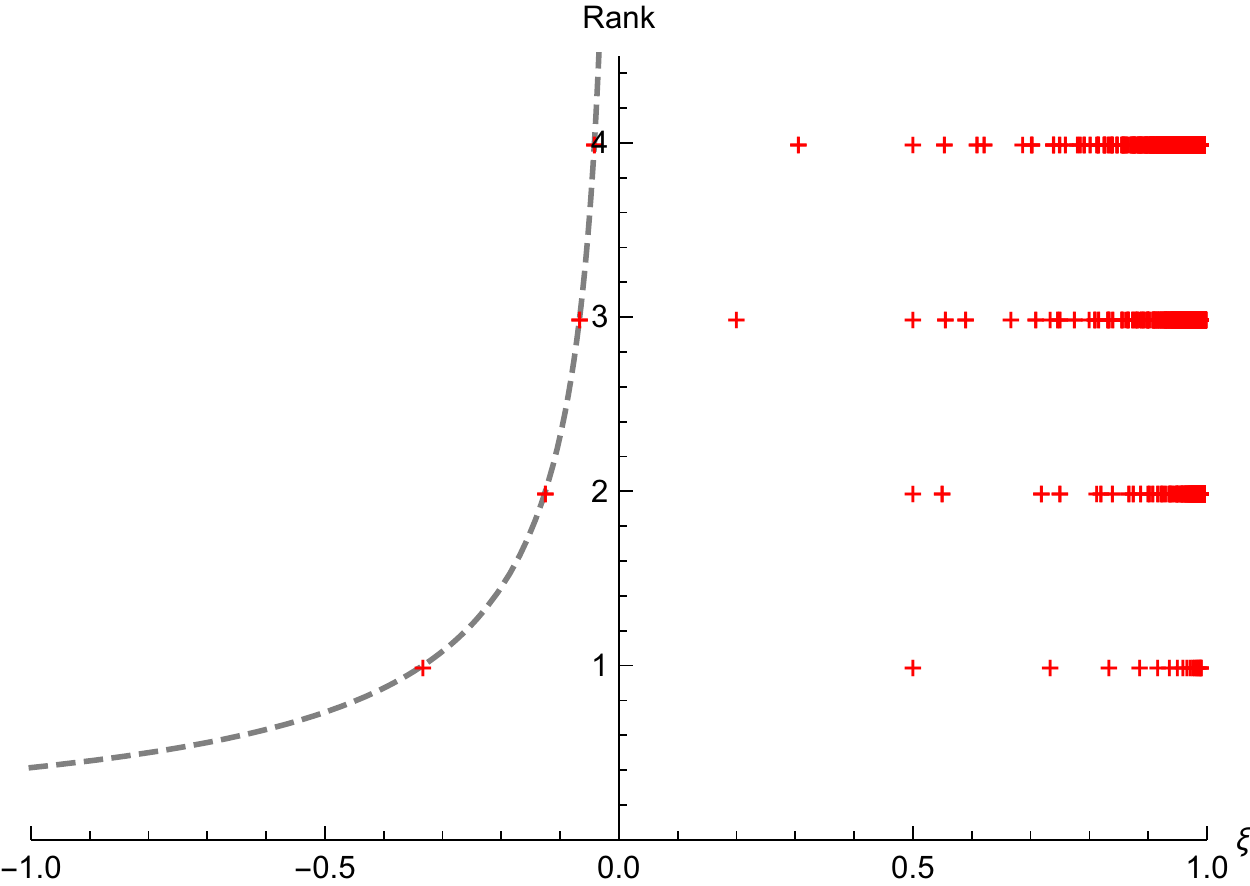}\qquad\quad    \includegraphics[scale=0.55]{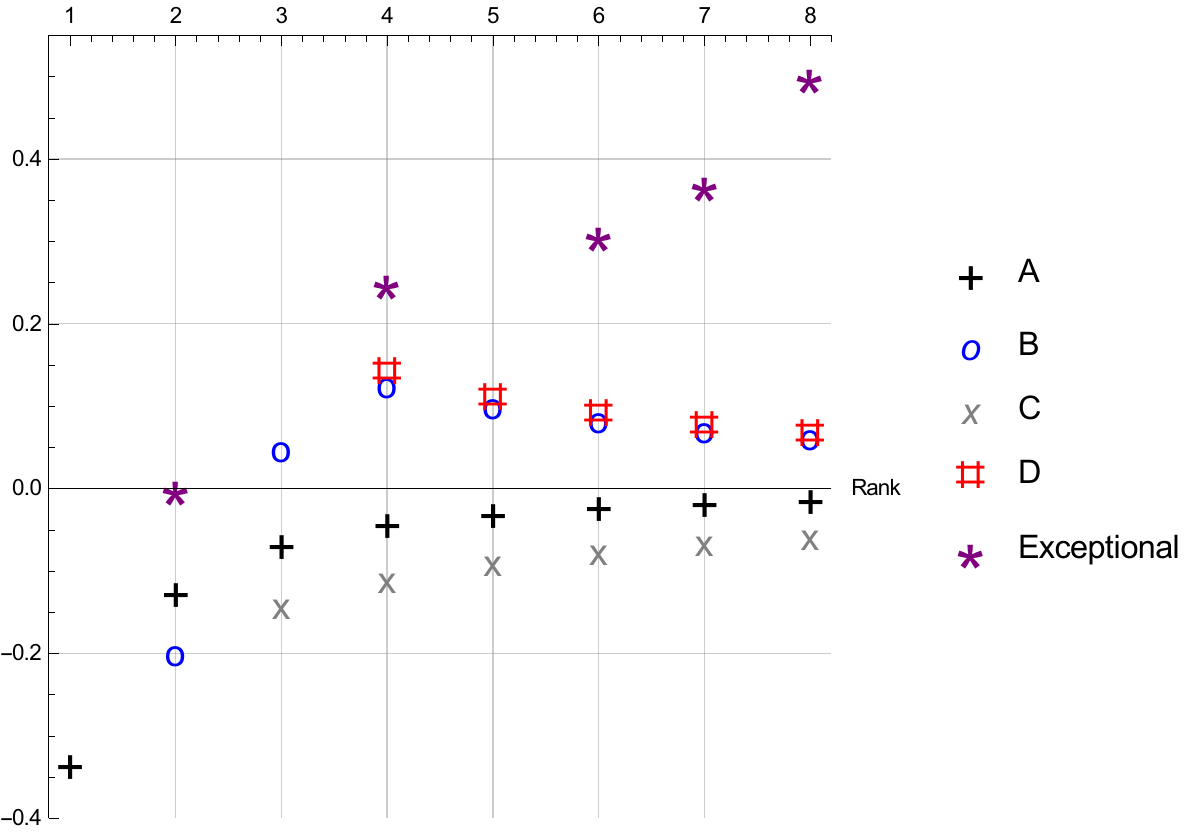}
              \caption{Left: We show the possible values of $\xi$ in the free theories with gauge groups $\SU(N)$. The minimal value $\xi=-(N^2-1)^{-1}$ is plotted in a dashed curve. Right: Minimal value of $\xi$ as a function of the  rank for various types of gauge groups. We remind of the identities: $B_2=C_2$, $A_3=D_3$.  Since $A_1=$SU(2)=SP(2) and $B_2=C_2$=SP(4), the minimal value is saturated by the symplectic groups.}
              \label{fig:xiRank}
            \end{center}
\end{figure}

\paragraph{Free gauge theory solutions.} We can connect the solution $\calA^{\text{analytic}}_{\{1,1,1,1\}}$ in \eqref{eq: most general analytic solution} to free gauge theory. It is obvious that in a free gauge theory, the normalized 4-pt function is
\beq
\vac{\calD_1(1)\calD_1(2) \calD_1(3) \calD_1(4)}=(12)(34)+(14)(23)+\xi (13)(24)\,,
\eeq
with the parameter $\xi$ given by ($\kappa_{ab}$ is the Killing form and the $T^a$ are appropriately normalized generators of the algebra)
\beq
\xi=\frac{\kappa_{ac}\kappa_{bd}\text{Tr}_\rep(T^a T^b T^c T^d)}{\kappa_{ab}\kappa_{cd}\text{Tr}_R(T^aT^bT^cT^d)}=1-\frac{1}{2} \frac{\text{Cas}_2(\text{Adj})}{\text{Cas}_2(\rep)}\,.
\eeq
One can use the index of a representation $\text{Ind}(\rep)$ to write $\text{Cas}_2(\rep)=\frac{\text{Ind}(\rep)}{\text{dim}(\rep)} \text{dim}(\text{Adj})$. Then a program such as  LieART \cite{Feger:2012bs} permits to compute $\xi$ for various algebras and representations and to in particular to find the minimal value of $\xi$. 
For example, for $\SU(N)$ and $\rep$ the fundamental representation, we get $\xi=-(N^2-1)^{-1}$. We show some allowed values in figure~\ref{fig:xiRank}. The smallest possible value of $\xi$ that we obtain is for the fundamental representation of $\SU(2)$, for which $\xi=-\tfrac{1}{3}$.

\begin{figure}[htbp!]
             \begin{center}       
              \includegraphics[scale=0.3]{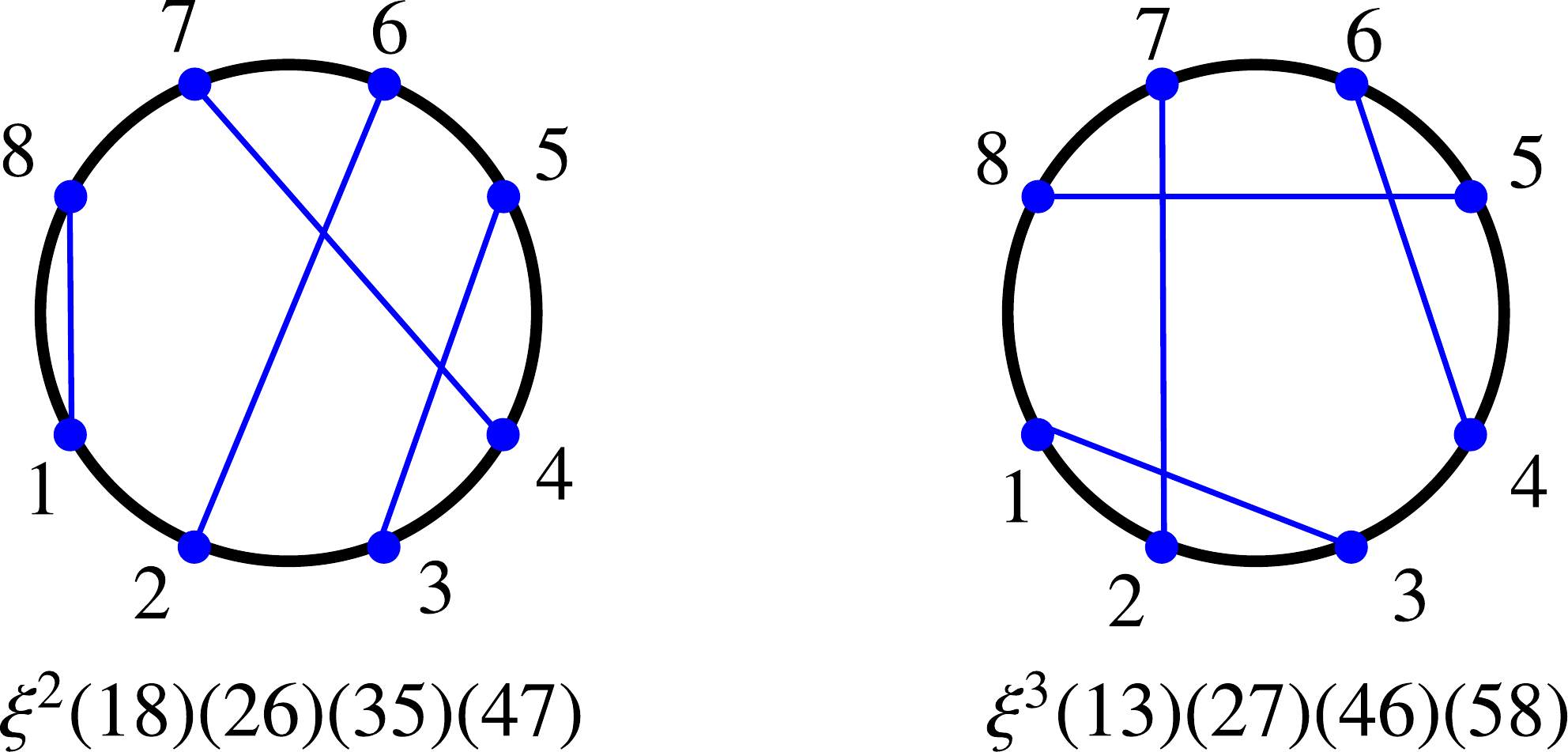}
              \caption{Examples of contributions to the 8-pt correlation function of $\calD_1$ using the modified Wick contraction rule depending on a parameter $\xi$. }
              \label{fig:modifiedWick}
            \end{center}
\end{figure}

\paragraph{A special solution with modified Wick contractions. }
We can obtain a special solution with $C_{2,2,2'}=0$ in which the multiplicity of the $\calD_\ell$ operators is equal to one. In this theory, we define $\calD_\ell=\tfrac{1}{\sqrt{\ell!}}:\calD_1^\ell:$ and compute the correlation function using a modified Wick contraction prescription. Specifically, we add one factor of the parameter $\xi$ to each crossing of the contraction lines when the operators are drawn on a circle, as shown in figure~\ref{fig:modifiedWick}.
The resulting solution has the OPE coefficients
\beq
\label{eq: OPE for the modified Wick prescription}
\begin{split}
&C_{1,1,2}=\sqrt{1+\xi}\,,\qquad C_{2,2,2}=(1+\xi)^{3/2}\,,\qquad C_{1,2,3}=\frac{\sqrt{1+\xi} \left(1+\xi+\xi ^2\right)}{\sqrt{1+2\xi+2\xi^2+\xi ^3}}\,,\\& C_{2,2,2'}=0\,,\qquad
C_{2,2,4}=\frac{(\xi +1) \left(\xi ^2+1\right) \left(\xi ^2+\xi +1\right)}{\sqrt{(\xi +1)^2 \left(\xi ^4+\xi ^3+2 \xi ^2+\xi +1\right)}}\,,
\end{split}
\eeq
and is unitary for all $\xi\in (-1,1]$. The point $\xi=-1$ technically cannot be included since in this case no operator of type $\mathcal{B}_2$  appears in the OPE of $\calD_1\times \calD_1$ and hence we cannot define $\calD_2$ as $\tfrac{1}{\sqrt{2}}:\calD_1^2:$. Nevertheless, we can get arbitrarily close to it. On the RHS of figure~\ref{fig:C112C222BoundsVaryDelta} this solution corresponds to a curve interpolating between the point $(0,0)$ and the point $(\sqrt{2},2\sqrt{2})$ as $\xi$ varies between $-1$ and $1$.

\section{First order perturbation of 
\texorpdfstring{$\langle\mathcal{D}_1\mathcal{D}_1\mathcal{D}_1\mathcal{D}_1  \rangle_{(1,0)}$}{D1 4-pt function around (1,0)}}
\label{app:10Perturbation}

In this appendix, we want to perform a first order perturbation like in section~\ref{subsec: first order perturbation} but these time starting from the point $(\Delta_{[0,0]},C_{1,1,2}^2)=(1,0)$ of figure~\ref{fig:DDBootstrapOPEshort}, which corresponds to $\xi=-1$ in \eqref{eq: most general analytic solution}. Specifically, the solution of crossing corresponding to the point $(1,0)$ is given by the free Wick contraction
\begin{equation}\label{4pt10}
\langle\mathcal{D}_1\mathcal{D}_1\mathcal{D}_1\mathcal{D}_1  \rangle_{(1,0)}=
(12)(34)-(13)(24)+(14)(23)\,,
\end{equation}
where  $(ij)$ denotes the super-propagator defined in \eqref{twoPOINT}. 
In the parametrization \eqref{eq:WI_solution}  this corresponds to $f^{(0)}=\tfrac{\chi(2\chi-1)}{\chi-1}$ and $F^{(0)}=1$
 from which one extracts the CFT data:
 \begin{equation}
(1,0):\qquad a^{(0)}_{\mathcal{B}_{2}}=0,\quad a^{(0)}_{\mathcal{C}_{[2,0]}}=1,\quad 
a^{(0)}_{\Delta}=\tfrac{\Gamma(\Delta+3)\Gamma(\Delta+1)(\Delta-1)}{\Gamma(2\Delta+2)},\qquad
  \mathsf{S}^{(0)}=\{3,5,7,\dots\}\,.
\end{equation}
In this case we will make use of the following identities:
\beq
\begin{split}
\label{gdiffeC}
f_{\mathcal{C}_{[2,0]}}(\tfrac{\chi}{\chi-1})-f_{\mathcal{C}_{[2,0]}}(\chi)&=0\,,\\
f^{(1)}_{\mathcal{C}_{[2,0]}}(\tfrac{\chi}{\chi-1})-f^{(1)}_{\mathcal{C}_{[2,0]}}(\chi)&=\log(1-\chi)\,f_{\mathcal{C}_{[2,0]}}(\chi)\,,
\end{split}
\eeq
together with
\begin{equation}
\label{gdiffe}
f^{(\ell)}_{\Delta}(\tfrac{\chi}{\chi-1})-f^{(\ell)}_{\Delta}(\chi)=\begin{cases}
0 & \ell=0\\
\log(1-\chi)\,f_{\Delta}(\chi) & \ell=1\\
2\log(1-\chi)\,f^{(1)}_{\Delta}(\chi)+\log(1-\chi)^2\,f_{\Delta}(\chi)  & \ell=2\\
\dots &
\end{cases}
\end{equation}
for $\Delta\in \{3,5,\dots\}$ and 
\begin{equation}
\label{gdiffeB2odd}
f_{\mathcal{B}_{2}}(\tfrac{\chi}{\chi-1})-f_{\mathcal{B}_{2}}(\chi)=h^{(1)}_{\mathcal{B}_{2}}(\chi)+ h^{(2)}_{\mathcal{B}_{2}}(\chi)\,\log(1-\chi)\,,
\end{equation}
with $h^{(1)}_{\mathcal{B}_{2}}(\chi)=\tfrac{(2-\chi)}{(\chi-1)\chi}(\chi^2+6\chi-6)$,  $h^{(2)}_{\mathcal{B}_{2}}(\chi)=\tfrac{12(1-\chi)}{\chi^2}$.
Using these identities, we can write the analogue of  \eqref{f1stransfproperties} as
\begin{equation}\label{f1stransfproperties10}
f^{(1)}_{\log}(\tfrac{\chi}{\chi-1})=f^{(1)}_{\log}(\chi)\,,
\quad
f^{(1)}_{\log^0}(\tfrac{\chi}{\chi-1})-f^{(1)}_{\log^0}(\chi)-\log(1-\chi)\,f^{(1)}_{\log}(\chi)
=a_{\mathcal{B}_{2}}^{(1)}\,h_{\mathcal{B}_{2}}(\chi)\,.
\end{equation}
Using the parametrization \eqref{f1form} and the same argument as in section~\ref{subsec: first order perturbation} we obtain the system of equations
\beq
\label{randqequations}
\begin{split}
r(\tfrac{\chi}{\chi-1})-r(\chi)\,&=\,0\,,
\qquad\quad\qquad
r(\chi)-\chi^2\,r(\tfrac{1}{1-\chi})=\left(\tfrac{\chi}{\chi-1}\right)^2 r(1-\chi)-a_{\mathcal{B}_{2}}^{(1)}\,h^{(2)}_{\mathcal{B}_{2}}(\chi)\,,\\
q(\tfrac{\chi}{\chi-1})-q(\chi)\,&=\,-a_{\mathcal{B}_{2}}^{(1)}\,h^{(1)}_{\mathcal{B}_{2}}(\chi)\,,
\qquad\quad
q(\chi)+\left(\tfrac{\chi}{\chi-1}\right)^2\,q(1-\chi)\,=\,0\,.
\end{split}
\eeq
The only rational solution to this system of equations appears to be
\beq
\begin{split}
r(\chi)&=R(\tfrac{\chi^2}{\chi-1})\,,
\qquad\qquad\qquad\quad\,\,\,
R(t)=6 a^{(1)}_{\mathcal{B}_{2}}\,t^2 (t^2 -5t +5)\,,\\
q(\chi)&=\tfrac{\chi(2\chi-1)}{\chi-1}\,Q(\chi(1-\chi))\,,
\qquad 
Q(t)=a^{(1)}_{\mathcal{B}_{2}}\,t^{-2} (3t^3+t^2+9t -6)\,,
\end{split}
\eeq
which implies that 
\begin{equation}
\gamma^{(1)}_{\Delta}=-\tfrac{1}{24}\, a^{(1)}_{\mathcal{B}_{2}}\,\prod_{k=-1}^{4}(\Delta+k)\,,
\qquad
 \gamma^{(1)}_{\mathcal{C}_{[2,0]}}=0
 \qquad
 a_{\mathcal{C}_{[2,0]}}^{(1)}=-5\, a^{(1)}_{\mathcal{B}_{2}}\,.
\end{equation}
Due to the fast growth of the anomalous dimension with $\Delta$, this perturbation seems reliable only for 
$ a^{(1)}_{\mathcal{B}_{2}}\sim10^{-6}$. 
As previously discussed, this region is hard to probe numerically and the $\Lambda =\infty$ rough extrapolation is still far away for $\Delta_{[0,0]}$ close to one.
As the gap $\Delta_{[0,0]}=1+\varepsilon$ and $\gamma_{\mathcal{C}_{[2,0]}}^{(1)}+\dots=1+\varepsilon \,0+\dots$,
the analysis just performed suggests that 
\begin{equation}
a_{\mathcal{B}_{2}}(\Delta_{[0,0]})\sim \sqrt{\Delta_{[0,0]}-1}+\dots \,,\qquad
\Delta_{[0,0]}\sim1\,.
\end{equation}
Thus, we expect the lower bound curve of figure~\ref{fig:DDBootstrapOPEshort} to follow a square root rather than a power law behavior in the vicinity of the point $(\Delta_{[0,0]},C_{1,1,2}^2)=(1,0)$.